\documentclass{article}
\usepackage[utf8]{inputenc}
\usepackage[margin=1in]{geometry}

\usepackage{microtype}
\usepackage{graphicx}
\usepackage{subfigure}
\usepackage{booktabs}
\usepackage{natbib}
\setcitestyle{open={(},close={)}}

\usepackage[colorlinks=true,citecolor=blue]{hyperref}

\usepackage{amsmath,amsthm,amsfonts,amssymb,mathdots,array,mathrsfs,bm,bbm,stmaryrd,graphicx,subfigure,xcolor}
\usepackage{algorithm,algorithmic}
\usepackage{breakcites}

\usepackage[T1]{fontenc}
\usepackage{enumerate}
\usepackage{inputenc}

\usepackage{graphicx} 
\usepackage{subfigure}

\usepackage{booktabs,balance}
\usepackage{rotating}
\usepackage{boldline}
\usepackage{makecell}
\usepackage{multirow}
\usepackage{balance}

\usepackage{tikz}

\newtheorem{theorem}{Theorem}[section]
\newtheorem{corollary}[theorem]{Corollary}
\newtheorem{lemma}[theorem]{Lemma}

\newtheorem{example}{Example}[section]
\newtheorem{remark}{Remark}[section]

\DeclareMathOperator*{\argmin}{argmin}

\newcommand{\llam}{\boldsymbol \lambda}

\newcommand{\bsmat}{\left[\begin{smallmatrix} }
	\newcommand{\esmat}{\end{smallmatrix}\right] }

\newcommand{\xx}{\mathbf x}
\newcommand{\bb}{\mathbf b}
\newcommand{\yy}{\mathbf y}

\usepackage{environ}
\NewEnviron{smallequation}{%
    \begin{equation}
    \scalebox{0.97}{$\BODY$}
    \end{equation}
    }
    \NewEnviron{smallalign}{%
    \begin{equation}
    \scalebox{0.97}{$\BODY$}
    \end{equation}
    }

\begin{document}

\title{\bf \huge Regression with Label Permutation in Generalized Linear Model }

\author{\vspace{0.5in}\\
\textbf{Guanhua Fang} and \textbf{Ping Li} \\\\
Cognitive Computing Lab\\
Baidu Research\\
10900 NE 8th St. Bellevue, WA 98004, USA\\\\
  \texttt{\{fanggh2018,\ pingli98\}@gmail.com}\\\\
}
\date{}
\maketitle

\begin{abstract}\vspace{0.2in}
\noindent\footnote{The initial version of this paper was submitted in May 2021.}The assumption that response and predictor belong to the same statistical unit may be violated in practice. Unbiased estimation and recovery of true label ordering based on unlabeled data are challenging tasks and have attracted increasing attentions in the recent literature. In this paper, we present a relatively complete analysis of label permutation problem for the generalized linear model with multivariate responses. The theory is established under different scenarios, with knowledge of true parameters, with partial knowledge of underlying label permutation matrix and without any knowledge. Our results remove the stringent conditions required by the current literature and are further extended to the missing observation setting which has never been considered in the field of label permutation problem. On computational side, we propose two methods, ``maximum likelihood estimation'' algorithm and ``two-step estimation'' algorithm, to accommodate for different settings. When the proportion of permuted labels is moderate, both methods work effectively. Multiple numerical experiments are provided and corroborate our theoretical findings.
\end{abstract}

\newpage

\section{Introduction}
A key assumption in regression problems is that response-predictor pairs correspond to the same statistical unit. In practice, this assumption may be violated when different subsets of variables are collected asynchronously and are merged together with certain label disagreements.
That is, responses and predictors may not be perfectly paired together so that the statistical inferences based on such label-contaminated data sets could be inaccurate and biased.
Research on the unlabeled problem has a long history and can be traced back to 1970s under the
name “broken sample problem”~\citep{degroot1971matchmaking, goel1975re, degroot1976matching,degroot1980estimation, chan2001file, bai2005broken, Slawski2020two}.
In recent years, we
have witnessed a renaissance of this problem due to its wide applications, such as data integration,
privacy protection, computer vision, robotics, sensor networks, etc.
See~\cite{unnikrishnan2015unlabeled, pananjady2017linear,pananjady2017denoising,slawski2019linear,zhang2019permutation,Slawski2020two,zhang2022benefits}
and the references therein for more explanations.

Important applications of label permutation problem include linkage record, data de-anonymization, and
header-free communication. In linkage record~\citep{newcombe1962record,fellegi1969theory}, people would like to integrate multiple databases,
where each contains different pieces of information about the same identity, into one comprehensive
database. In this process, the biggest challenge is how to find the matching across different databases.
For data de-anonymization~\citep{nazarov2018sparse}, the task is to identify
the labels, which aims to preserve privacy, with public data. It can be seen as the inverse problem of privacy protection.
For the header-free communication~\citep{pananjady2017denoising, shi2019learning}, we have a sensor network where the sensor identity is omitted during communication to reduce the transmission cost and latency. In this scenario, reconstruction of signal involves recovering the unknown correspondence.

In the literature, for the sake of simplicity, linear models are often considered for studying the label permutation problem.
However, the linear model assumption may be violated when the error distribution is skewed or heavy tailed.
Additionally, linear models cannot capture the structure of count data which is another popular data type and is becoming increasingly ubiquitous (e.g. survival analysis~\citep{fleming2011counting}, online streaming services~\citep{cugola2012processing}, educational testing~\citep{templin2010diagnostic}, etc).
With these in mind, in this paper, we specifically adopt the formulation of generalized linear model (GLM) to take care of multivariate non-Gaussian responses.
The problem is formulated as follows,
\begin{eqnarray}
Y = \Pi^{\sharp} Y^{\sharp},
\end{eqnarray}
where $Y^{\sharp}$ is the response matrix when the data are labeled correctly. For each entry of $Y^{\sharp}$,  $Y^{\sharp}[i,l]$ admits the density function
\[f_{il}(y) = \exp\{y \lambda_{il} - \psi(\lambda_{il}) + c(y)\}\]
for $i \in [n]$ and $l \in [m]$ with $\lambda_{il} = \xx_i^T \bb_l^{\sharp}$.
Here, $n$ is the number of units/individuals, $m$ is the number of observations for each unit/individual, $\lambda_{il}$ refers to the natural parameter. $c(y)$ is a nuisance function free of parameter.
Unobserved $\Pi^{\sharp} \in \mathbb R^{n \times n}$ denotes an underlying row permutation matrix.
In other words, we only observe the response matrix up to certain label permutations.
$X := (\xx_i) \in \mathbb R^{n \times p}$ represents the covariate/design matrix which is fully observed; $B^{\sharp} := (\bb_l^{\sharp}) \in \mathbb R^{p \times m}$ is the underlying true parameter coefficient matrix, which may be unknown. Our task is to recover the label permutation matrix $\Pi^{\sharp}$ based on permuted data $Y$ and design matrix $X$.

\vspace{0.1in}
\textbf{Related work.} There is a rapidly growing body of literature on regression problems with unknown label permutation,
starting from~\citealp{unnikrishnan2015unlabeled, unnikrishnan2018unlabeled, pananjady2017linear}. Paper~\citep{pananjady2017linear} presents necessary and sufficient conditions for
permutation recovery for linear models with Gaussian design. Extensions to multivariate linear models are considered in~\cite{pananjady2017denoising, zhang2019permutation, zhang2020optimal, zhang2022benefits}. The papers~\citep{abid2017linear, hsu2017linear} show that consistent estimation of
the regression parameter is impossible without substantial additional assumptions.~\cite{tsakiris2018algebraic, tsakiris2018eigenspace} have studied important theoretical aspects such as well-posedness from an
algebraic perspective, and have also put forth practical computational schemes such as a branch-and-bound algorithm~\citep{emiya2014compressed} and concave maximization~\citep{peng2020linear}. An approximate EM scheme
with a Markov-Chain-Monte-Carlo (MCMC) approximation of the E-step is discussed in~\cite{abid2018stochastic}.
Additionally, approaches to linear and multivariate linear regression with sparsely mismatched data are studied in~\cite{slawski2019linear, Slawski2020two,slawski2021pseudo, slawski2020sparse}.
A tight analyses on sparse regression problem is provided in~\cite{zhang2021sparse}.
Nevertheless, on the other hand, a relatively small amount of papers have considered regression with unlabeled/permuted data outside the standard linear model. The topics of those papers include spherical regression~\citep{shi2020spherical}, univariate isotonic regression and statistical seriation~\citep{carpentier2016learning,rigollet2019uncoupled,flammarion2019optimal,ma2020optimal,balabdaoui2021unlinked}, and binary regression~\citep{wang2018signal}.

The most related work is~\cite{wang2020estimation}, where they consider a generalized linear regression models within exponential family.
The difference is that they only consider the case $m = 1$ and the corresponding theoretical analyses are established when the true parameter $B^{\sharp}$ is assumed to be known.
The case $m > 1$ should be of independent interest for the following reasons. First,
in the context of record linkage, it is natural to assume that both predictor matrix $X$ and response $Y$ are multi-dimensional. Second, the availability of multiple responses affected by the same permutation is expected to facilitate estimation.
Stronger assumptions are required for label recovery when $m = 1$, while such assumptions can be relaxed when $m$ grows as $n$ grows. In addition, the estimation problem is more challenging when the true parameter $B^{\sharp}$ is unknown.
The theory under unknown parameter settings is waiting to be developed. To be reader-friendly, Table~\ref{tab:review} summarizes the key differences of our setting from the existing ones.

\begin{table}[h!]
\vspace{0.1in}
	\centering
	\small
	\begin{tabular}{ c|c|c}
		\hline
		\hline
		& Linear model & GLM \\
		\hline
		Uni-dim ($m=1$)  &~\cite{pananjady2017linear},~\cite{unnikrishnan2018unlabeled}  & ~\cite{wang2020estimation}  \\
		\hline
		Multi-dim ($m>1$)  &~\cite{zhang2019permutation, zhang2020optimal},~\cite{Slawski2020two} &  Ours \\
		\hline
		\hline
	\end{tabular}
	\caption{A summary of literature review.}\label{tab:review}\vspace{0.1in}
\end{table}

\textbf{Contributions.} In this paper, we study the label permutation problem under generalized linear model framework which is different from the classical linear model in the following ways. The response could be discrete instead of continuous such that the resulting estimator does not admit a nice closed form. In the linear model, the signal-to-noise ratio (SNR) plays an important role in recovering the underlying labels. In contrast, there is no such unified criteria in the generalized linear model.
The existing work of~\cite{wang2020estimation} for generalized linear model only considers the case of $m = 1$. When the regression parameter $B$ is assumed to be known, the sufficient condition for label permutation recovery requires
$\min_{1 \leq i_1 \neq i_2 \leq n}|\psi'(\lambda_{i_1}) - \psi'(\lambda_{i_2})|$ grows as $n$ grows.
There are no theoretical guarantees under the situation $m > 1$ or the situation when regression parameter $B$ is unknown.
In this paper, we bridge this gap and show the perfect label recovery results under different scenarios.
Moreover, we also consider the situation of missing observations, i.e., each entry in response $Y$ may be missing completely at random with certain probability.
The corresponding theory has also been established in this paper.
Such missing observation case has not been studied yet in the literature for unlabeled regression problems.
On the technical side, we also want to point out the analysis of generalized linear models is harder than that of linear models due to the existence of exponential function whose second order derivative may not be bounded. Additionally, we need to take special care of analyzing the maximum likelihood estimator which admits no closed form and involves $n!$ different possible permutations.
The results in current paper add theoretical values in many applications, e.g., data integration, privacy protection, etc.

\vspace{0.1in}

\textbf{Outline.}~ The rest of paper is organized as follows. Section~\ref{sec:backgound} introduces the generalized linear model setting and useful notations.
Section~\ref{sec:known} presents the permutation recovery analysis when parameters are assumed to be known.
In Section~\ref{sec:unknown}, the main theory is established when parameters are unknown and underlying permutation matrix is partially known or unknown.
We further extend the results to the missing observation setting in Section~\ref{sec:miss}.
Simulation experiments are provided in Section~\ref{sec:sim} to support our theoretical findings.
Finally, the concluding remarks are given in Section~\ref{sec:disscussion}.

\newpage

\section{Permutation Problem} \label{sec:backgound}

\subsection{Toy example}

An illustrative example is shown in Table~\ref{tab:toy}. On the left side, it presents the personal information (i.e. salary and age) of five individuals in the correct label order.
On the right side, information of salary and age is coupled in a wrong label order due to the privacy reasons or errors caused by merging data from different sources. Obviously, salary and age are \textbf{impossible} to follow Gaussian distributions. Our goal is to recover the true label order given the permuted data set and under certain model assumptions in the framework of generalized linear model.

\begin{table}[ht]
	\centering
	\caption{A toy example of label permutation problem. }
	\label{tab:toy}
	\begin{tabular}{ccc|cc}
		\hline
		\hline
		\multicolumn{3}{c}{Original} & \multicolumn{2}{c}{Permuted} \\
		\hline
		Label & Salary &  Age &  Salary (Label) &  Age (Label) \\
		\hline
		1 & 6500 & 50 & 6500 (1) & 45 (3)  \\
		2 & 4300 & 30 & 5000 (3) & 30 (2)  \\
		3 & 5000 & 45 & 3200 (4) & 50 (1)  \\
		4 & 3200 & 25 & 4300 (2) & 25 (4)  \\
		5 & 8000 & 55 & 8000 (5) & 55 (5)   \\
		\hline
		\hline
	\end{tabular}
\end{table}

\subsection{Model}

In this paper, we specifically study the generalized linear model with label permutation. The problem can be written in the following matrix form,
\begin{eqnarray}
Y = \Pi^{\sharp} Y^{\sharp},
\end{eqnarray}
where $Y^{\sharp} \sim f(X B^{\sharp})$, i.e., $Y^{\sharp}[i,l] \sim f_{il}(y)$ with
\[f_{il}(y) = \exp\{y \lambda_{il} - \psi(\lambda_{il}) + c(y)\}, ~ \lambda_{il} = \mathbf x_i^T \mathbf b_l^{\sharp}\] for $i \in [n]$ and $l \in [m]$.
Unobserved $\Pi^{\sharp} \in \mathbb R^{n \times n}$ denotes an underlying row permutation matrix (i.e. a binary matrix with each row/column containing one and only one non-zero entry), $X \in \mathbb R^{n \times p}$ represents the covariate/design matrix, and $B^{\sharp} \in \mathbb R^{p \times m}$ is the underlying true parameter coefficient matrix.
Here $\psi(\cdot)$ is a smooth uni-variate convex function over $\mathbb R$.
When we take function $\psi(x) = x^2$, then the density is a normal density.
When we take $\psi(x) = \exp\{x\}$, then it becomes a Poisson distribution.
Our goal is to recover the underlying permutation matrix $\Pi^{\sharp}$ given mislabeled observations $Y$ and $X$.

For a fixed permutation $\Pi$, the log-likelihood function after removing the nuisance parts (the term not related to the parameters) is given by
\begin{eqnarray}
L(\Pi,B) = \sum_{i=1}^n \sum_{l = 1}^m \bigg\{ Y[i,l] (\xx_{\Pi(i)}^T \bb_l) - \psi(\xx_{\Pi(i)}^T \bb_l) \bigg\}
= \langle - \psi(\Pi X B) + Y \circ \Pi X B \rangle \label{eq:likelihood}.
\end{eqnarray}
In the rest of paper, we consider to recover $\Pi^{\sharp}$ by maximizing the above log-likelihood function. When the true parameter matrix $B^{\sharp}$ is known, the estimator will be
$\hat \Pi := \arg\max_{\Pi} L(\Pi, B^{\sharp})$.
When the parameter matrix $B$ is unknown, the estimator will be
\[ \hat \Pi := \arg\max_{\Pi} \max_{B} L(\Pi, B).\]

\newpage

\textbf{Notation.}
We use $\sharp$ to denote the true value and  use $a \gtrsim b ~ (a \lesssim b)$ to represent $a \geq K b ~ (a \leq b/K)$ for some sufficiently large constant $K$.
$\|\mathbf a\|$ and $\|A\|$ represent $\ell_2$ norm of vector $\mathbf a$ and spectral norm of matrix $A$, respectively.
For any $\mathcal S$, its cardinality is denoted by $|\mathcal S|$.
For two positive real numbers $a$ and $b$,  $b = O(a), b = \Omega(a)$ and $b = \Theta(a)$ indicate the relations, $b \leq C_2 a$, $b \geq a/C_1$ and $a/C_1 \leq b \leq C_2 a$, correspondingly, where $C_1, C_2$ are some constants.
For random sequences, $x_n = o_p(1)$ means $x_n$ converges to $0$ in probability and $x_n = O_p(y_n)$ means $x_n/y_n$ is stochastically bounded as $n \rightarrow \infty$.
For an arbitrary univariate function $f$, $f(\mathbf a)$ and $f(A)$ are obtained via applying $f$ to vector $\mathbf a$ and matrix $A$ elementwisely.
We let $\psi'(\lambda)$ and $\psi''(\lambda)$ be the first and second order derivative of $\psi(\lambda)$.
A more complete notation list is given in Table~\ref{tab:notation}.

\begin{table}[t]
	\centering
	\caption{Notation List.\vspace{0.1in} }
	\label{tab:notation}
	{
		\begin{tabular}{ll}
			\hline
			\hline
			Notation &  Definition \\
			\hline
			$n$ & the number of individuals  \\
			$m$ & the number of observations for each individual   \\
			$p$ & the number of covariates/predictors \\
			$X$ & the covariate/design matrix ($n$ by $p$)  \\
			$Y$ & the observed/response matrix ($n$ by $m$) \\
			$\yy_i$ & the response vector for individual $i$ \\
			$B$ & the coefficient/parameter matrix  \\
			$\xx_i^T$ & the $i$th row of $X$ \\
			$\bb_l$ & the $l$th column of $B$  \\
			$\Pi$ & the row permutation matrix ($n$ by $n$)\\
			$\mathbf I$ & the identity of row permutation matrix \\
			$\Pi(i)$ & the permuted label for individual $i$ \\
			$d(\Pi_1,\Pi_2)$ & the Hamming distance, i.e., $\sum_{i=1}^n \mathbf 1\{\Pi_1(i) \neq \Pi_2(i)\}$ \\
			$\llam_i$ & $(\mathbf x_i^T \mathbf b_1, \ldots, \mathbf x_i^T \mathbf b_m)$, the  vector of linear component for individual $i$ \\
			$\langle \mathbf a \rangle$ / $\langle A \rangle$  & the sum of all entries in vector $\mathbf a$ / matrix $A$ \\
			$\mathbf a[l]$ & the $l$th element of vector $\mathbf a$ \\
			$A[i,j]$ & the element of matrix $A$ in $i$th row and $j$th column  \\
			$A[\mathcal S,:] / A[:, \mathcal S]$ & the sub-matrix of $A$ with row/column indices from set $\mathcal S$  \\
			$A_1 \circ A_2$  & the Hadamard product of $A_1$ and $A_2$ \\
			$[K]$ & $\{1,\ldots, K\}$ for any positive integer $K$. \\
			$\|\xx\|$ / $\|\xx\|_1$ & $\ell_2$-norm / $\ell_1$-norm for vector $\xx$. \\
			$|\Omega|$ & the cardinality of set $\Omega$. \\
			\hline
			\hline
		\end{tabular}
	}
\end{table}

\newpage

\section{Recovery Analysis when $B^{\sharp}$ is known}\label{sec:known}

\subsection{Permutation Recovery}
When $B^{\sharp}$ is known, we only need to estimate $\Pi$ by maximizing the likelihood function without estimating $B$.
In other words, the best estimator should be
\begin{eqnarray*}
	\hat \Pi &:=& \arg\max_{\Pi} L(\Pi, B^{\sharp}) = \arg\max_{\Pi}
	\langle - \psi(\Pi X B^{\sharp}) + Y \circ \Pi X B^{\sharp} \rangle.
\end{eqnarray*}
Successful recovery of label permutation matrix (i.e. $\hat \Pi = \Pi^{\sharp}$) means that
\begin{eqnarray}
\langle - \psi(\Pi X B^{\sharp}) + Y \circ \Pi X B^{\sharp} \rangle \leq \langle - \psi(\Pi^{\sharp} X B^{\sharp}) + Y \circ \Pi^{\sharp} X B^{\sharp} \rangle \nonumber
\end{eqnarray}
holds for any $\Pi \neq \Pi^{\sharp}$.
In other words, for any fixed $\Pi \neq \Pi^{\sharp}$, we need to identify certain sufficient conditions to ensure that the following probability
\begin{eqnarray}
& & P\big(\sup_{\Pi \neq \Pi^{\sharp}} \langle - \psi(\Pi X B) + Y \circ \Pi X B \rangle
\geq \langle - \psi(\Pi^{\sharp} X B) + Y \circ \Pi^{\sharp} X B \rangle \big) \nonumber
\end{eqnarray}
is vanishing as both $n$ and $m$ go to infinity in a suitable asymptotic regime.

\vspace{0.1in}
Before moving to our main results, we first introduce the following row-wise quantities.

\vspace{0.1in}
\noindent \textbf{Information Gap} For each pair of individuals $i$ and $j$, we define
\begin{eqnarray}\label{eq:gap:known}
\Delta_{ij} := \langle \psi'(\llam_i) \circ \llam_i - \psi(\llam_i) \rangle - \langle \psi'(\llam_i) \circ \llam_j - \psi(\llam_j) \rangle,
\end{eqnarray}
where $\llam_i = (\xx_i^T \bb_1^{\sharp}, \ldots, \xx_i^T \bb_m^{\sharp})$ for $i \in [n]$.

\vspace{0.1in}
\noindent \textbf{Variance} For each pair of individuals $i$ and $j$, we define
\begin{eqnarray}\label{eq:var:known}
v_{ij} :=  \langle \psi''(\llam_i) \circ (\llam_i - \llam_j)^2 \rangle = \sum_{l=1}^m \psi''(\llam_i[l]) (\llam_i[l] - \llam_j[l])^2,
\end{eqnarray}
where $\psi'$ and $\psi''$ are the first and second order derivative of function $\psi$.

It can be checked that $\Delta_{ij}$ and $v_{ij}$ are the expectation and variance of $\langle \yy_i \circ \llam_i - \psi(\llam_i) \rangle - \langle \yy_i \circ \llam_j - \psi(\llam_j) \rangle$, respectively.
We can show that all labels are distinguishable when $\Delta_{ij}$ is relatively large compared with $v_{ij}$ for all pairs of $i$ and $j$.
In other words, $\min_{i,j \in [n]} \Delta_{ij}^2 / v_{ij}$ can be viewed as the counterpart of \textit{signal-to-noise ratio} (SNR,~\citealp{pananjady2017linear}) in the linear regression setting.
\begin{theorem}\label{thm:1}
	Assume $B^{\sharp}$ is known and suppose $X$, $B^{\sharp}$ and $\Pi^{\sharp}$ satisfy that
	\begin{eqnarray}
	\Delta_{ij} \gtrsim \sqrt{(\log n) v_{ij}},  ~~ \forall i,j \in [n]. \label{cond:rmk}
	\end{eqnarray}
	We have the perfect label recovery with high probability,
	\begin{eqnarray}
	P(\hat \Pi \neq \Pi^{\sharp}) \leq n^2 \max_{i \neq j} \exp\{- \frac{\Delta_{ij}^2}{16 v_{ij}}\} \rightarrow 0. \nonumber
	\end{eqnarray}
\end{theorem}
By comparing the orders of two sides in \eqref{cond:rmk}, it is easier for larger value of $m$ to satisfy \eqref{cond:rmk}.

\subsection{Examples}

In this section, multiple examples along with intuitive explanations are given to illustrate the relationship between $m$ and $n$ for perfect permutation recovery.
Examples~\ref{eg1} -~\ref{eg4} given below describe four different data generation mechanisms. We can observe that, in those cases, it is impossible to recover $\Pi^{\sharp}$ when $m = 1$.
Hence, the sufficient conditions,
\begin{align}
& \min_{i_1 \neq i_2} |\psi'(\lambda_{i_1}) - \psi'(\lambda_{i_2})| \gtrsim \sqrt{\log n} ~ (\textrm{linear model}), \nonumber \\
& \min_{i_1 \neq i_2} |\sqrt{\psi'(\lambda_{i_1})} - \sqrt{\psi'(\lambda_{i_2})}| \gtrsim \sqrt{\log n} ~ (\textrm{poisson model}) \nonumber
\end{align}
given by~\cite{wang2020estimation}
are too restrictive.

\begin{example}\label{eg1}
	Consider the scenario $p=1$, then $X = \mathbf x$ is a vector. Assume $\mathbf x[i] \sim \textrm{Uniform}[a_1,a_2]$ for all $i$ and assume each entry of $B^{\sharp}$ is bounded between $b_1$ and $b_2$ ($0 < b_1 < b_2$).
	Without loss of generality, $a_1, a_2, b_1, b_2$ are all positive.
	We define $x_{gap} := \min_{i,j} x_{gap,ij} := \min_{i,j} |\mathbf x[i] - \mathbf x[j]|$, which is the minimum difference between any pair, $\mathbf x[i]$ and $\mathbf x[j]$.
	It is not hard to see that $x_{gap} = \Theta_p(\frac{1}{n^2})$.	
	Moreover, when $m \gtrsim n^4 \log n$, it is sufficient for recovery of $\Pi^{\sharp}$.
\end{example}

\vspace{0.1in}

\begin{example}\label{eg2}
	Consider the scenario $p= log_2(n) + 1$ and $n = 2^{n_1}$ ($n_1$ is a positive integer) and the design matrix $X$ is complete in the sense that it satisfies
	\[X = \begin{pmatrix}
	1 & 0 & 0 & \cdots & 0\\
	1 & 1 & 0 & \cdots & 0 \\
	1& 0 & 1 & \cdots & 0\\
	\vdots & \vdots & \vdots  & \ddots & \vdots  \\
	1 & 1 & 1 & \cdots & 1\\
	\end{pmatrix}.\]
	For instance, in the educational testing~\citep{templin2010diagnostic},
	each of $n$ rows corresponds to a student with certain skill sets. The first column represents the intercept and rest of columns represent $p$ different skills. Entries are binary-valued indicating the possess (1) or non-possess (0) of certain skills for different students.
	Without loss of generality, we assume that each entry of $B^{\sharp}$ is generated from the standard normal distribution. In this case, it suffices to require $m \geq \log n$ for correctly estimating $\Pi^{\sharp}$ for any strictly convex $\psi$ with bounded second derivative.
\end{example}

\vspace{0.1in}

\begin{example}\label{eg3}
	Consider the scenario $p$ and $n$ with $C_p^s \geq n$. ($C_p^s$ is the $s$th coefficient in polynomial $(1 + x)^p$) with $s$ being a fixed constant. The design matrix $X$ is sparse and bounded, that is, $X$ satisfies that $\|\xx_i\|_0 \leq s$ for any $i \in [n]$ and $a_1 \leq |X| \leq a_2$.
	Each entry of matrix $B^{\sharp}$ is generated by a standard normal distribution.
	We also assume that each row of $X$ has different support.
	Under this setting, $m \gtrsim \log n$ suffices for permutation recovery for any strictly convex $\psi$ with bounded second derivative.
\end{example}

\vspace{0.1in}

\begin{example}\label{eg4}
	Consider the scenario that each  entry of $X$ follows $N(0,1/p)$ independently and entry of $B^{\sharp}$ is generated from $N(0,1)$ independently.
	Under this setting, it can be shown that $m \gtrsim \log n$ suffices for permutation recovery for any strictly convex $\psi$ with bounded second derivative.
\end{example}

\vspace{0.1in}
\subsection{On the lower bound of $m$}

In this section, we discuss the minimum required number of $m$ for permutation recovery.
In particular, we consider the case that $B^{\sharp}$ is known, which is an easier task compared with the problem when $B^{\sharp}$ is unknown.
To start with, we first recall the following Fano's lemma~\citep{assouad1996fano}.
\begin{lemma}[Fano's Lemma]\label{lem:fano}
	Let $X$ be a random variable following probability distribution $f$, where $f$ is from set $\{f_1, \ldots, f_{r+1}\}$ which satisfies that
	\[KL(f_i \| f_j) \leq \beta  ~~ \textrm{for all}~~ i \neq j.\]
	Let $\psi(X) \in \{1, \ldots, r+1\}$ be an estimate of index of distribution. Then
	\begin{eqnarray}
	\inf_{\psi} \sup_i P(\psi(X) \neq i) \geq 1 - \frac{\beta + \log 2}{\log r}.
	\end{eqnarray}
\end{lemma}

\newpage

We consider the following $n!$ models. For any permutation matrix $\Pi_k$ $(k = 1, \ldots, n!)$, we define $f_k$ as the probability distribution of $Y = \Pi_k Y^{\sharp}$.
Therefore, the KL divergence between $f_{k_1}$ and $f_{k_2}$ is
\begin{align}\label{eq:KL}
KL(f_{k_1} \| f_{k_2})
=& \mathbf E_{f_{k_1}} \langle Y \circ \boldsymbol \Lambda_{k_1} - \psi(\boldsymbol \Lambda_{k_1}) \rangle - \mathbf E_{f_{k_1}} \langle Y \circ \boldsymbol \Lambda_{k_2} - \psi(\boldsymbol \Lambda_{k_2}) \rangle \nonumber \\
:=& \Lambda_{k_1}(\Pi_{k_1}, B^{\sharp}) - \Lambda_{k_1}(\Pi_{k_2}, B^{\sharp}),
\end{align}
where $\boldsymbol \Lambda_{k} = \Pi_k X B^{\sharp}$.
We let $\Delta(X, B^{\sharp}) := \max_{k_1 \neq k_2} \Lambda_{k_1}(\Pi_{k_1}, B^{\sharp}) - \Lambda_{k_1}(\Pi_{k_2}, B^{\sharp})$.
Therefore, by Fano's lemma, we have
$
\inf_{\hat \Pi} \sup_{\Pi_k} P(\hat \Pi \neq \Pi_k) \geq 1 - \frac{ \Delta(X,B^{\sharp}) + \log 2}{\log (n!)}$.
In particular, if $\Delta(X,B^{\sharp}) \leq C m n $, we consequently have
\begin{eqnarray}\label{eq:lowerbound2}
\inf_{\hat \Pi} \sup_{\Pi_k} P(\hat \Pi \neq \Pi_k) \geq 1 - \frac{C m n + \log 2}{n \log n} \geq 1/2,
\end{eqnarray}
when $m \lesssim \log n$.
In other words, $m = \log n$ is the minimal number for perfect permutation recovery up to a multiplicative constant in Example~\ref{eg4}.

On the other hand, $m \gtrsim \log n$ may not be tight under some situations.
For instance, in Example 1, we can see that there exist $i_1$ and $i_2$ such that
$|\xx_{i_1} - \xx_{i_2}|$ is $\Theta(1/n^2)$.
We let $\Pi_1 = \mathbf I$ and set $\Pi_2(i) = i$ for $i \neq i_1, i_2$ and $\Pi_2(i_1) = i_2$ and $\Pi_2(i_2) = i_1$.
Under such case, we have the following result.
\begin{corollary}\label{prop:lowerbound:ex1}
	By the constructions of $\Pi_1$ and $\Pi_2$, we have
	\begin{eqnarray}\label{eq:lowerbound:ex1}
	\inf_{\hat \Pi} \sup_{\Pi_k \in \{\Pi_1, \Pi_2\}} P(\hat \Pi \neq \Pi_k) \geq 1 - \frac{ c m / n^4 + \log 2}{\log 2}.
	\end{eqnarray}
\end{corollary}
Thus, the minimum requirement of $m$ is at least of order $n^4$ ($\gg 1$) in Example~\ref{eg1}.

\section{Recovery Analysis when $B^{\sharp}$ is unknown}\label{sec:unknown}


However, in practice, we have no prior knowledge of $B$ and need to estimate the parameter matrix $B$ and permutation matrix $\Pi$ simultaneously.
For a fixed $\Pi$, we define $\hat B(\Pi)$ to be the best estimator maximizing the likelihood function, that is,
\begin{eqnarray}
\hat B(\Pi) = \arg \max_{B} \langle - \psi(\Pi X B) + Y \circ \Pi X B\rangle.
\end{eqnarray}
On the computational side, this is a concave optimization and $\hat B(\Pi)$ can be solved efficiently.
On theoretical side, $\hat B(\Pi)$ does not admit explicit form which makes analysis harder.
In the following, we discuss situations under which the labels can be recovered perfectly.

First of all,  we note that the model is not identifiable when $p \geq n$ and $X$ has full row rank.
This is because, there exists a $p$ by $n$ matrix $P_x$ such that $I = X P_x$.
We can find that
\[\Pi^{\sharp} X B^{\sharp} = \Pi^{\sharp} \Pi \Pi  X B^{\sharp}  = (\Pi^{\sharp} \Pi) X (P_x \Pi  X B^{\sharp}).\]
Thus, the underlying $\Pi^{\sharp}$ and $B^{\sharp}$ are again not identifiable.
Such non-identifiability means that we have no chance to recover the true label permutation matrix, since there exist multiple global optimal values. In~\cite{unnikrishnan2015unlabeled}, it is further shown that $n \geq 2p$ is a necessary condition for perfect permutation recovery under the linear model setting with $m = 1$ and zero noise.

In the rest of paper, we only consider the case that $p < n$ for the generalized linear model. Moreover, we only focus on the situation that $p$ is $n^{a}$ with $a < 1/2$ so that the estimator has nice asymptotic properties even if we do not know the true $\Pi^{\sharp}$.

We recall the definition of log-likelihood function
$L(\Pi, B) =  \langle - \psi(\Pi X B) + Y \circ (\Pi X B) \rangle$
and introduce its corresponding population version
\begin{eqnarray}\label{def:lam1}
\Lambda(\Pi, B) := \mathbb E \langle - \psi(\Pi X B) + Y \circ (\Pi X B) \rangle
=  \langle - \psi(\Pi X B) + \psi'(\Pi^{\sharp} X B^{\sharp}) \circ (\Pi X B) \rangle,
\end{eqnarray}
where the expectation is taken with respect to $Y$.

\subsection{Scenario 1: $d(\mathbf I, \Pi^{\sharp})$ is small}

If we have the prior knowledge that the underlying permutation matrix $\Pi^{\sharp}$ is close to $\mathbf I$ (i.e. $d(\mathbf I, \Pi^{\sharp})$ is small), we then
consider a ``two-step estimation'' computational method for dealing such case.
In the first step, ``two-step'' algorithm aims to find a reasonable estimator of $B^{\sharp}$ by treating $\Pi = \mathbf I$. In the second step, we plug in this estimator to the objective and obtain the permutation matrix by maximizing the log-likelihood function.
The implementation details are given in Algorithm~\ref{alg:two_step}.
\begin{algorithm}[ht]
	\caption{Two-step Estimation.}
	\label{alg:two_step}
	\begin{algorithmic}
		\STATE {\bfseries Input:}
		Observation matrix $Y$ and design matrix $X$.
		\STATE {\bfseries Output:}
		Estimated permutation matrix $\hat \Pi$ and estimated coefficient matrix
		$\hat B$.
		\STATE 1. Solve $\hat B := \arg\max_{B} \{ \langle - \psi(X B) +  Y \circ X B \rangle \}$.
		\STATE 2. Solve $\hat \Pi := \arg\max_{\Pi} \{ \langle - \psi(\Pi X \hat B) +  Y \circ \Pi X \hat B \rangle \}$.
	\end{algorithmic}
\end{algorithm}

In below, we provide a theoretical analysis of the proposed two-step estimator.
Under mild conditions, we show that $\hat \Pi$ returned by Algorithm~\ref{alg:two_step} perfectly matches $\Pi^{\sharp}$ with high probability.
To start with, we first assume the following assumptions on function $\psi$ and design matrix, $X$.
\begin{itemize}
	\item[] \textit{A0} We assume that $\psi''(\cdot)$ is either monotonic or bounded.
	\item[] \textit{A1} Each entry of $X$ is bounded (i.e. $|X[i,j]| \leq C_0$ for universal constant $C_0$).
	\item[] \textit{A2} There exist constants $c_1 > 0$, and $\gamma_{1p}$ (which may depend on $p$) such that
	$ \sharp\{i: X[i,:] \mathbf b \geq c_1\} \geq n/\gamma_{1p}$
	and $ \sharp\{i: X[i,:] \mathbf b \leq - c_1\} \geq n/\gamma_{1p}$ hold for any $\mathbf b$ with $\|\mathbf b\| = 1$.
\end{itemize}

\vspace{0.1in}

\begin{remark}\label{rmk:0}
	Assumption A0 is satisfied by most generalized linear models.
	For examples, $\psi''$ is bounded for Gaussian or Bernoulli distribution;
	$\psi''$ is monotonic for Poisson or Gamma distribution.
\end{remark}

\vspace{0.1in}

\begin{remark}\label{rmk:1}
	For a general $n$ by $p$ matrix $X$, its largest singular value is bounded by $\sqrt{n} \sqrt{p} \max_{i,j}|X[i,j]| = O(\sqrt{np})$.
	For an $n \times p$ matrix $X$ with each entry being sampled from sub-Gaussian distribution, then its largest singular value is $O_p(\sqrt{n} + \sqrt{p})$.
	(Here we say $Z$ is a sub-Gaussian random variable if $\mathbb E \exp\{t Z\} \leq \exp\{t^2 \sigma^2/2\}$ holds for all $t > 0$ and fixed constant $\sigma$.)
\end{remark}

\vspace{0.1in}

\begin{remark}\label{rmk:2}
	It can be checked that Assumption \textit{A2} is satisfied with high probability when $X$ is a matrix with i.i.d sub-Gaussian random variables as its entries. Under such case, $\gamma_{1p}$ is reduced to some constant.  Furthermore, \textit{A2} tells us that the smallest singular value of $X$ is bounded from below. In fact, $\sigma_{min}(X) \geq \sqrt{c_1^2 n/\gamma_{1p}} = c_1 \sqrt{n} / \sqrt{\gamma_{1p}}$.
\end{remark}

\vspace{0.1in}

We further introduce the following notations:
$x_{max} := \max_{i} \|\xx_i\|$;
$\psi_{max}^{'\sharp} := \max_{i, j} \psi(\xx_i^T \bb_j^{\sharp})$, $\psi_{max}^{''\sharp} := \max_{i, j} \exp\{\xx_i^T \bb_j^{\sharp}\}$,
and
$\psi_{min}^{''\sharp} := \min_{i, j} \exp\{\xx_i^T \bb_j^{\sharp}\}$
which represent the maximum expected value of $Y_{ij}$'s, maximum variance/minimum variance of $Y_{ij}$'s respectively.
We also write $\psi_{cb}^{\sharp} = \psi_{max}^{'\sharp} + \psi_{max}^{''\sharp}$
and define permutation-wise variance term
$v_{\Pi,partial} = \sum_{i: \Pi^\sharp(i) \neq \Pi(i)} \sum_{l=1}^m \psi''(\llam_{\Pi^{\sharp}(i)}^{\sharp}[l]) (\llam_{\Pi(i)}^{\sharp}[l] - \llam_{\Pi^{\sharp}(i)}^{\sharp}[l])^2$
and minimum pairwise variance term
$v_{min} = \min_{i,j} \sum_{l=1}^m \psi''(\llam_{i}^{\sharp}[l]) (\llam_{i}^{\sharp}[l] - \llam_{j}^{\sharp}[l])^2$
to quantify the differences between $X$'s rows.

\newpage

\begin{theorem}\label{thm:known}
	With the knowledge that $d(\mathbf I, \Pi^{\sharp}) \leq h_{max}$ and assumptions \textit{A0} - \textit{A2}, we also assume that $p = O(n^{a})$ ($a < \frac{1}{2}$) and $h_{max} \lesssim n / (p \gamma_{1p} \log n)$.
	Then it holds that
	\begin{eqnarray}\label{bound:case1}
	\|\hat \bb_l - \bb_l^{\sharp}\| =: O_p(\delta^{\ast})
	= O_p(\frac{\sqrt{p} (\sqrt{\psi_{cb}^{\sharp}} \sqrt{n - h_{max}} + \psi_{cb}^{\sharp} h_{max} \log n)}{n \psi_{min}^{''\sharp}} \gamma_{1p})
	\end{eqnarray}
	for $l \in [m]$.
	Furthermore, if
	\begin{eqnarray}\label{condition:1}
	\Lambda(\Pi^{\sharp}, B^{\sharp}) - \Lambda(\Pi, B^{\sharp}) \gtrsim v_{\Pi,partial} \cdot \sqrt{\log n / v_{min}}
	\end{eqnarray}
	and
	\begin{eqnarray}\label{condition:2}
	\Lambda(\Pi^{\sharp}, B^{\sharp}) - \Lambda(\Pi, B^{\sharp}) \gtrsim
	m d(\Pi, \Pi^{\sharp}) \psi_{cb}^{\sharp} x_{max} \delta^{\ast},
	\end{eqnarray}
	then it holds that
	$P(\hat \Pi \neq \Pi^{\sharp})  \rightarrow 0$.
\end{theorem}

Here,  $\delta^{\ast}$ defined in \eqref{bound:case1} is the estimation error for regression parameter. The first term can be viewed as the variance term and the second term is the bias term. Conditions \eqref{condition:1} and \eqref{condition:2} require that the information gap should dominate the errors caused by variance and bias, respectively.
The requirement $h_{max} \leq n/(p \gamma_{1p} \log n)$ restricts the number of permuted labels.
When $p$ is fixed, then $n / (\log n)$ is similar to the condition $h_{max} \leq c n / (\log(n/h_{max}))$ required in~\cite{Slawski2020two} for linear models. The additional $p$ in the denominator appears since that we do not have the access to the explicit form of $\hat B(\Pi)$ in generalized linear model setting.

Especially, we can find that $\Lambda(\Pi^{\sharp}, B^{\sharp}) - \Lambda(\Pi, B^{\sharp})$ is $\Omega(m d(\Pi, \Pi^{\sharp}))$,
$v_{\Pi,partial}$ is $O(m d(\Pi, \Pi^{\sharp}))$ and $v_{min} = \Omega(m)$ when $\llam_{i}^{\sharp}[l]$ is bounded and $\min_{i,j} \|\llam_{i}^{\sharp} - \llam_{j}^{\sharp}\|_1$ is $\Omega(m)$.
Then condition \eqref{condition:1} can be simplified to
\[m d(\Pi, \Pi^{\sharp}) \gtrsim m d(\Pi, \Pi^{\sharp}) \sqrt{\log n / m}, \]
which requires $m \geq K \log n$ for some large constant $K$. This leads to the following corollary.

\begin{corollary}
	Under the sub-Gaussian design matrix $X$ with knowledge that $d(\mathbf I, \Pi^{\sharp}) \leq h_{max}$, we assume $h_{max} p = n / \log n$, $p = O(n^{a})$ ($a < \frac{1}{2}$), and $m \gtrsim \log n$.
	We have
	\[P(\hat \Pi \neq \Pi^{\sharp}) \rightarrow 0 \quad \text{as}~ n \rightarrow \infty.\]
\end{corollary}

\subsection{Scenario 2: no knowledge of $d(\mathbf I, \Pi^{\sharp})$}

For general $\Pi^{\sharp}$, we also aim to recover the underlying permutation matrix without any knowledge of $d(\mathbf I, \Pi^{\sharp})$ and $B^{\sharp}$.
To facilitate the theoretical analysis, we first define $B(\Pi) := \arg\max_{B} \Lambda(\Pi, B)$.
Then it is straightforward to check that $B(\Pi^{\sharp}) = B^{\sharp}$.
We also define
\begin{eqnarray}\label{def:lam2}
\Lambda(\Pi) := \max_{B} \Lambda(\Pi, B) = \Lambda(\Pi, B(\Pi)).
\end{eqnarray}
We additionally introduce the following permutation-wise quantities.

\noindent\textbf{Information Gap} For each permutation $\Pi$, we define
\begin{eqnarray}\label{eq:gap:unknown}
\Delta(X, B^{\sharp}, \Pi^{\sharp}, \Pi) &=& \Lambda(\Pi^{\sharp}) - \Lambda(\Pi).
\end{eqnarray}
This quantity can be interpreted as the information gap between permutation $\Pi^{\sharp}$ and $\Pi$. It can be seen that $\Delta(X, B^{\sharp}, \Pi^{\sharp}, \Pi)$ is always non-negative.

\noindent\textbf{Variance} For each fixed permutation $\Pi$, we define
\begin{eqnarray}\label{eq:var:unknown}
v_{\Pi,B}&=& \sum_{i=1}^n \sum_{l=1}^m \psi''(\llam_{\Pi^{\sharp}(i)}[l]) (\llam_{\Pi(i)}[l])^2.
\end{eqnarray}
It can be easily checked that $v_{\Pi,B}$ is the variance of $L(\Pi, B)$.


\begin{theorem}\label{thm:unknown:main}
	Under assumptions \textit{A0} - \textit{A2} and with no knowledge of $B^{\sharp}$ and $\Pi^{\sharp}$, we assume that $p = O(n^{a})$ ($a < \frac{1}{2}$) and there exists a constant $c_0$ such that
	\begin{eqnarray}\label{cond:case1}
	\Delta(X,B^{\sharp},\Pi^{\sharp},\Pi) \gtrsim \max\{\sqrt{(n + m p) m n \psi_{max}^{''\sharp} x_{max}^2 \log n},  (n \log n + mp) x_{max}\}
	\end{eqnarray}
	for any $\Pi$ satisfying $d(\Pi, \Pi^{\sharp}) > c_0 \frac{n}{p \gamma_{1p} \log n}$, and
	\begin{eqnarray}\label{cond:case2}
	\Lambda(\Pi^{\sharp}, B^{\sharp}) - \Lambda(\Pi, B^{\sharp})  \gtrsim \max\{v_{\Pi,partial} \cdot \sqrt{\log n / v_{min}},  m d(\Pi, \Pi^{\sharp}) \psi_{cb}^{\sharp} x_{max} \delta^{\ast}\}
	\end{eqnarray}
	for any $\Pi$ satisfying $d(\Pi, \Pi^{\sharp}) \leq c_0 \frac{n}{p \gamma_{1p} \log n}$.
	Then it holds that
	\begin{eqnarray}
	P(\hat \Pi \neq \Pi^{\sharp}) \rightarrow 0
	\end{eqnarray}
	as $n \rightarrow \infty$.
	Furthermore,
	$
	\|\hat \bb_l - \bb_l^{\sharp}\| = O_p(\frac{\gamma_{1p} \sqrt{p
			\psi_{cb}^{\sharp}}}{\sqrt{n} \psi_{min}^{\sharp}})
	$
	for all $l \in [m]$.
\end{theorem}

Based on Theorem~\ref{thm:unknown:main}, we have the following corollary.
\begin{corollary}
	Without any knowledge of $B^{\sharp}$ and $\Pi^{\sharp}$, we assume that
	$
	\Delta(X,B^{\sharp},\Pi^{\sharp},\Pi) \geq c_1 m d(\Pi, \Pi^{\sharp}) $ for $d(\Pi, \Pi^{\sharp}) > \frac{c_0 n}{p \log n}$
	and
	$
	\Lambda(\Pi^{\sharp}, B^{\sharp}) - \Lambda(\Pi, B^{\sharp}) \geq c_2 m d(\Pi, \Pi^{\sharp})$ for $d(\Pi, \Pi^{\sharp}) \leq \frac{c_0 n}{p \log n}$
	($c_0$, $c_1$, $c_2$ are universal constants).
	Then it holds that
	$
	P(\hat \Pi \neq \Pi^{\sharp}) \rightarrow 0,
	$
	as long as $m / (x_{max}^2 \log n ) \rightarrow \infty$, $\gamma_{1p} = O(1)$ and $p = n^{a}$ ($0 < a < \frac{1}{2}$).\end{corollary}

\subsection{On Computation of Maximum Likelihood Estimator}

We consider to compute the maximum likelihood estimator, which is
\begin{eqnarray}
(\hat \Pi, \hat B) = \arg\max_{\Pi, B} \langle - \psi(\Pi X B) + Y \circ \Pi X B \rangle.
\end{eqnarray}
Unfortunately, when $p > 1$, the above optimization problem (even for the linear models) is NP-hard.
We consider a coordinate ascent method, i.e., alternatively maximizing $\Pi$ given $B$ and maximizing $B$ given $\Pi$. The method will always converge to some critical point but not necessarily to the global optimal solution.

For the choice of a good initial permutation matrix $\Pi_{ini}$,
we consider the following heuristic objective,
\begin{eqnarray}\label{obj:heuristic}
\Pi_{ini} = \arg\max_{\Pi} \langle \Pi, Y_{\psi} Y_{\psi}^T X  X^T\rangle,
\end{eqnarray}
where $Y_{\psi} = (\psi')^{-1} (Y+1)$ be the inverse transformation of the original data through link function $\psi'$.
The intuition is that $Y_{\psi} \approx \Pi^{\sharp} X B^{\sharp}$.
It can be calculated that
\begin{eqnarray}
\langle \Pi, Y_{\psi} Y_{\psi}^T X  X^T\rangle
\approx \langle \Pi, \Pi^{\sharp} X B^{\sharp} (\Pi^{\sharp} X B^{\sharp})^T X  X^T\rangle = \|B^{\sharp}\|^2 \langle X, \Pi^{\sharp} X \rangle \langle \Pi X, \Pi^{\sharp} X \rangle, \label{eq:intuition}
\end{eqnarray}
when $m=1$ and $p=1$. If term $\langle X, \Pi^{\sharp} X \rangle$ is positive, then the maximum value of \eqref{eq:intuition} is achieved when $\Pi = \Pi^{\sharp}$.
Such warm start computational scheme is presented in Algorithm~\ref{alg:warm_start}
and maximum likelihood estimation scheme is given in Algorithm~\ref{alg:mle}.

\begin{algorithm}[]
	\caption{Warm start for maximum likelihood estimation.}
	\label{alg:warm_start}
	\begin{algorithmic}
		\STATE {\bfseries Input:}
		Response matrix $Y$ and design matrix $X$
		\STATE {\bfseries Output:}
		A good initial permutation matrix $\Pi_{ini}$.
		\STATE 1. Compute the matrix $Y_{\psi} = (\psi')^{-1}(Y + 1)$.
		\STATE 2. Compute the matrix $C = Y_{\psi} Y_{\psi}^T X X^T$.
		\STATE 3. Solve $\Pi_{ini} := \arg\max_{\Pi} \langle \Pi, C\rangle$.
		\STATE 4. Return $\Pi_{ini}$ as the initial $\Pi$.
	\end{algorithmic}
\end{algorithm}

\begin{algorithm}[]
	\caption{Maximum likelihood (ML) estimation. }
	\label{alg:mle}
	\begin{algorithmic}
		\STATE {\bfseries Input:} Response matrix $Y$, design matrix $X$ and initial permutation matrix $\Pi_{ini}$.
		\STATE {\bfseries Output:} Estimated permutation matrix $\hat \Pi$ and estimated coefficient matrix $\hat B$.
		\STATE Let $\hat \Pi = \Pi_{ini}$.
		\WHILE{the likelihood is not converged}
		\STATE 1. Solve $\hat B := \arg\max_{B} \{ \langle - \psi(
		\hat \Pi X B) + Y \circ \hat \Pi X B \rangle \}$.
		\STATE 2. Solve $\hat \Pi := \arg\max_{\Pi} \{ \langle - \psi(\Pi X \hat B) + Y \circ \Pi X \hat B \rangle \}$.
		\ENDWHILE
		\STATE Return $\hat B$ and $\hat \Pi$.
	\end{algorithmic}
\end{algorithm}

\subsection{Remarks}\label{sec:remark}

The technical challenge in the generalized linear model setting lies in the facts that the second derivative of likelihood function is associated with exponential functions which are not bounded. To be more specific, the Hessian matrix can be written as
\begin{eqnarray}
\nabla^2 L(\mathbf I,\bb) = - X^T D X, \nonumber
\end{eqnarray}
with $D = \textrm{diag}(d_1, \ldots, d_n)$ and $d_i = \psi''(\xx_i^T \bb)$ when $m = 1$.
For example, $\psi''(x) = \exp\{x\}$ is obviously unbounded for Poisson model.
Moreover, for any fixed $\Pi$, the maximizer,
\[\hat \bb(\Pi) = \arg\max_{\bb} L(\Pi, \bb), \]
does not admit the closed form. This brings additional difficulty while there is no such issue in the standard linear models.

In~\cite{wang2020estimation}, their analysis only considers the case when $d(\mathbf I, \Pi^{\sharp})$ is small and $B^{\sharp}$ is known. Our results include the most general cases, i.e., we have no prior knowledge of $\Pi$ and $B^{\sharp}$. Therefore, we need to take special care of uniform bound over all possible permutations in our analyses.

From the computational perspective, note that both "two-step estimation" and "ML" methods require computing
$\hat \Pi := \arg\max_{\Pi} \{ \langle - \psi(\Pi X \hat B) + Y \circ \Pi X \hat B \rangle \}$. Since the number of candidates for the permutation matrix is n!, we cannot directly solve this optimization problem.
However, we can reformulate this problem into a linear assignment problem which can be solved efficiently by specialized techniques such as Hungarian algorithm~\citep{kuhn1955hungarian} or the Auction algorithm~\citep{bertsekas1992forward}.
To be more specific, we define an $n$ by $n$ cost matrix $C = (C[i,j])$, with
\[C[i,j] = \langle \psi(\xx_i B) - \yy_j \circ (\xx_i B) \rangle. \]
Note that permutation matrix $\Pi$ has only one non-zero element "1" in each row and column. It is then equivalent to solve the assignment problem,
\[\min_{\tau} \sum_{i \in [n]} C[\tau(i), i], \]
where $\tau$ is an one-to-one mapping from $[n]$ to $[n]$.
Hence $\hat \Pi$ can be solved via using Hungarian algorithm or Auction algorithm.

Compared with~\cite{wang2020estimation}, they adopt an $\ell_1$ regularization framework for computing $\hat \Pi$ and $\hat B$, while our method does not.
This is due to the fact that~\cite{wang2020estimation} assume a sparsely mismatch regime that $d(\Pi^{\sharp}, \mathbf I)$ is small. That is the reason they introduce
$\ell_1$ regularizer to encourage recovering a sparse permutation. By contrast, we do not require $d(\Pi^{\sharp}, \mathbf I)$ to be small in our theory.

\newpage

\section{Extension to Missing Observation Case}\label{sec:miss}

In this section, we generalize our results to the situations when data are not fully observed.
To be specific, we consider the following model
\begin{eqnarray}
Y_{miss} =  E \circ Y = E \circ (\Pi^{\sharp}Y^{\sharp}),
\end{eqnarray}
where $E$ is a binary matrix such that ``1'' means the entry is observed and ``0'' means the entry is missing.
The elements in $E$ are independent Bernoulli($q$) random variables and $q$ $(0 < q < 1)$ is the observation rate.

The likelihood function with missing observations can be written as
\begin{eqnarray}
L(\Pi, B, E) &=& \{\sum_{i,l: E[i,l] = 1}
Y[i,l] (\xx_{\Pi(i)}^T\bb_l) - \psi(\xx_{\Pi(i)}^T \bb_l) \} \nonumber \\
&=& \langle E \circ (Y \circ (\Pi X B) - \psi(\Pi X B) )  \rangle
\end{eqnarray}
and its expectation can be computed as
\begin{eqnarray}
\Lambda(\Pi, B, q) &=& \mathbb E \{\sum_{i,l: E[i,l] = 1}
Y[i,l] (\xx_{\Pi(i)}^T\bb_l) - \psi(\xx_{\Pi(i)}^T\bb_l) \} \nonumber \\
&=& q \cdot \Lambda(\Pi,B),
\end{eqnarray}
where the expectation is taken over both $E$ and $Y$. We also define
\[\Lambda(\Pi,q) := \max_{B} \Lambda(\Pi,B,q) = q \cdot \Lambda(\Pi).\]

\subsection{When $B^{\sharp}$ is known }
In this scenario, we can similarly define the following terms, the row-wise information gap,
\begin{eqnarray}\label{eq:row:gap:q}
\Delta_{ij}(q) &:=& \mathbb E \sum_{l} \mathbf 1\{E[\Pi^{\sharp}(i),l] = 1\}\{ ( \psi'(\llam_{i}^{\sharp}[l]) \llam_{i}^{\sharp}[l] - \psi(\llam_{i}^{\sharp}[l])) - ( \psi'(\llam_{i}^{\sharp}[l]) \llam_{j}^{\sharp}[l] - \psi(\llam_{j}^{\sharp}[l])) \} \nonumber \\
&=& q \sum_l \{ (\psi'(\llam_{i}^{\sharp}[l]) \llam_{i}^{\sharp}[l] - \psi(\llam_{i}^{\sharp}[l])) - (\psi'(\llam_{i}^{\sharp}[l]) \llam_{j}^{\sharp}[l] - \psi(\llam_{j}^{\sharp}[l])) \} = q \Delta_{ij}
\end{eqnarray}
and the row-wise variance,
\begin{eqnarray}
v_{ij}(q)&:=&
q \sum_{l=1}^m \psi''(\llam_{i}^{\sharp}[l]) (\llam_{i}^{\sharp}[l] - \llam_{i}^{\sharp}[l])^2 \nonumber \\
& & + ~ q(1-q)\sum_{l=1}^m( \psi'(\llam_{i}^{\sharp}[l])( \llam_{i}^{\sharp}[l] - \llam_{j}^{\sharp}[l]) - (\psi(\llam_{i}^{\sharp}[l]) - \psi(\llam_{j}^{\sharp}[l])))^2,
\end{eqnarray}
which is the variance of $
\mathrm{var}(\langle E[\Pi^{\sharp}(i),:] \circ (\yy_i \circ \llam_i^{\sharp} - \psi(\llam_i^{\sharp})) \rangle - \langle E[\Pi^{\sharp}(i),:] \circ (\yy_i \circ \llam_j^{\sharp} - \psi(\llam_j^{\sharp})) \rangle)$.

\begin{theorem}
	Assume $B^{\sharp}$ is known and suppose $X$, $B^{\sharp}$, $\Pi^{\sharp}$ satisfy that
	\begin{eqnarray}
	\Delta_{ij}(q) \gtrsim \sqrt{(\log n) v_{ij}(q)} ~~ \forall i,j \in [n],
	\end{eqnarray}
	then it holds that
	\begin{eqnarray}
	P(\hat \Pi \neq \Pi^{\sharp}) \leq n^2 \max_{i \neq j} \exp\{- \frac{\Delta_{ij}^2(q)}{16 v_{ij}(q)}\}.
	\end{eqnarray}
\end{theorem}

\begin{remark}
	Especially when $\lambda_{il}^{\sharp}$'s are bounded and $\min_{i,j} \sum_{l \in [m]} (\lambda_{il}^{\sharp} - \lambda_{jl}^{\sharp})^2 = \Omega(m)$, then $q \geq \frac{\log n}{m}$ is required for the perfect permutation recovery. In other words, the number of required observations ($m$) for each individual is reciprocal to observation rate ($q$).
\end{remark}

\subsection{When $B$ is unknown and $d(\mathbf I, \Pi^{\sharp})$ is small}

Under this scenario, we further define the partial variance term
\begin{eqnarray}
v_{\Pi,partial,q} &=& q \sum_{i: \Pi(i) \neq \Pi^{\sharp}(i)} \sum_{l=1}^m \bigg\{  \psi''(\llam_{i}^{\sharp}[l]) (\llam_{\Pi(i)}[l]^{\sharp} - \llam_{i}[l]^{\sharp})^2 \nonumber \\
& & + q(1-q) (\psi'(\llam_{i}[l]^{\sharp}) (\llam_{\Pi(i)}[l]^{\sharp} - \llam_{i}[l]^{\sharp}) - \psi(\llam_{\Pi(i)}^{\sharp}[l]) + \psi(\llam_{i}[l]^{\sharp}) )^2 \bigg\} \nonumber
\end{eqnarray}
and minimum variance gap
\[v_{min,q} = \min_{i,j} \{q \sum_{l=1}^m \psi''(\llam_{i}^{\sharp}[l]) (\llam_{i}[l]^{\sharp} - \llam_{j}[l]^{\sharp})^2 + q(1-q) \sum_{l=1}^m (\psi'(\llam_{i}^{\sharp}[l]) (\llam_{i}[l]^{\sharp} - \llam_{j}[l]^{\sharp}) - \psi(\llam_{i}[l]^{\sharp}) + \psi(\llam_{j}[l]^{\sharp}) )^2 \}. \]
We also assume the following assumptions on design matrix.
\begin{itemize}
	\item[] \textit{E1} Entries of $X$ are bounded by some constant $C_0$.
	\item[] \textit{E2} Let $\mathcal S_l = \{i: E[i,l] = 1\}$ ($l = 1, \ldots, m$). There exist constants $c_2 > 0$ and $\gamma_{2p}$ such that
	$ \sharp\{i: \xx_i^T \mathbf b \geq c_2, i \in \mathcal S_l \} \geq |\mathcal S_l|/\gamma_{2p}$
	and
	$ \sharp\{i: \xx_i^T \mathbf b \leq - c_2, i \in \mathcal S_l \} \geq |\mathcal S_l|/\gamma_{2p}$ hold for any $\mathbf b$ with $\|\mathbf b\| = 1$.
\end{itemize}

\begin{remark}
	Assumptions \textit{E1} and \textit{E2} are parallel to \textit{A1} and \textit{A2}.
	They put the restrictions on sub-design matrices $X[\mathcal S_l,:]$'s.
	In particular, \textit{E2} holds by taking $\gamma_{2p} = \Theta(1)$ with high probability, when each entry of $X$ follows i.i.d. standard Normal distribution and $qn \gtrsim p$.
\end{remark}

\begin{theorem}
	With the knowledge that $d(\mathbf I, \Pi^{\sharp}) \leq h_{max}$ and assumptions \textit{A0}, \textit{E1} and \textit{E2}, we assume $p = O((qn)^{a})$ ($a < \frac{1}{2}$) and $h_{max} \lesssim nq/(p \gamma_{2p} \log n)$.
	We then have that
	\begin{eqnarray}
	\|\hat \bb_l - \bb_l^{\sharp}\| = O(\delta_q^{\ast}) :=
	O(\frac{\sqrt{p}(\sqrt{q \psi^{''\sharp}_{max} + q(1-q) \psi^{\sharp 2}_{cb}} \sqrt{n - h_{max}} + \psi_{max}^{''\sharp}h_{max} \log n)}{qn \psi_{min}^{\sharp}/\gamma_{2p}}) .
	\end{eqnarray}
	Furthermore, if
	\begin{eqnarray}
	\Lambda(\Pi^{\sharp}, B^{\sharp}) - \Lambda(\Pi, B^{\sharp}) \gtrsim
	\frac{1}{q} v_{\Pi,partial,q} \cdot \sqrt{\log n / v_{min,q}}
	\end{eqnarray}
	and
	\begin{eqnarray}
	\Lambda(\Pi^{\sharp}, B^{\sharp}) - \Lambda(\Pi, B^{\sharp}) \gtrsim
	\frac{1}{q} m d(\Pi, \Pi^{\sharp}) \psi_{cb}^{\sharp} x_{max} \delta_q^{\ast},
	\end{eqnarray}
	then it holds that
	\begin{eqnarray}
	P(\hat \Pi \neq \Pi^{\sharp}) \rightarrow 0.
	\end{eqnarray}
\end{theorem}

\begin{remark}
	Under the setting of sub-Gaussian design, it is sufficient to have $m \geq \log n / q$ for permutation recovery when $d(\mathbf I, \Pi^{\sharp}) \leq c_0 \frac{nq}{p \log n}$ for some constant $c_0$.
\end{remark}

\subsection{Without knowledge of $B$ and $d(\mathbf I, \Pi^{\sharp})$}

In this situation, we further assume the following conditions.
\begin{itemize}
	\item[] \textit{E2'}  There exist constants $c_3 > 0$ and $\gamma_{3p}$ such that
	$ \sharp\{i: \xx_i^T \mathbf b \geq c_3, i \in \mathcal S \} \geq q n/\gamma_{3p}$ and
	$ \sharp\{i: \xx_i^T \mathbf b \leq -c_3, i \in \mathcal S \} \geq q n/\gamma_{3p}$ hold for any $\mathbf b$ with $\|\mathbf b\| = 1$ and $\mathcal S$ with $|\mathcal S| \geq qn/2$.
	(It is a modified and stronger version of \textit{E2}.)
\end{itemize}

Additionally, we let $\Delta_q(X,B^{\sharp},\Pi,\Pi^{\sharp}) := \Lambda(\Pi^{''\sharp},q) - \Lambda(\Pi,q)$ which is equal to $q \Delta(X,B^{\sharp},\Pi,\Pi^{\sharp})$,
and define the following variance-related quantity,
\begin{eqnarray*}
	V_{2}(q) = (q n \psi_{max}^{\sharp} + q(1-q) n \psi_{cb}^{\sharp 2})(\psi(x_{max}))^2.
\end{eqnarray*}

\begin{theorem}
	Under assumptions \textit{A0}, \textit{E1} and \textit{E2'}, we assume that there exists $c_0$ such that
	\begin{eqnarray}
	\Delta_q(X,B^{\sharp},\Pi,\Pi^{\sharp}) \gtrsim  \max\{\sqrt{(n \log n + m p \log(n)) m V_2(q)}, \psi_{cb}^{\sharp} (n \log^2n + mp\log n)\} \label{cond:1:q}
	\end{eqnarray}
	holds for any $\Pi$ with $d(\Pi, \Pi^{\sharp}) > c_0 \frac{nq}{p \gamma_{3p} \log n}$, and
	\begin{eqnarray}
	\Lambda(\Pi^{\sharp}, B^{\sharp}) - \Lambda(\Pi, B^{\sharp}) \gtrsim
	\max\{\frac{1}{q} v_{\Pi,partial,q} \cdot \sqrt{\log n / v_{min,q}},
	\frac{1}{q} m d(\Pi, \Pi^{\sharp}) \psi_{cb}^{\sharp} x_{max} \delta_q^{\ast}
	\}
	\end{eqnarray}
	holds for any $\Pi$ with $d(\Pi, \Pi^{\sharp}) \leq c_0 \frac{nq}{p \gamma_{3p} \log n}$.
	
	Then it holds that
	\begin{eqnarray}
	P(\hat \Pi \neq \Pi^{\sharp}) \rightarrow 0
	\end{eqnarray}
	as $n \rightarrow \infty$.
	Furthermore,
	\begin{eqnarray}
	\|\hat \bb_l - \bb_l^{\sharp}\| = O_p(\frac{\sqrt{pn}(\sqrt{q \psi^{''\sharp}_{max} + q(1-q) \psi^{\sharp 2}_{cb}} )}{qn \psi_{min}^{''\sharp}/{\color{black} \gamma_{3p}}})
	\end{eqnarray}
	for all $l \in [m]$.
\end{theorem}

Especially, when $\gamma_{3p}$, $\psi_{min}^{''\sharp}, \psi_{max}^{''\sharp}$, $\psi_{cb}^{\sharp}$ are $O(1)$, and $\min_{i \neq j} \sum_{l \in [m]}|\llam_{i}^{\sharp}[l] - \llam_j^{\sharp}[l]| = \Omega(m)$, it suffices to require
\[m \gtrsim \frac{(\psi(x_{max}))^2 \log n}{q} ~, ~ q \gtrsim \frac{p (\psi(x_{max}))^2 \log n}{n} \]
for exact permutation recovery.

\subsection{ML Estimator with Missing Observations}

The warm-start stage of ML estimation algorithm is modified in the missing observation setting. In particular, we use SoftImpute~\citep{hastie2015matrix} method to impute the missing entries of the data matrix. The procedure is given as below in Algorithm~\ref{alg:mle:miss}.

\begin{algorithm}[]
	\caption{ML estimation with warm start for missing observations.}
	\label{alg:mle:miss}
	\begin{algorithmic}
		\STATE {\bfseries Input:}
		Observations with missing entries $Y_{miss}$, design matrix $X$
		\STATE {\bfseries Output:}
		A good initial permutation matrix $\Pi_{ini}$.
		\STATE 1. Compute the matrix $Y_{\psi,miss} = (\psi')^{-1}(Y_{miss} + 1)$.
		\STATE 2. Use SoftImpute method to complete the matrix, $Y_{\psi,miss}$, to get $Y_{\psi,comp}$.
		\STATE 3. Compute the matrix $C = Y_{\psi,comp} Y_{\psi,comp}^T X X^T$.
		\STATE 4. Solve $\Pi_{ini} := \arg\max_{\Pi} \langle \Pi, C\rangle$.
		\STATE 5. Set $\hat \Pi = \Pi_{ini}$ as the initial permutation matrix.
		\WHILE{The likelihood not converged}
		\STATE 6.a. Solve $\hat B := \arg\max_{B} \{ \langle - E \circ \psi(
		\hat \Pi X B) + E \circ Y_{miss} \circ \hat \Pi X B \rangle \}$.
		\STATE 6.b. Solve $\hat \Pi := \arg\max_{\Pi} \{ \langle - E \circ \psi(\Pi X \hat B) + E \circ Y_{miss} \circ \Pi X \hat B \rangle \}$.
		\ENDWHILE
		\STATE 7. Return $\hat B$ and $\hat \Pi$.
	\end{algorithmic}
\end{algorithm}

\newpage
\subsection{Two-step Estimator with Missing Observations}

\noindent Similarly, we introduce the two-step estimator under the missing observation setting. The procedure is given as follows in Algorithm~\ref{alg:two-step:miss}.

\begin{algorithm}[h]
	\caption{Two-step Estimation with missing observations.}
	\label{alg:two-step:miss}
	\begin{algorithmic}
		\STATE {\bfseries Input:}
		Observations with missing entries $Y_{miss}$, indicator matrix $E$ and design matrix $X$
		\STATE {\bfseries Output:}
		Estimated permutation matrix $\hat \Pi$ and estimated coefficient matrix
		$\hat B$.
		\STATE 1. Solve $\hat B := \arg\max_{B} \{ \langle - E \circ \psi(X B) + E \circ Y_{miss} \circ X B \rangle \}$.
		\STATE 2. Solve $\hat \Pi := \arg\max_{\Pi} \{ \langle - E \circ \psi(\Pi X \hat B) + E \circ Y_{miss} \circ \Pi X \hat B \rangle \}$.
	\end{algorithmic}
\end{algorithm}

\section{Simulation Studies}\label{sec:sim}

\begin{figure}[h]
	\begin{center}
	\mbox{
		\includegraphics[width=3in]{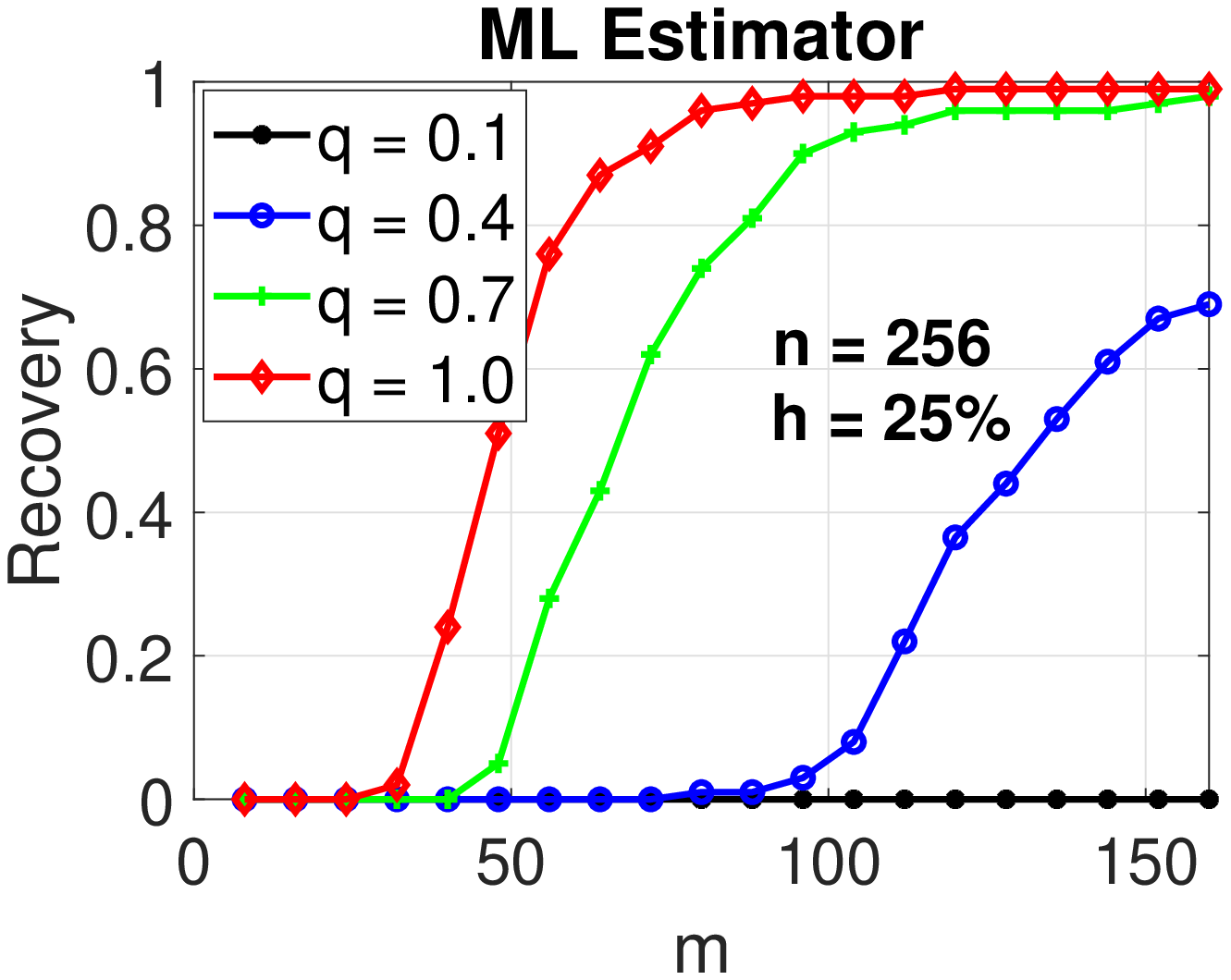}
		\includegraphics[width=3in]{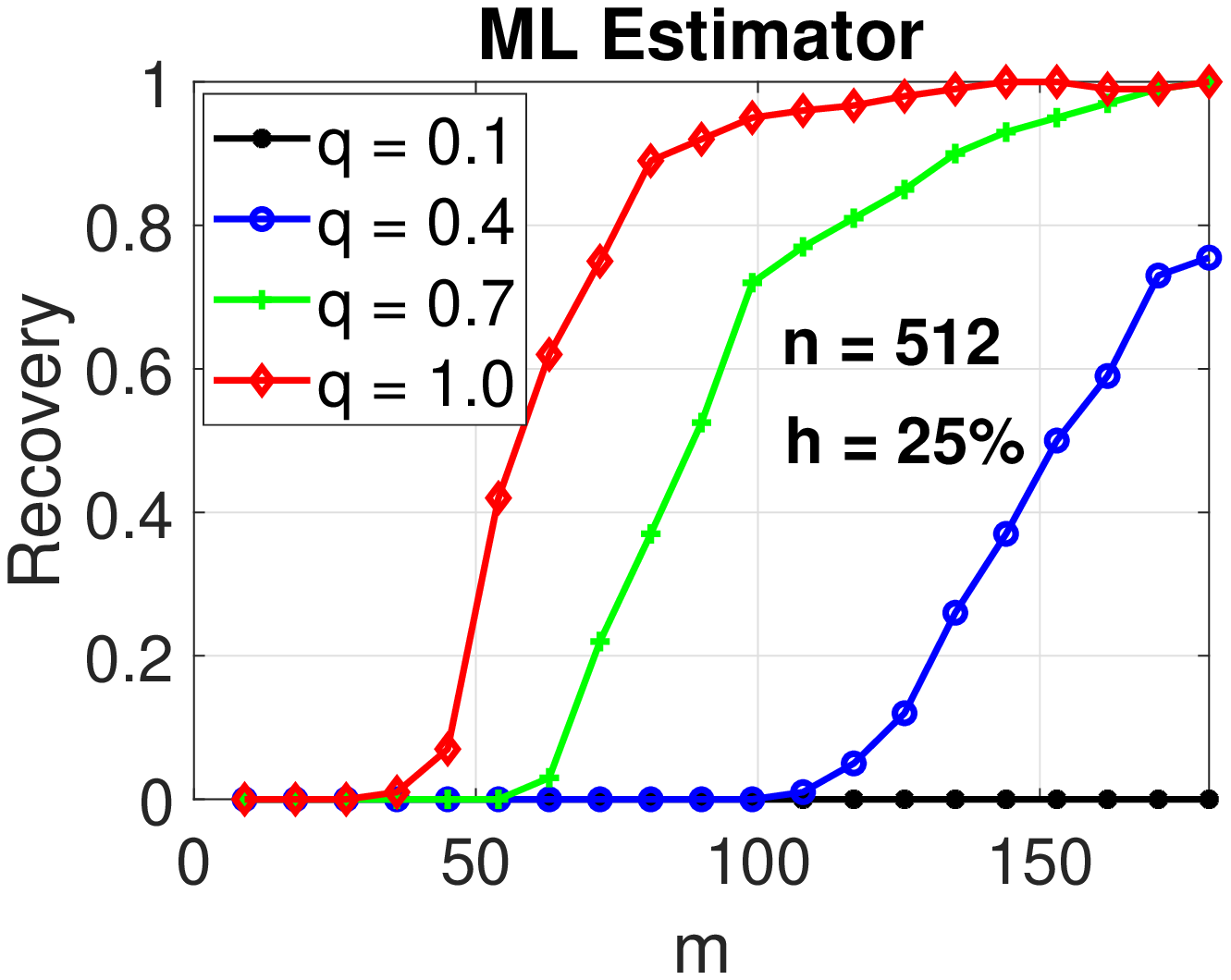}
	}
	\mbox{
		\includegraphics[width=3in]{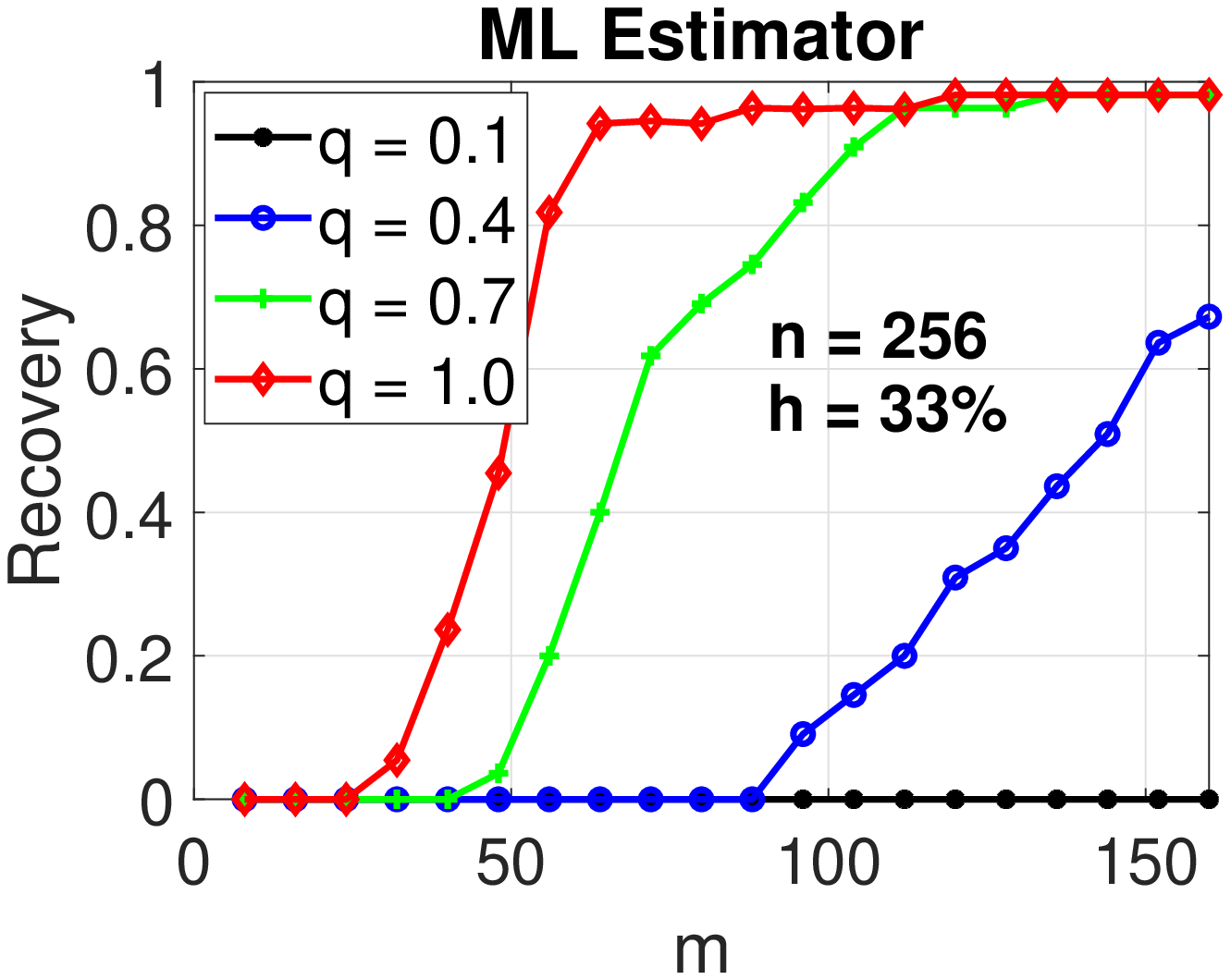}
		\includegraphics[width=3in]{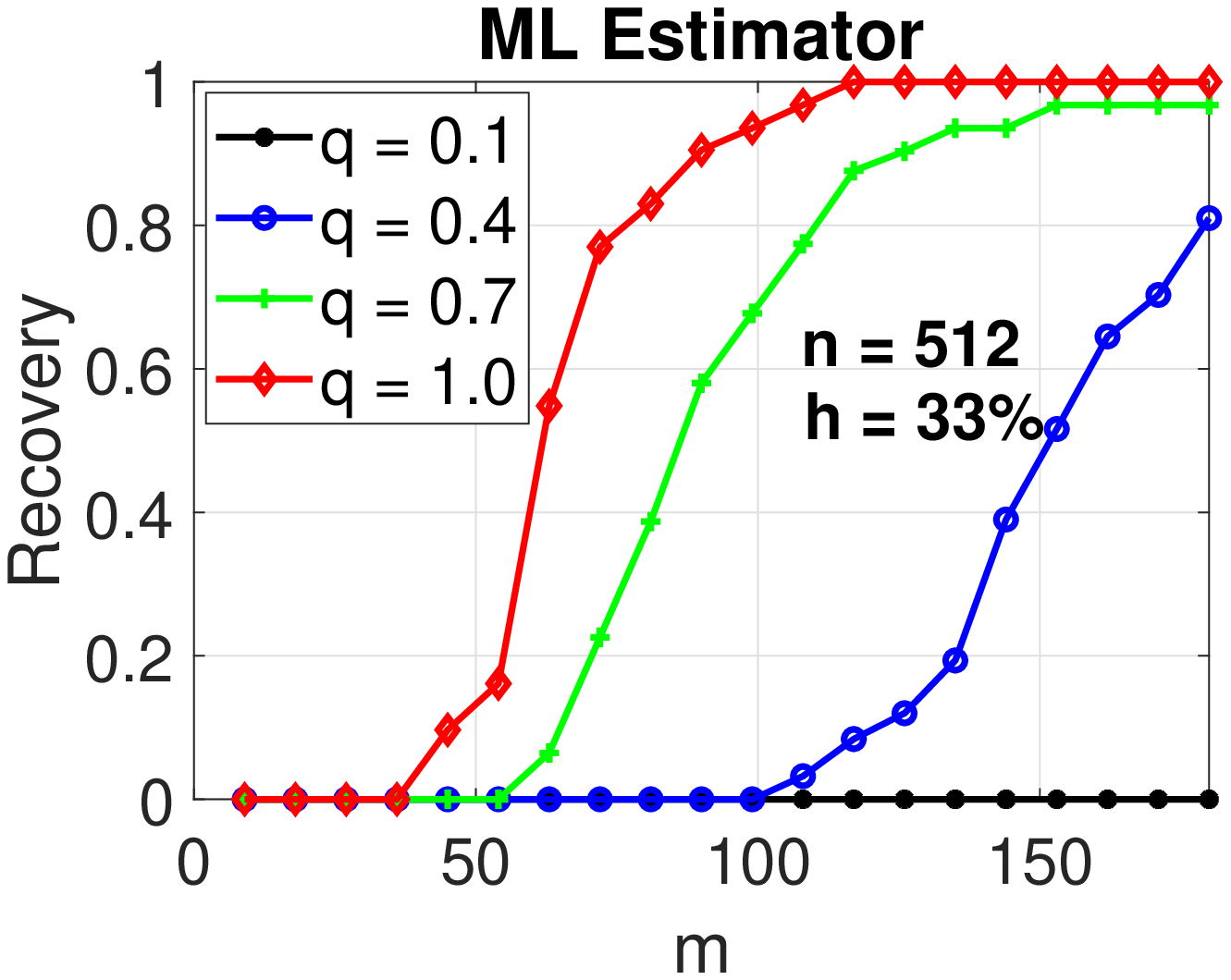}
	}	
	
	\end{center}
	\vspace{-0.2in}
	\caption{The curves of label permutation recovery under different $m$, $q$ and $h$ by using maximum likelihood estimation algorithm. Upper left: $n = 256, h = 0.25 n$; Upper right: $n = 512, h = 0.25 n$; Bottom left: $n = 256, h = 0.33 n$; Bottom right: $n = 512, h = 0.33 n$.  Each point is the average of 500 replications. }\label{fig:mle}
\end{figure}

\noindent\textbf{Setting 1} \  In the first simulation setting, we consider to evaluate the performance of maximum likelihood estimation method.
	We set $n$ to be 256 and 512 and let 25\% or 33 \% labels be permuted. We vary $m$ from $\{\log_2 n, 2 \log_2 n, \ldots, 20 \log_2 n\}$ and set observation rate $q$ at different levels.
	For design matrix $X$, each row independently follows a multivariate Gaussian distribution $N(\mathbf 0, I_p/p)$ ($p = 10$).
	For coefficient matrix $B$, each element is i.i.d. standard Gaussian random variable.
	The curves of probability for successful permutation recovery are plotted in Figure~\ref{fig:mle}.
	
\newpage

\begin{figure}[h]
	\begin{center}
	\mbox{
		\includegraphics[width=3in]{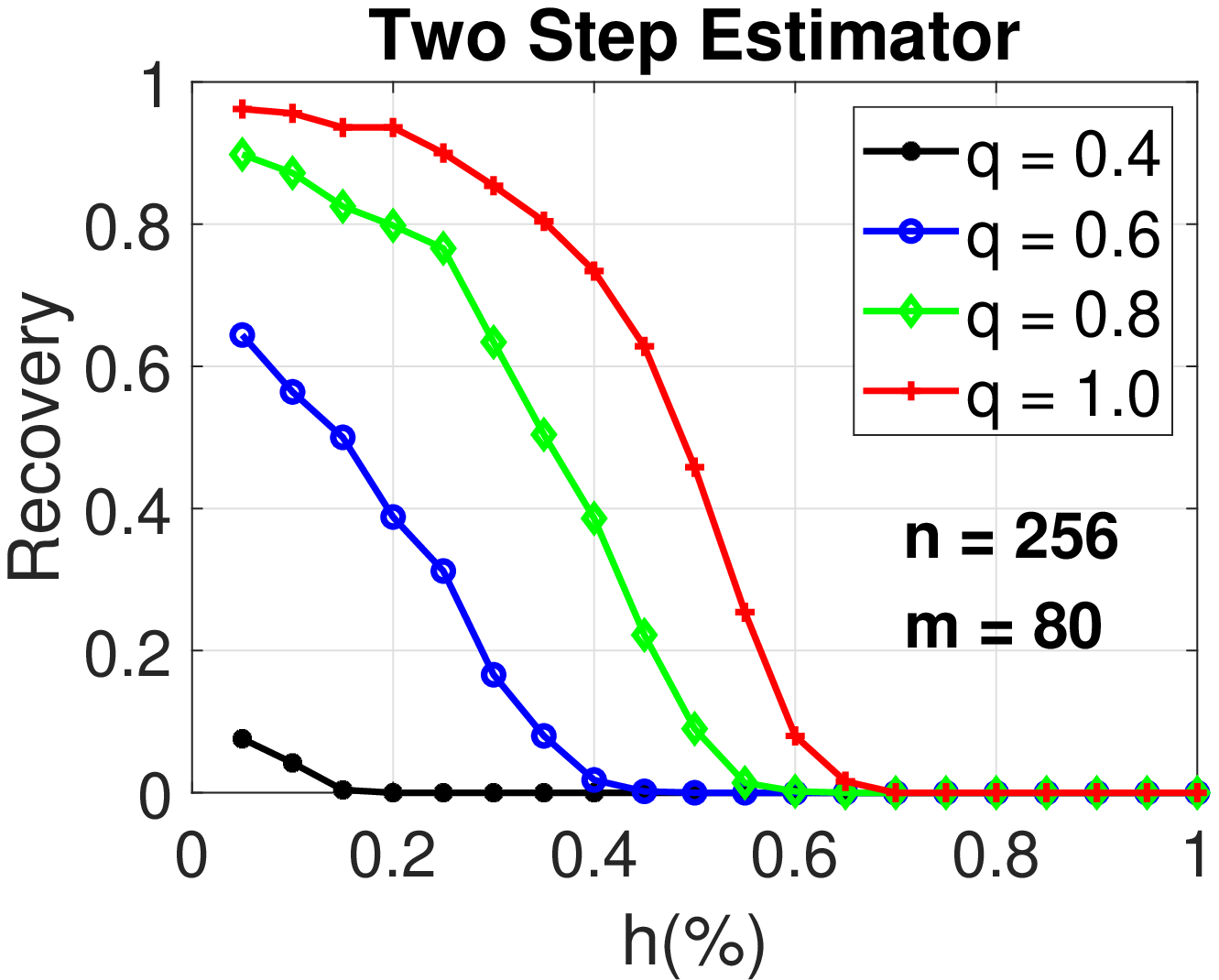}
		\includegraphics[width=3in]{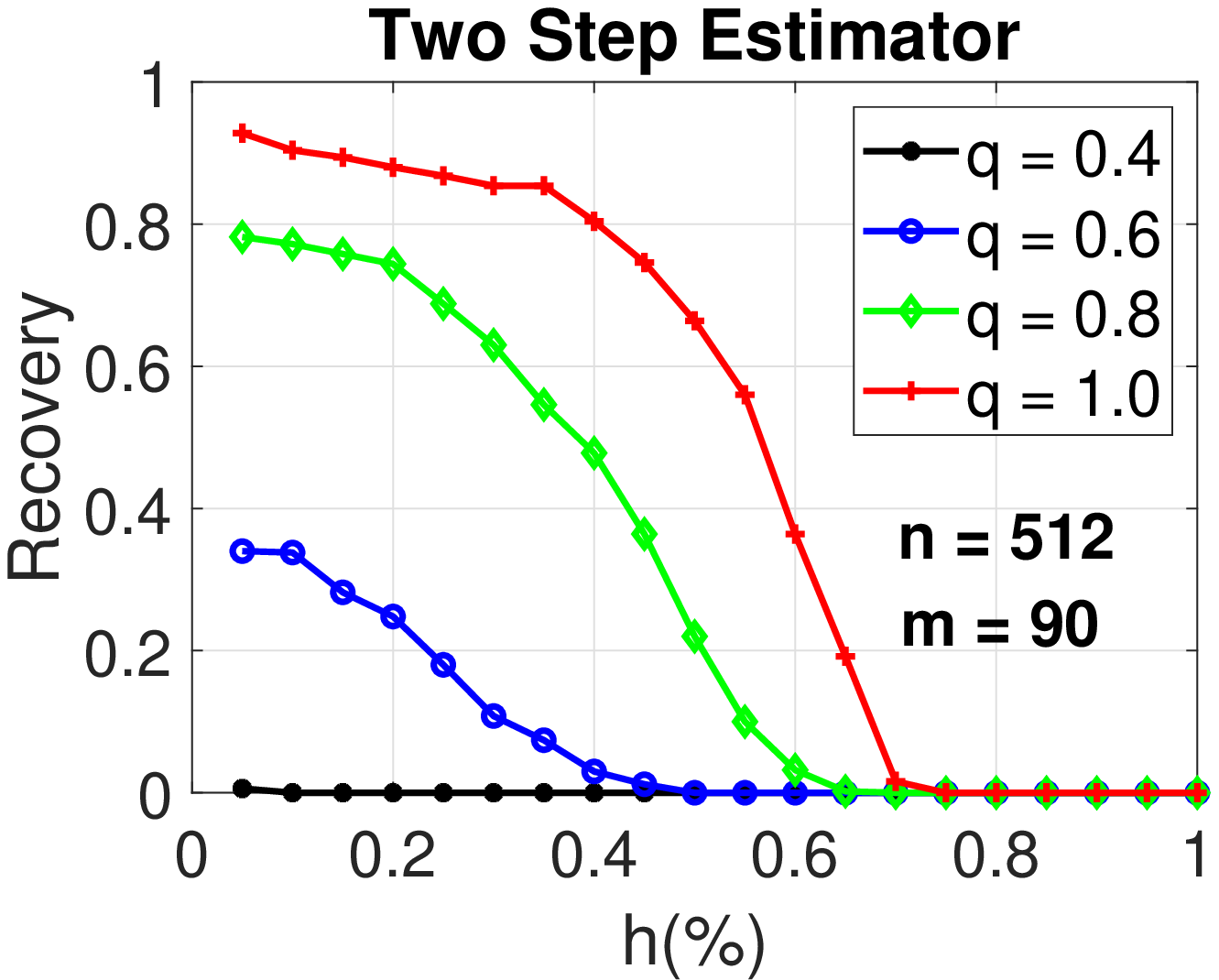}
		}
	\mbox{
		\includegraphics[width=3in]{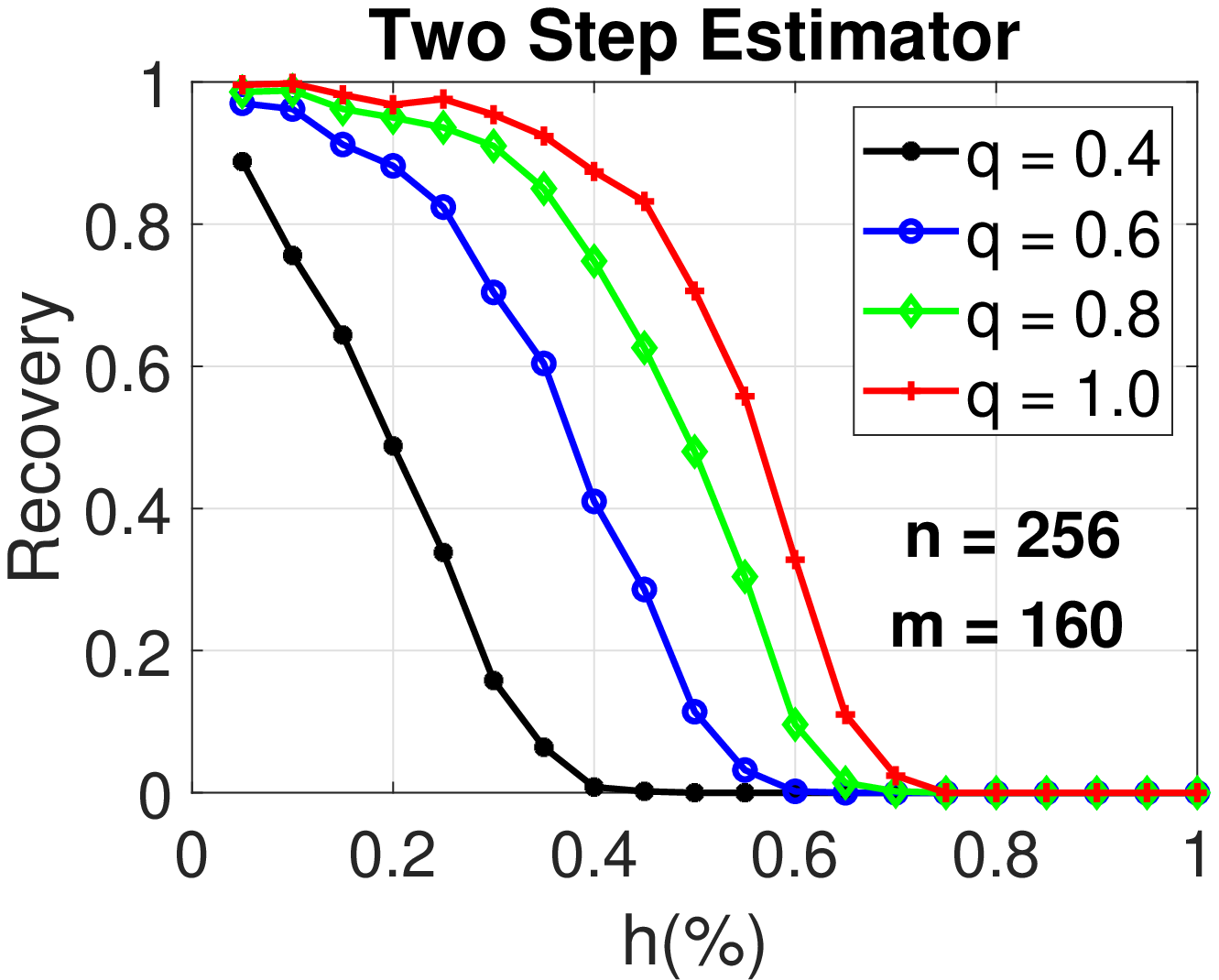}
		\includegraphics[width=3in]{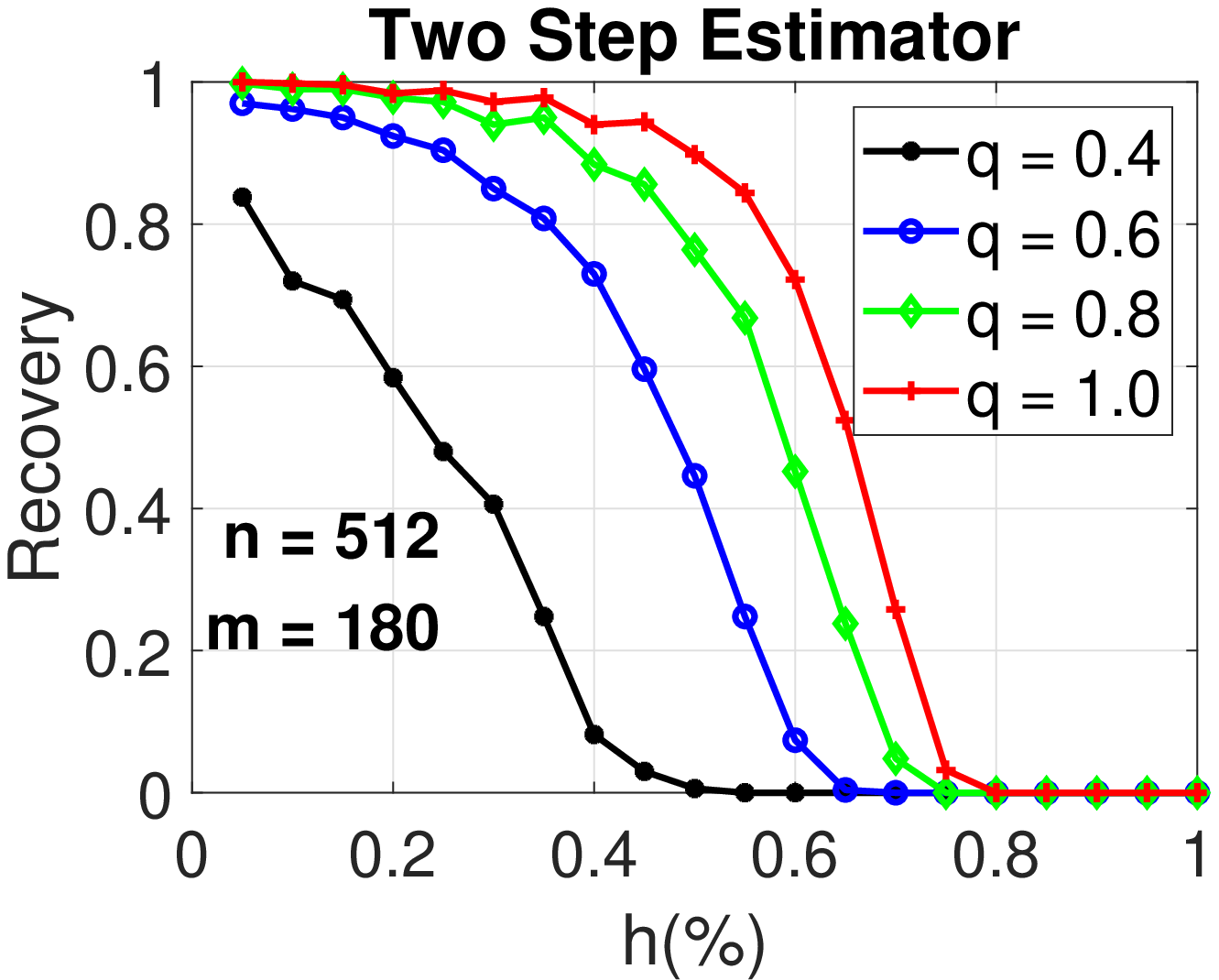}
		}		
		
	\end{center}
	\vspace{-0.2in}
	\caption{The curves of label permutation recovery under different $n$, $m$ and $h$ by using two-step estimation algorithm. Upper left: $n = 256, m = 80$; Upper right: $n = 512, m = 90$; Bottom left: $n = 256, m = 160$; Bottom Right: $n = 512, m = 180$. Each point is the average of 500 replications.}\label{fig:2step}
\end{figure}

\noindent\textbf{Setting 2} \ In the second simulation setting, we illustrate the performance of two-step estimation method. We deliberately permute the true label by some proportions (5\%, 10\%, \ldots, 100 \%).
	We set $n$ to be 256 / 512 and set $m = 10 \log_2 n$ / $ 20 \log_2 n$. The observation rate $q$ varies from $0.4$ to 1.0. The design matrix and coefficient matrix remains the same as in the first setting.
	The curves of probability for successful permutation recovery are plotted in Figure~\ref{fig:2step}.
	
\vspace{0.1in}

\begin{figure}[t]
	\begin{center}
	
		\includegraphics[width=3in]{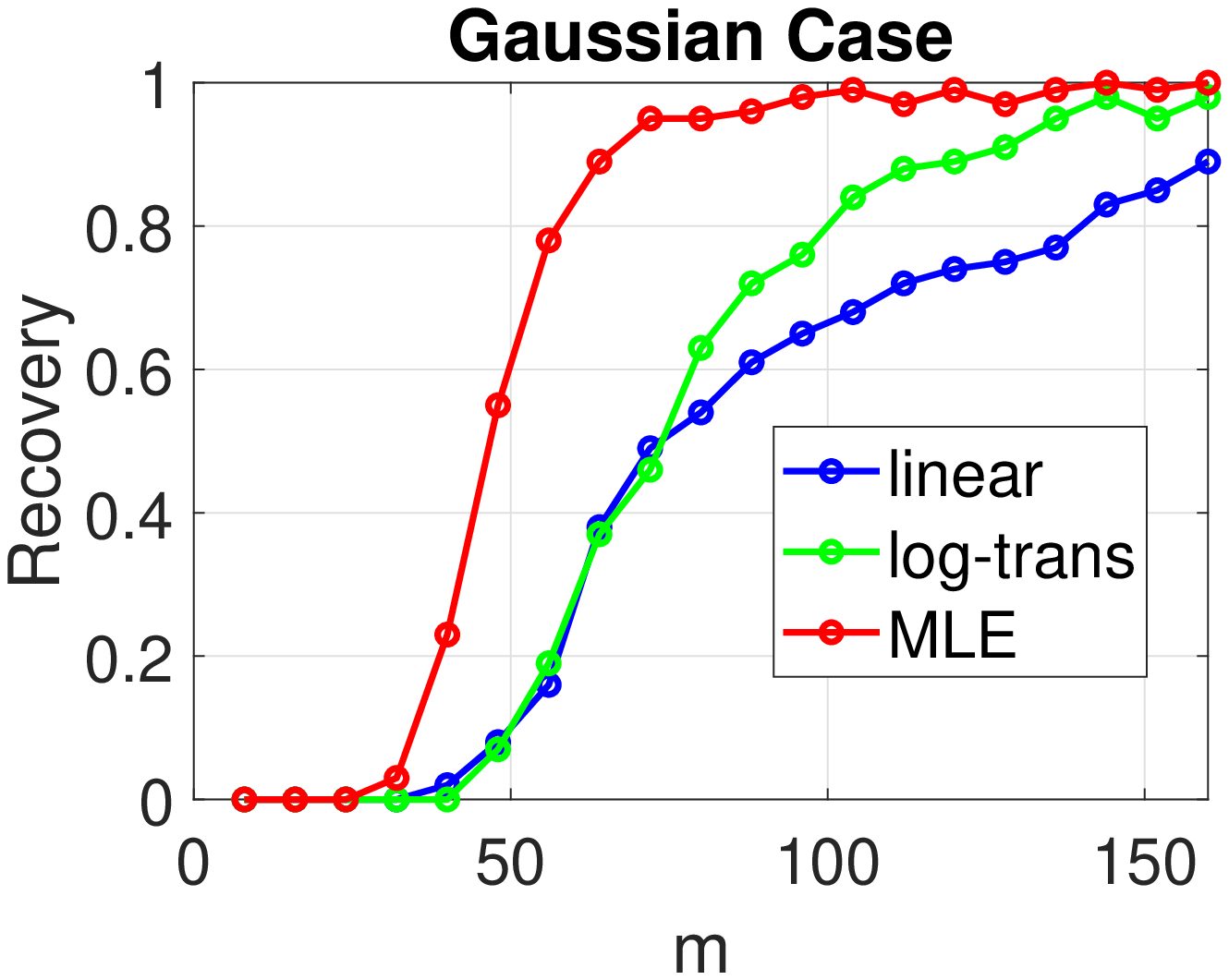}
		\includegraphics[width=3in]{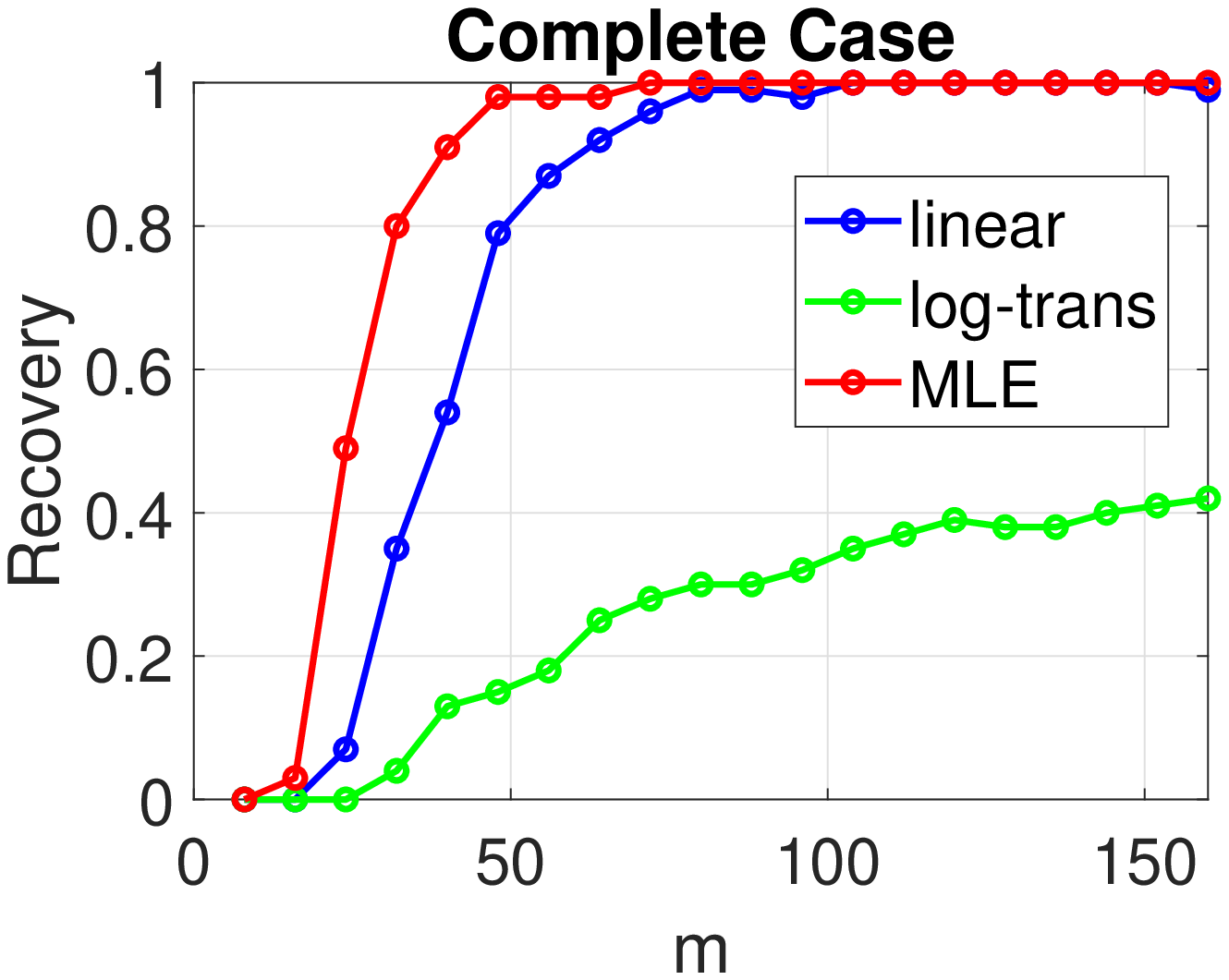}
		\includegraphics[width=3in]{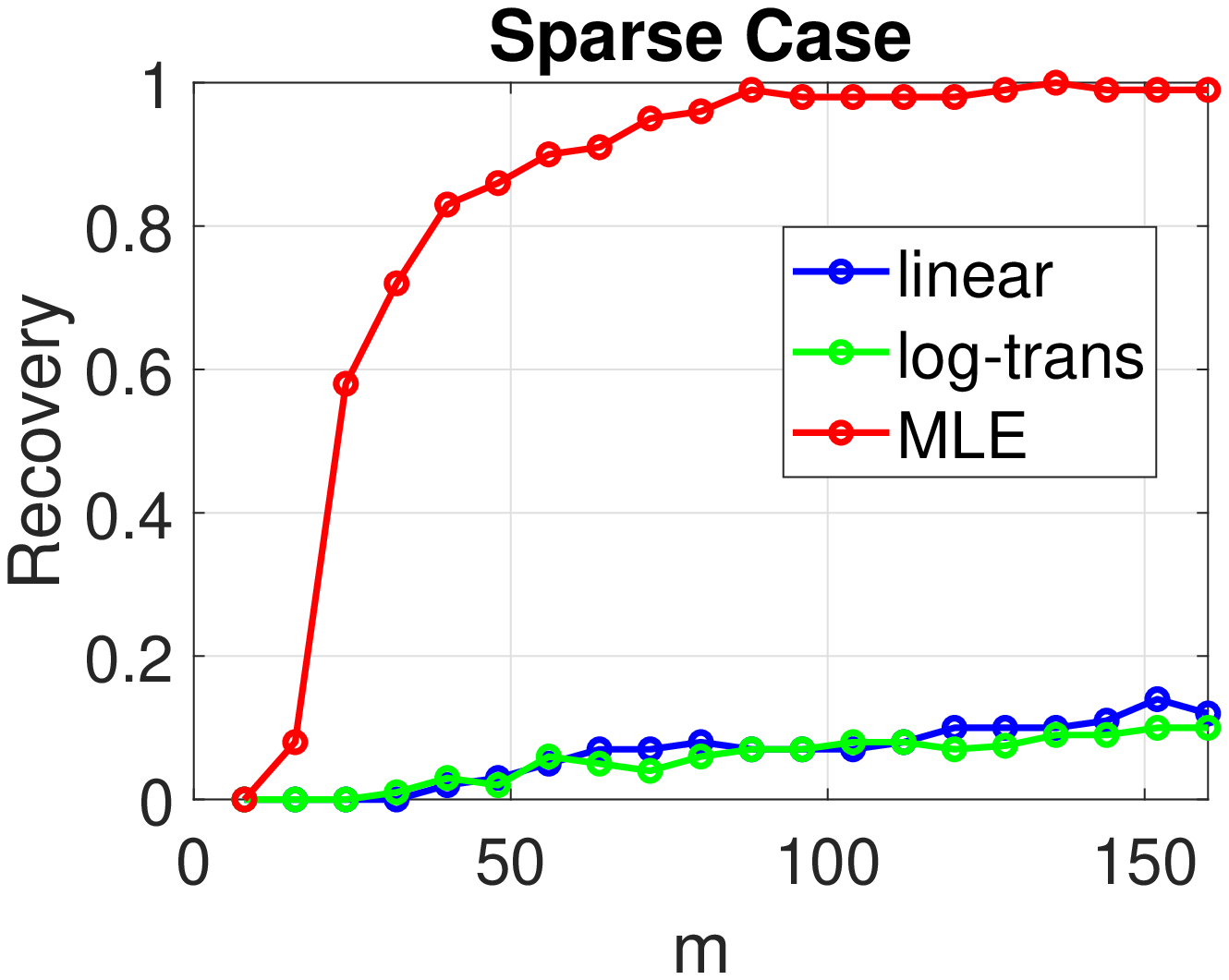}
		\includegraphics[width=3in]{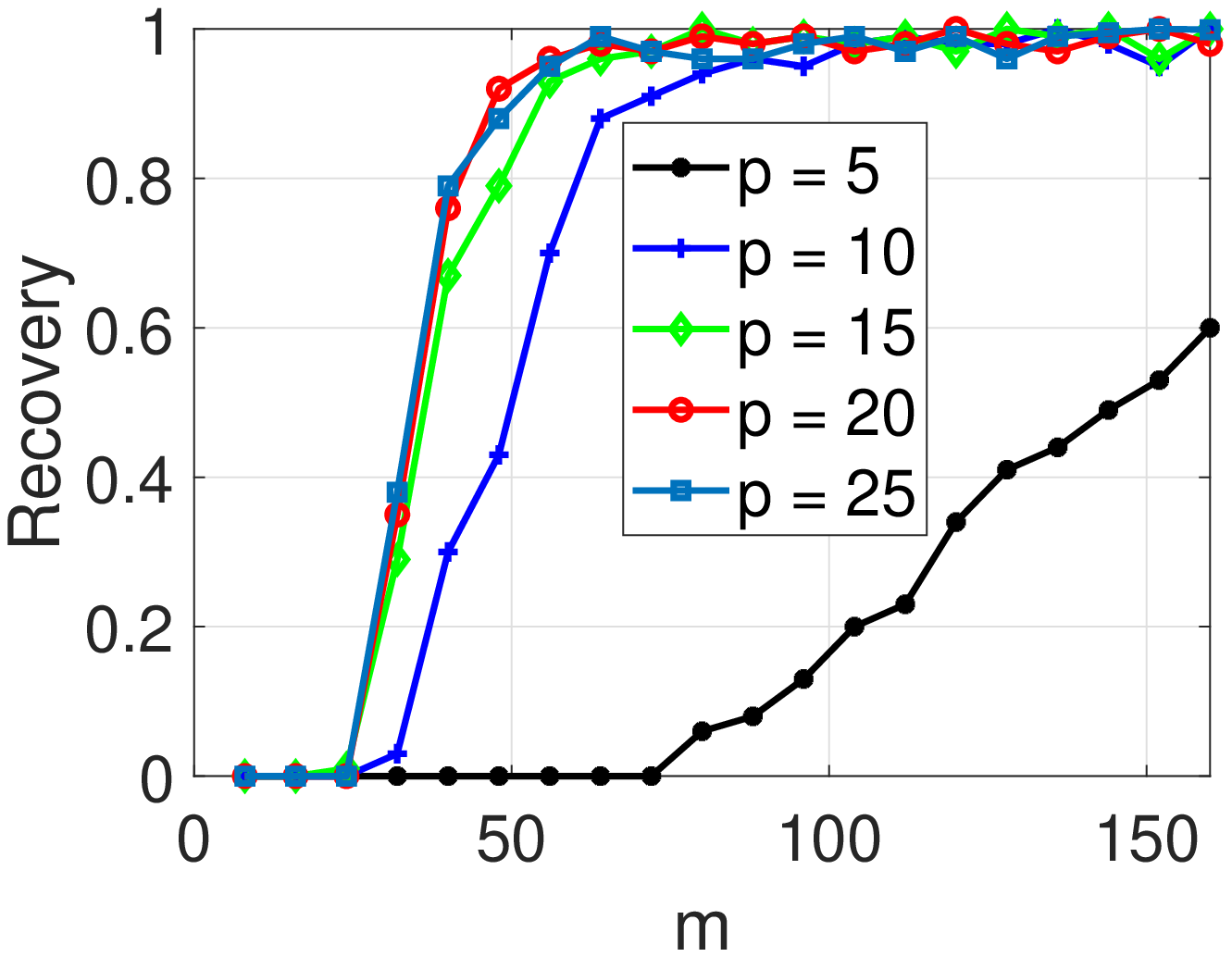}

	\end{center}
	\vspace{-0.2in}
	\caption{The top left, top right and bottom left plots show the curves of label permutation recovery under different design settings by using different estimation methods. The bottom right plot shows the permutation recovery curves under different $p$. }\label{fig:comparison}
\end{figure}

\noindent\textbf{Setting 3} \ In the third simulation setting, we compare with the results by fitting a linear model directly to the original data (``linear'') or to the log-transform of data (``log-trans'') under different generation schemes. (We use the ADMM-based algorithm in Section~\ref{sec:comp_mtd} for implementation.)
	\begin{enumerate}
		\item For design matrix $X$, each row independently follows a multivariate Gaussian distribution $N(\mathbf 0, I_p/p)$.
		For coefficient matrix $B$, each element is i.i.d. standard Gaussian random variable. In this case, we set $n = 256$ and $p = 10$.
		\item Matrix $X$ is a complete design matrix .
		For coefficient matrix $B$, each element is i.i.d. uniform random variable on $U(0,2)$. In this case, $n = 256$, $p = 1 + \log_2 n$.
		\item For sparse design matrix $X$, each row has at most $s$ non-zero entries and positions of non-zero elements are sampled uniformly.
		For coefficient matrix $B$, each element is i.i.d. uniform random variable on $U(0,2)$. In this case, we set $n = 256$, $p = 20$ and $s = 5$.
	\end{enumerate}
	Such comparisons under model mis-specification are shown in Figure~\ref{fig:comparison}.
	

\newpage

\noindent\textbf{Setting 4} \ In the fourth simulation setting, we consider to evaluate the performance of maximum likelihood estimator when $p$ varies.
	We set $n$ to be 256 and 512 and let 25\% of labels be permuted. We vary $m$ from $\{\log_2 n, 2 \log_2 n, \ldots, 20 \log_2 n\}$.
	For design matrix $X$, each row independently follows a multivariate Gaussian distribution $N(\mathbf 0, I_p/p)$ with $p = 5, 10, 15, 20$ or $25$.
	For coefficient matrix $B$, each element is i.i.d. standard Gaussian random variable.
	The curves of probability for successful permutation recovery are shown in the bottom-right plot in Figure~\ref{fig:comparison}.

\vspace{0.1in}
From Figure~\ref{fig:mle}, we can see that the probability of successful label recovery increases as $m$ increases. The probability changes drastically from 0 to 1 when $m \approx 10 \log_2 n$. This matches our theory. In addition, we can see that $m$ required for perfect permutation recovery gets larger as observation rate $q$ decreases.
From Figure~\ref{fig:2step}, we can observe that the probability of successful label recovery decreases as proportion of wrong label increases. The probability changes drastically from 1 to 0 when $20 \%$ of individuals are given with wrong labels.
Additionally, as the observation rate decreases, the successful recovery probability also decreases.
From Figure~\ref{fig:comparison}, we can see that the recovery results will get worse if we fit the data generated from log-linear model by using linear methods. Thus, model mis-specification (i.e. non-Gaussian setting) may lead to bad recovery results.
Furthermore, we can see that the value of $p$ does not effect the recovery result when it is in the suitable regime of $n$, i.e., $p \gtrsim \log n$ and $p \lesssim n^{1/2}$. This matches our findings in theory and Example 4.

\section{Conclusion}\label{sec:disscussion}

In this paper, we provide theoretical analyses of label permutation problem for the generalized linear model.
The theory takes multivariate responses into account and is established under three different scenarios, with knowledge of $B^{\sharp}$, with knowledge of $d(\mathbf I, \Pi^{\sharp})$ and without any knowledge.
Our results are more general and remove the stringent conditions which are required by the case when $m = 1$.
A detailed comparison with existing literature are also provided to emphasize the technical challenges of considered setting.
We further extend our results to the missing observation setting which has never been considered in the literature of label permutation problem.
We also propose two computational methods, ``maximum likelihood estimation'' algorithm with warm start and ``two-step estimation'' algorithm.
When the proportion of permuted labels is not too large, both methods work effectively under different settings of generating design matrix $X$.
Experimental results match our theoretical findings.
In practice,
our computational methods sometimes may fail to find the global optimum when the proportion of permuted labels is large. Developing more efficient estimation methods constitutes a further promising direction.

\vspace{0.8in}

\bibliographystyle{plainnat}
\bibliography{labelref}

\clearpage

\appendix

\section{Explanation of Four Examples}\label{sec:fourexample}

In this appendix, we provide detailed explanations for examples given in Section~\ref{sec:known}.

\vspace{0.1in}
\noindent\textbf{Example~\ref{eg1}}.
In this case, by convexity of $\psi$, we find that
\[\Delta_{i j} = \langle \psi'(\llam_i) \circ \llam_i - \psi(\llam_i) \rangle - \langle \psi'(\llam_i) \circ \llam_{j} - \psi(\llam_{j}) \rangle \geq  \frac{1}{2} \kappa_{0} \sum_{l} (x[i]B^{\sharp}[l] - x[j]B^{\sharp}[l]|)^2 \geq \frac{m \kappa_0}{2} x_{gap,ij}^2 b_1^2,\]
where $\kappa_0$ is the minimum $\psi''(x)$ over the range of $x = z_1 z_2$ with $z_1 \in [a_1,a_2]$ and $z_2 \in [b_1,b_2]$.
We also have
\[v_{i j} = \sum_{l=1}^m \psi''(\llam_i[l]) (\llam_{j}[l] -  \llam_{i}[l])^2 \leq m \kappa_1 b_2^2 x_{gap,ij}^2,\]
where $\kappa_0$ is the maximum $\psi''(x)$ over the range of $x = z_1 z_2$ with $z_1 \in [a_1,a_2]$ and $z_2 \in [b_1,b_2]$.

Therefore, even with the knowledge of true parameter matrix $B^{\sharp}$, we still need large $m$ to ensure the recovery of $\Pi^{\sharp}$. That is,
\begin{eqnarray}
\frac{m \kappa_0}{2} x_{gap,ij}^2 b_1^2 \gtrsim \sqrt{(\log n) m \kappa_1 b_2^2 x_{gap,ij}^2} \nonumber
\end{eqnarray}
for any pair of $i,j$.
By simplification, we require $m \geq \max_{i,j} K \frac{\log n}{x_{gap,ij}^2} \approx n^4 \log n$ with some constant $K$.

\vspace{0.1in}
\noindent\textbf{Example~\ref{eg2}}.
In this case, we define $w_{l,ij} := \xx_i^T \bb_l - \xx_j^T \bb_l$ and $d_{ij} := \sum_{k} \mathbf 1\{X[i,k] \neq X[j,k]\}$ ($d_{ij} \geq 1$). We can find that $w_{l,ij}^2 / d_{ij}$ has mean $1$ and variance $O(1)$. Then we have $\sum_{l=1}^m w_{l,ij}^2 = \Theta_p(m d_{ij})$.

Note that $\psi(x)$ is strictly convex, then there is a constant $\kappa_0$ such that $\psi(y) \geq \psi(x) + \psi'(x)(y-x) + \frac{\kappa_0}{2} (y - x)^2$ for any $x, y$.

By the same reason in Example~\ref{eg1}, for any $i \neq j$, we have
\[\Delta_{i j} = \langle \psi'(\llam_i) \circ \llam_i - \psi(\llam_i) \rangle - \langle \psi'(\llam_i) \circ \llam_{j} - \psi(\llam_{j}) \rangle \geq \frac{\kappa_0}{2} \sum_{l=1}^m w_{l,i j}^2\]
held for some constant $c$ and
also have
\[v_{i j} = \sum_{l=1}^m \psi''(\llam_i[l]) (\llam_{j}[l] - \llam_i[l])^2 \leq
\kappa_1 \sum_{l=1}^m w_{l,ij}^2.\]

Therefore, when the true parameter matrix $B^{\sharp}$ is known, we need large $m$ to ensure the recovery of $\Pi^{\sharp}$. That~is,
\begin{eqnarray}
\frac{\kappa_0}{2} \sum_{l=1}^m w_{l,i j}^2 \gtrsim \sqrt{(\log n) \kappa_1 \sum_{l=1}^m w_{l,ij}^2}. \nonumber
\end{eqnarray}
By simplification, we require
\[\sum_{l=1}^m w_{l,i j}^2 \geq \frac{2 \kappa_1}{\kappa_0} \log n. \]
Using $\sum_{l=1}^m w_{l,ij}^2 = \Theta_p(m d_{ij})$, it suffices to require $m \gtrsim \max_{ij} \frac{\log n}{ d_{ij}} = O(\log n)$.

\newpage

\noindent\textbf{Example~\ref{eg3}}.
In this case, by repeating the same procedure as in Example~\ref{eg2}, we require
\[\sum_{l=1}^m w_{l,i j}^2 \geq \frac{2 \kappa_1}{\kappa_0} \log n. \]
It suffices to find the lower bound of $\sum_{l=1}^m w_{l,i j}^2$.
In this case, with high probability, $w_{l,ij}^2$ is lower bounded by $ c a_1^2 d_{ij}$, where $d_{ij} = \sum_{l} \mathbf 1\{X[i,l] - X[j,l] \neq 0\} \geq 1$ according to the assumption that each row of $X$ has different support.
Thus $\sum_{l=1}^m w_{l,i j}^2$ is bounded below by $c a_1^2 m$. We hence require that $m \gtrsim \log n$ for perfect recovery.

\vspace{0.1in}
\noindent\textbf{Example~\ref{eg4}}.
Following the same reason in Example~\ref{eg2}, we require
\[\sum_{l=1}^m w_{l,i j}^2 \geq \frac{2 \kappa_1}{\kappa_0} \log n. \]
It also suffices to find the lower bound of $\sum_{l=1}^m w_{l,i j}^2$.
In this case, given fixed $\xx_i, \xx_j$, $w_{l,ij}^2$, $w_{l,ij}^2 / \|\xx_i - \xx_j\|^2$ follows a Chi-square distribution with degree 1  bounded by $ c a_1^2 d_{ij}$.
With high probability, it holds $\sum_l w_{l,ij}^2 = \Theta_p(m \|\xx_i - \xx_j\|^2)$
Therefore, it suffices to have $m \gtrsim \max_{i,j} \frac{\log n}{\|\xx_i - \xx_j\|^2} = \Theta_p(\log n)$.

\section{On $\Delta(X,B^{\sharp},\Pi^{\sharp}, \Pi)$}\label{app:delta}

By the definition, we can see that there is no explicit form for $\Delta(X,B^{\sharp}, \Pi^{\sharp}, \Pi)$.
In this appendix, we provide a discussion on the lower bound of $\Delta(X,B^{\sharp},\Pi^{\sharp},\Pi)$.

Note that we can always rewrite $\Pi^{\ast} X$ as $X$ and treat $\Pi(\Pi^{\sharp})^{-1}$ as $\Pi$.
We then assume that $\Pi^{\sharp} = \mathbf I$ for the sake of simplicity.
Moreover, we only need to consider $m = 1$ by noticing that $\Lambda(\Pi, B)$ can be written as the separate function of each column of $B$.

Take any $\Pi \neq \mathbf I$ and let $h:= d(\Pi, \mathbf I)$.
Here, without loss of generality, we assume that $\Pi(i) = i$ for any $i > h$.
By the definition that $\Lambda(\Pi) = \max_{\mathbf b} \Lambda(\Pi, \mathbf b) = \max_{\mathbf b} \sum_{i=1}^n \Lambda_i(\Pi, \mathbf b)$, and $\mathbf b^{\sharp}$ is the true parameter, we then have that
\begin{eqnarray}\label{eq:condition:gap}
\Lambda(\mathbf I) - \Lambda(\Pi) &=& \Lambda(\mathbf I, \mathbf b^{\sharp}) - \max_{\mathbf b} \Lambda(\Pi, \mathbf b) \nonumber \\
&=& \sum_{i=1}^n \Lambda_i(\mathbf I, \mathbf b^{\sharp}) - \max_{\mathbf b} \sum_{i=1}^n \Lambda_i(\Pi, \mathbf b) \nonumber \\
&\geq& \sum_{i=1}^n \Lambda_i(\mathbf I, \mathbf b^{\sharp}) - (\max_{\mathbf b} \sum_{i=1}^h \Lambda_i(\Pi, \mathbf b) + \max_{\mathbf b} \sum_{i= h+1}^n \Lambda_i(\Pi, \mathbf b)) \nonumber \\
&\geq& \sum_{i=1}^h \Lambda_i(\mathbf I, \mathbf b^{\sharp}) - \max_{\mathbf b} \sum_{i=1}^h \Lambda_i(\Pi, \mathbf b)  \nonumber \\
&\geq& \min_{\mathbf b} \sum_{i=1}^h \{ \Lambda_i(\mathbf I, \mathbf b^{\sharp}) - \Lambda_i(\Pi, \mathbf b)\} \nonumber \\
&\geq& \lambda_0 \min_{\mathbf b} \sum_{i=1}^h  (X[i,:] \mathbf b^{\sharp} - X[\Pi(i),] \mathbf b)^2 \nonumber \\
&\geq& \lambda_0  \min_{\mathbf b} d_{gap}^2,
\end{eqnarray}
where $d_{gap} := \min_{\mathbf b}\| X_{\Pi,1} \mathbf b - X_1 \mathbf b^{\sharp}\|$
with $X_{1} := X[1:h,]$ and $X_{\Pi,1} = (\Pi X)[1:h,]$ and $\lambda_0$ is equal to \\ $\min_i \{\psi''(\xx_i^T \mathbf b^{\sharp}), \psi''(\xx_i^T \mathbf b(\Pi))\}$.
Moreover, $d_{gap}$ admits an explicit form, which is,
\begin{eqnarray}
d_{gap} = \|\mathbf P_{X_{\Pi,1}} X_{1} \mathbf b^{\sharp}\|, \nonumber
\end{eqnarray}
where $\mathbf P_{X_{\Pi,1}} = I - X_{\Pi,1}(X_{\Pi,1}^T X_{\Pi,1})^{-1}X_{\Pi,1}^T$.

When $p = 1$, we know that $X_{\Pi,1}^T X_{\Pi,1}$ is equal to $ \|X_{\Pi,1}\|^2$ .
Thus
\begin{eqnarray}
\|(X_{\Pi,1}^T X_{\Pi,1})^{-1}X_{\Pi,1}^T X_1\| &\leq& \|X_{\Pi,1}^T X_1\|/ \|X_{\Pi,1}\|^2 = 1 -  \frac{\|X_{\Pi,1}\|^2 - \|X_{\Pi,1}^T X_1\|}{\|X_{\Pi,1}\|^2} \nonumber \\
&=&  (1 - \frac{\|X_{\Pi,1}\|^2 - \|X_{\Pi,1}^T X_1\|}{\|X_{\Pi,1}\|^2}) \nonumber \\
&=&  (1 - \frac{\|X_{\Pi,1}\|^2 - \|X_{\Pi,1}^T X_1\| + \|X_{1}\|^2 - \|X_{1}^T X_{\Pi,1}\|}{2 \|X_{\Pi,1}\|^2}) \nonumber \\
&\leq&  (1 - \frac{\|X_{\Pi,1} - X_1\|^2}{2 \|X_{\Pi,1}\|^2}). \nonumber
\end{eqnarray}
Thus
\begin{eqnarray}
d_{gap} = \|X_1 \mathbf b^{\sharp} - X_{\Pi,1}(X_{\Pi,1}^T X_{\Pi,1})^{-1}X_{\Pi,1}^T X_1 \mathbf b^{\sharp} \| \geq \|X_1 \mathbf b^{\sharp}\| \frac{\|X_{\Pi,1} - X_1\|^2}{2 \|X_{\Pi,1}\|^2} = \Omega(\sqrt{h}). \nonumber
\end{eqnarray}
Therefore $\Lambda(\mathbf I) - \Lambda(\Pi) \geq c_0 m d(\mathbf I, \Pi)$ for some constant $c_0$ which is related to the design matrix $X$.

When $p > 1$, there exists a rotation matrix $W$ such that $W \mathbf b^{\sharp} = \mathbf e_1 \|\mathbf b^{\sharp}\|$ ($\mathbf e_1$ is a vector with all entries being zero but the first entry being 1). Write $X_1 = \tilde X_1 W$. Then,
\[d_{gap} = \min_{\mathbf b} \|X_{\Pi,1} \mathbf b - X_1 \mathbf b^{\sharp}\|
= \min_{\mathbf b} \| \tilde X_{\Pi,1} \mathbf b - \tilde X_1 \mathbf e_1 \|\mathbf b^{\sharp}\| \|
= \|\mathbf b^{\sharp}\| \min_{\mathbf b} \| \tilde X_{\Pi,1} \mathbf b - \tilde X_1 \mathbf e_1 \|.\]
Thus, we have $d_{gap} = \|\mathbf b^{\sharp}\| \|(I - \tilde X_{\Pi,1}(\tilde X_{\Pi,1}^T \tilde X_{\Pi,1})^{-1} \tilde X_{\Pi,1}^T) \tilde X_1 \mathbf e_1\|
\geq c_0 \|\mathbf b^{\sharp}\| \|\tilde X_1 e_1\| = c_0 \Omega(\sqrt{h})$,
where $c_0$ is the distance from $\tilde X_1 e_1/ \|\tilde X_1 e_1\|$ to the space spanned by $\tilde X_{\Pi,1}$.
Thus $\Lambda(\mathbf I) - \Lambda(\Pi) \geq c_0' m d(\mathbf I, \Pi)$ by adjusting the constant $c_0'$.

\section{Proof of Results when $B$ is Known: Theorem~\ref{thm:1}}\label{app:known}

In this section, we prove the results when $B$ is known.
We additionally use $A[i,:]$/$A[:,j]$ to represent the $i$th row/$j$th column of matrix $A$;
$\textrm{diag}(\mathbf a)$ is the diagonal matrix with $l$th diagonal element being $\mathbf a[l]$; $\|A\|_F$ is the Frobenius norm of matrix $A$;
$\sigma_{min}(A)/\sigma_{max}(A)$ represents the minimum/maximum positive singular value of matrix $A$.

To prove the result, we only need to show the following probability,
\begin{eqnarray}
P(\sup_{\Pi \neq \Pi^{\sharp}} L(\Pi, B) \geq  L(\Pi^{\sharp}, B)), \label{uni-bound}
\end{eqnarray}
goes to zero as $n$ and $m$ increase.
The naive union bound will give an upper bound,
\begin{eqnarray}
\sum_{\Pi \neq \Pi^{\sharp}} P(L(\Pi, B) \geq  L(\Pi^{\sharp}, B)). \nonumber
\end{eqnarray}
Note that there are $n !$ possible permutations. The above quantity could be exponentially large.
However, we can find that log-likelihood $L(\Pi, B) = \sum_{i=1}^n \langle - \psi((\Pi X B)[i,:]) + Y \circ (\Pi X B)[i,:] \rangle $ is an additive function of $X$'s rows.
Therefore, \eqref{uni-bound} is bounded by
\begin{eqnarray}
&\leq & P(\max_{j \neq i} \langle - \psi((XB)[j,:]) + (Y[i,:] \circ (XB)[j,:]) \rangle \geq \langle - \psi((XB)[i,:]) + (Y \circ XB)[i,:]\rangle) \nonumber \\
&\leq& \sum_{j \neq i} P( \langle - \psi((XB)[j,:]) + (Y[i,:] \circ (XB)[j,:]) \rangle \geq \langle - \psi((XB)[i,:]) + (Y \circ XB)[i,:]\rangle) \nonumber
\end{eqnarray}
Next we bound each term, $P( \langle - \psi((XB)[j,:]) + (Y[i,:] \circ (XB)[j,:]) \rangle \geq \langle - \psi((XB)[i,:]) + (Y \circ XB)[i,:]\rangle)$, in above inequality.

Recall the definition of $\llam_i$, we thus have
\begin{eqnarray}
& & \mathbb E \langle \yy_i \circ \llam_i - \psi(\llam_i) \rangle \nonumber \\
&=& \langle \psi'(\llam_i) \circ \llam_i - \psi(\llam_i) \rangle.
\end{eqnarray}
It can be checked that
\begin{eqnarray}
\Delta_{ij} &=& \langle \psi'(\llam_i) \circ \llam_i - \psi(\llam_i) \rangle - \langle \psi'(\llam_i) \circ \llam_j - \psi(\llam_j) \rangle
\geq 0 
\end{eqnarray}
for any convex function $\psi$.

For any $i \neq j$, we next calculate the variance of $\langle \yy_i \circ \llam_i - \psi(\llam_i) \rangle - \langle \yy_i \circ \llam_j - \psi(\llam_j) \rangle$.
\begin{eqnarray}
& & \mathrm{var}(\langle \yy_i \circ \llam_i - \psi(\llam_i) \rangle - \langle \yy_i \circ \llam_j - \psi(\llam_j) \rangle) \nonumber  \\
&\leq& \sum_{l=1}^m \mathrm{var} (Y[i,l] (\llam_i[l] - \llam_j[l])) \nonumber \\
&=& \sum_{l=1}^m \psi''(\llam_i[l]) (\llam_i[l] - \llam_j[l])^2 =: v_{ij}.
\end{eqnarray}

To characterize the difference between $\Delta_{ij}$ and $\langle Y[i,:] \circ \llam_i - \psi(\llam_i) \rangle - \langle Y[i,:] \circ \llam_j - \psi(\llam_j) \rangle$, we can show that
\begin{eqnarray}
& & P(|\langle \yy_i \circ \llam_i - \psi(\llam_i) \rangle - \langle \yy_i \circ \llam_j - \psi(\llam_j) \rangle - \Delta_{ij}|  \geq v_{ij} x) \nonumber \\
& = & P(|\langle (\yy_i - \psi'(\llam_i)) \circ (\llam_i - \llam_j) \rangle|  \geq v_{ij} x) \nonumber \\
&\leq& \exp\{- t v_{ij}  x\} \exp\{v_{ij} t^2\} \label{pois:mgf}\\
&\leq& \exp\{ - v_{ij} x^2 / 4\} \label{concen:1},
\end{eqnarray}
where \eqref{pois:mgf} utilizes the property of moment generating function of generalized linear model. That is,
it is well known that
$\mathbb E[\exp\{tY\}] = \exp\{\psi(\lambda + t) - \psi(\lambda)\}$,
where the density of $Y$ is proportional to
$\exp\{y \lambda - \psi(\lambda)\}$.
Then we have
\begin{eqnarray*}
	& & \exp\{t \langle (\yy_i - \psi'(\llam_i)) \circ (\llam_i - \llam_j) \rangle\} \\
	&=& \prod_{l=1}^m \exp\{\psi(\llam_i[l] + t(\llam_i[l] - \llam_j[l])) - \psi(\llam_i[l]) - t \psi'(\llam_i[l])(\llam_i[l] - \llam_j[l])\} \\
	&\leq& \prod_{l=1}^m \exp\{\psi''(\llam_{i}[l]) (\llam_i[l] - \llam_j[l])^2 t^2\} \\
	&=& \exp\{ t^2 \sum_{l=1}^m \psi''(\llam_{i}[l]) (\llam_i[l] - \llam_j[l])^2\} \\
	&=& \exp\{ v_{ij} t^2 \}
\end{eqnarray*}
for any small $t$ satisfying
$\frac{1}{2}\psi''(\lambda_i[l]) > |\psi'''(\lambda_i[l]) (\lambda_i[l] - \lambda_j[l])t|$.

By taking $x = \Delta_{ij}/2v_{ij}$, yhis gives
\begin{eqnarray}
& & P( \langle - \psi((XB)[j,:]) + (Y[i,:] \circ (XB)[j,:]) \rangle \geq \langle - \psi((XB)[i,:]) + (Y \circ XB)[i,:]\rangle) \nonumber \\
& \leq & P( \langle - \psi((XB)[j,:]) + (Y[i,:] \circ (XB)[j,:]) \rangle \geq \langle - \psi((XB)[i,:]) + (Y \circ XB)[i,:]\rangle -\frac{\Delta_{ij}}{2})  \nonumber \\
&\leq& \exp\{- \Delta_{ij}^2/(16 v_{ij})\}.
\end{eqnarray}
Finally, by union bound over all possible pairs of $i$ and $j$, this completes the proof of Theorem~\ref{thm:1}.

\section{Proof of Results with Knowledge that $d(\mathbf I, \Pi^{\sharp})$ is Small: Theorem~\ref{thm:known}}\label{app:twostep}

In this section, we prove the results when we have the prior knowledge that
$d(\mathbf I, \Pi^{\sharp})$ is small.
In order to prove the recovery consistency, we need to control the following quantities,
$\|B - B^{\sharp}\|_F$
and
$(L(\mathbf I, B) - L(\Pi, B)) - (\Lambda(\mathbf I, B) - \Lambda(\Pi, B))$.

Suppose we have already known that the estimator $\hat {B}$ which is close to the truth $B^{\sharp}$, i.e., in the $\delta$-neighborhood of $B^{\sharp}$.
Then we can show that
\begin{eqnarray}\label{eq:goal:concentration2}
P( \sup_{B \in B_{\delta}(B^{\sharp})} |(L(\mathbf I, B) - L(\Pi, B)) - (\Lambda(\mathbf I, B) - \Lambda(\Pi, B))| \geq 2 v_{\Pi, partial} x )
\end{eqnarray}
vanishes for any fixed $\Pi$, where $B_{\delta}(B^{\sharp}) := \{B : \|\bb_l - \bb_l^{\sharp}\|_{2} \leq \delta, ~ l \in [m]\}$ and $\delta$ is a sufficiently small constant which is determined later.
Therefore,
\begin{eqnarray}\label{eq:goal:concentration:single2}
& & P( \sup_{ B \in B_{\delta}(B^{\sharp})}  |(L(\mathbf I, B) - L(\Pi, B)) - (\Lambda(\mathbf I, B) - \Lambda(\Pi, B))| \geq 2 v_{\Pi,partial} x ) \nonumber \\
&\leq& P( |(L(\mathbf I, B^{\sharp}) - L(\Pi, B^{\sharp})) - (\Lambda(\mathbf I, B^{\sharp}) - \Lambda(\Pi, B^{\sharp}))| \geq v_{\Pi,partial}x ) \nonumber \\
&\leq& \exp\{ - \frac{1}{4} v_{\Pi,partial} x^2 \}.
\end{eqnarray}
Hence, we get
\begin{eqnarray}\label{eq:goal:concentration:combine2}
& & P( \max_{\Pi} \sup_{B \in B_{\delta}(B^{\sharp})}  |(L(\mathbf I, B) - L(\Pi, B)) - (\Lambda(\mathbf I, B) - \Lambda(\Pi, B))| \geq 2 v_{\Pi,partial} x ) \nonumber \\
&\leq& \sum_{\Pi} P(  |(L(\mathbf I, B^{\sharp}) - L(\Pi, B^{\sharp})) - (\Lambda(\mathbf I, B^{\sharp}) - \Lambda(\Pi, B^{\sharp}))| \geq v_{\Pi,partial} x ) \nonumber  \\
& = & \sum_h \sum_{\Pi: d(\Pi,\mathbf I) = h} \cdot \exp\{ - v_{\Pi,partial} x^2 / 4 \} \nonumber \\
&\leq& \sum_h n!/(n-h)! \cdot \exp\{ - v_{\Pi,partial} x^2 / 4 \} \nonumber \\
&\leq& \sum_h n^h \cdot \exp\{ - h v_{min} x^2 / 4 \} \nonumber \\
&=& \sum_h \exp\{ - h (v_{min} x^2 - \log n)\}  \nonumber \\
&\leq& \frac{\exp\{ - 2 (v_{min} x^2 - \log n)\}}{1 - \exp\{ - (v_{min} x^2 - \log n)\}}, \label{eq:close:rate:combine}
\end{eqnarray}
where $v_{min} := \min_{i,j} \sum_l \llam_{il}^{\sharp} \log^2(\llam_{jl}^{\sharp}/\llam_{il}^{\sharp})$ and we use the fact that $\sum_{\Pi} = \sum_{h} \sum_{\Pi: d(\Pi, \mathbf I) = h}$
and $|\{\Pi: d(\Pi,\mathbf I) = h\}| \leq n!/(n-h)! \leq n^h$.

By \eqref{eq:close:rate:combine}, we know that $L(\Pi, \hat{B}) \leq
L(\mathbf I, \hat{B}) - \Lambda(\mathbf I, \hat{B}) + \Lambda(\Pi, \hat{B}) + O_p(x v_{\Pi,partial})$ with $x = \sqrt{\log n / v_{min}}$.
This tells us that if we can show that
\begin{eqnarray}\label{eq:goal:onestep}
\Lambda(\mathbf I, \hat B) - \Lambda(\Pi, \hat{B}) - O_p(x v_{\Pi,partial}) > 0
\end{eqnarray}
for any $\Pi$ with $d(\mathbf I, \Pi^{\sharp}) \leq h_{max}$.
Then we can conclude that $\hat \Pi = \mathbf I$ which gives the desired result.

\subsection{First bound of $\|B - B^{\sharp}\|_F$}
Since the likelihood function can be written as the sum of separate functions of $\bb_1, \ldots, \bb_m$, we only need to focus on single $\bb_l$ for $l \in [m]$.
Without loss of generality, we can assume $m=1$. Then the estimator $\hat {\mathbf b}$ is
\begin{eqnarray}
\hat {\mathbf b} = \arg \max_{\mathbf b} \{ \langle - \psi(X \mathbf b) + Y \circ X \mathbf b \rangle \}.
\end{eqnarray}
When $d(\mathbf I, \Pi^{\sharp}) \leq h_{max}$, we aim to show $\hat{\mathbf b}$ is a consistent estimator of $\mathbf b^{\sharp}$.

For simplicity, we can assume $\Pi^{\sharp}(i) = i$ for $i > h_{max}$
and let $L(\mathbf b) = \langle - \psi(X \mathbf b) + Y \circ X \mathbf b \rangle $.
In the following, we aim to find a $\delta_n$ such that
for any $\mathbf b$ with $\|\mathbf b - \mathbf b^{\sharp}\| \geq \delta_n$, it holds
$L(\mathbf b) < L(\mathbf b^{\sharp})$.
By the definition that
$L(\hat {\mathbf b}) \geq L(\mathbf b^{\sharp})$, we will arrive at $\|\hat {\mathbf b} - \mathbf b^{\sharp}\| \leq \delta_n$.

By computation, we can see
\begin{eqnarray}
& & L(\mathbf b^{\sharp}) - L(\mathbf b) \nonumber \\
&=& (L(\mathbf b^{\sharp}) - \Lambda(\mathbf b^{\sharp})) - (L(\mathbf b) - \Lambda(\mathbf b)) +  \Lambda(\mathbf b^{\sharp}) - \Lambda(\mathbf b) \nonumber \\
&=& (L(\mathbf b^{\sharp}) - \Lambda(\mathbf b^{\sharp})) - (L(\mathbf b) - \Lambda(\mathbf b)) +  \Lambda_1(\mathbf b^{\sharp}) - \Lambda_1(\mathbf b)
+ \Lambda_2(\mathbf b^{\sharp}) - \Lambda_2(\mathbf b) \\
&\geq& - O_p(\sqrt{v_{b^{\sharp}}}) - O_p(\sqrt{v_b}) - h_{max} C_{max}  + \sigma_{min}(H) \delta_n^2. \label{eq:small1}
\end{eqnarray}
For the last inequality, we use the following facts, $L(\mathbf b) - \Lambda(\mathbf b) = O_p( \sqrt{v_b})$ with $v_b = \sum_{i=1}^n \psi''(\lambda_{i}^{\sharp}) (\lambda_i(\mathbf b))^2$ for any $\mathbf b$. In addition,
\begin{eqnarray}
& & |\Lambda_1(\mathbf b^{\sharp}) - \Lambda_1(\mathbf b)| \nonumber \\
&=& |\sum_{i \leq h_{max}} \{- \psi(\xx_i^T \mathbf b^{\sharp}) + \psi'(\xx_i^T \mathbf b^{\sharp}) \xx_i^T \mathbf b^{\sharp} - (- \psi(\xx_i^T \mathbf b) + \psi'(\xx_i^T \mathbf b^{\sharp}) \xx_i^T \mathbf b) \} | \nonumber \\
&\leq& \sum_{i \leq h_{max}} |\lambda_{max}(\delta_n)| + \psi'(\xx_i^T \mathbf b^{\sharp}) |\xx_i^T(\mathbf b^{\sharp} - \mathbf b)| \nonumber \\
&\leq& 2 h_{max} \psi_{max}(\delta_n) +  h_{max} \psi_{max}^{'\sharp} \max_{i,\mathbf b} |\xx_i^T(\mathbf b^{\sharp} - \mathbf b)| \nonumber \\
&=& h_{max} C_{max},
\end{eqnarray}
where $C_{max} := 2 \psi_{max}(\delta_n) + \psi_{max}^{'\sharp} \max_{i, \mathbf b} |\xx_i^T(\mathbf b^{\sharp} - \mathbf b)|$, $\psi_{max}(\delta) = \max_{i, \mathbf b: \|\mathbf b - \mathbf b^{\sharp}\| \leq \delta_n} \psi(\xx_i^T \mathbf b)$ and
$\psi_{max}^{'\sharp} = \max_{i} \psi'(\mathbf x_i^T \mathbf b^{\sharp})$.
Moreover, via using the fact that $f(y) = f(x) + f'(x)(y-x) + \frac{1}{2}f''(z)(y-z)^2$ ($z$ is some point between $x$ and $y$)for any smooth function $f$, we have
\begin{eqnarray}
& & \Lambda_2(\mathbf b) - \Lambda_2(\mathbf b^{'}) \nonumber \\
&=& \sum_{i > h_{max}} \{- \psi(\xx_i^T \mathbf b) + \psi'(\xx_i^T \mathbf b) \xx_i^T \mathbf b - (- \psi(\xx_i^T \mathbf b^{'}) + \psi'(\xx_i^T \mathbf b) \xx_i^T \mathbf b^{'}) \} \nonumber \\
&=& \sum_{i > h_{max}} (\psi''(\mathbf x_i^T \tilde{\mathbf b}) (\mathbf x_i^T \mathbf b - \mathbf x_i^T \mathbf b^{'} ) )^2/2 \nonumber \\
&=&  (\mathbf b - \mathbf b^{'})^T \sum_{i \geq h_{max}} \psi''(\xx_i^T \tilde {\mathbf b}) \{\xx_i \xx_i^T\} (\mathbf b - \mathbf b^{'}) / 2 \nonumber \\
&=& (\mathbf b - \mathbf b^{'})^T H (\mathbf b - \mathbf b^{'}) / 2,
\end{eqnarray}
where $\tilde {\mathbf b}$ is between $\mathbf b$ and $\mathbf b^{'}$, and
\begin{eqnarray}
H = \sum_{i > h_{max}} \psi''(\xx_i^T \tilde{\mathbf b})\xx_i \xx_i^T.
\end{eqnarray}
Again we know that
\[(\mathbf b - \mathbf b^{\sharp})^T H (\mathbf b - \mathbf b^{'}) \geq c_1 (n - h_{max}) \psi_{min}^{\sharp} / {\color{black} \gamma_{1p}}\|\mathbf b - \mathbf b^{\sharp}\|^2\]
holds for some constant $c_1$ by using the curvature technique (see \eqref{eq:curve}).
Therefore, we can specifically take
\[\delta_n^2 = C {\color{black} \gamma_{1p}} \frac{\sqrt{v_{b^{\sharp}}} + \psi_{max}^{\sharp} h_{max}}{(n - h_{max}) \psi_{min}^{\sharp}}\]
with some large constant $C$. With this choice of $\delta_n$, from \eqref{eq:small1}, we can check that
\[L(\mathbf b^{\sharp}) - L(\mathbf b) > 0\]
for any $\mathbf b$ with $\|\mathbf b - \mathbf b^{\sharp}\| = \delta_n$ when $p = O(n^{a})$ ($a < \frac{1}{2}$). By the concavity of likelihood function, we then know that the two-step estimator $\hat{\mathbf b}$ must lie in the ball $\{\mathbf b: \|{\mathbf b} - \mathbf b^{\sharp}\| \leq \delta_n\}$.

\subsection{Second bound of $\|B - B^{\sharp}\|_F$}
We consider the Taylor expansion of $L(\mathbf b)$ at value $\mathbf b^{\sharp}$. Then, it can be computed that
\begin{eqnarray}
\mathbf 0 = \nabla L(\hat {\mathbf b}) = \nabla L(\mathbf b^{\sharp}) + \nabla^2 L(\bar{\mathbf b}) (\hat{\mathbf b} - \mathbf b^{\sharp})
\end{eqnarray}
where $\bar{\mathbf b}$ is some point between $\hat{\mathbf b}$ and $\mathbf b^{\sharp}$.
By the formula $\nabla^2 L(\mathbf b) = \sum_{i=1}^n \psi''(\xx_i^T \mathbf b) \xx_i \xx_i^T$, we then know that
\begin{eqnarray}
|\nabla^2 L(\mathbf b^{\sharp}) - \nabla^2 L(\bar{\mathbf b})| &=& |\sum_{i=1}^n \psi''(\xx_i^T \mathbf b^{\sharp}) \xx_i \xx_i^T (1 - \psi''(\xx_i^T \bar{\mathbf b})/ \psi''(\xx_i^T \mathbf b^{\sharp}))| \nonumber \\
&\leq& |\sum_{i=1}^n \psi''(\xx_i^T \mathbf b^{\sharp})\cdot o_p(1) \cdot \xx_i \xx_i^T |,
\end{eqnarray}
since $\|\bar{\mathbf b} - \mathbf b^{\sharp}\| \leq \|\hat{\mathbf b} - \mathbf b^{\sharp}\| = o_p(1)$.
Then we know that $ \sigma_{min}(\nabla^2 L(\mathbf b^{\sharp}))
\geq \sigma_{min}(\partial^2 L(\mathbf b^{\sharp}))/2$.
We thus have
\begin{eqnarray}
\|\hat{\mathbf b} - \mathbf b^{\sharp}\| = \| (\nabla^2 L(\bar{\mathbf b}))^{-1} \partial L(\mathbf b^{\sharp})\| \leq \frac{2}{\sigma_{min}(\partial^2 L(\mathbf b^{\sharp}))} \|\partial L(\mathbf b^{\sharp})\|.
\end{eqnarray}
For $l$-th ($l = 1,\ldots, p$) element of $\nabla L(\mathbf b^{\sharp})$, we can find that $\nabla L(\mathbf b^{\sharp})[l] = \sum_{i \leq h_{max}}
\nabla L_i(\mathbf b^{\sharp})[l] + \sum_{i > h_{max}} \nabla L_i(\mathbf b^{\sharp})[l]$. The first term is bounded by
\[b_s := |\sum_{i \leq h_{max}} (Y[i] -\psi'(\xx_i^T \mathbf b^{\sharp})) X[i,l] \}|,\]
which is order of $\log n \sum_{i \leq h_{max}} (\psi_{max}^{''\sharp} \log n + \psi_{max}^{'\sharp}) |X[i,l]|$.
Here we use the observation that
\begin{eqnarray}\label{Y:bound}
Y[i,j] = O_p(\psi_{max}^{''\sharp} \log n + \psi_{max}^{'\sharp} )
\end{eqnarray}
for all $i \in [n], j \in [m]$.
The second term is bounded by $ C \sqrt{\textrm{var}(\sum_{i > h_{max}} \nabla L_i(\mathbf b^{\sharp})[l])}$ and the upper bound of $\textrm{var}(\sum_{i > h_{max}} \nabla L_i(\mathbf b^{\sharp})[l])$ can be computed explicitly, i.e.,
\[v_s := \max_{l \in [p]} \sum_{i > h_{max}} \psi''(\xx_i^T\mathbf b^{\sharp}) (X[i,l])^2.\]
Putting all above together, we get
\begin{eqnarray}
\|\hat{\mathbf b} - \mathbf b^{\sharp}\| &\leq& \sqrt{p} (\sqrt{v_s}+b_s) / \sigma_{min}(\partial^2 L(\mathbf b^{\sharp})) =
O_p(\frac{\sqrt{p} (\sqrt{\psi_{max}^{''\sharp}}\|X\|_{2,\infty} + (\psi_{max}^{''\sharp} \log n + \psi_{max}^{'\sharp}) h_{max})}{\sigma_{min}^2(X)}) \nonumber \\
&=& O_p(\frac{\sqrt{p} (\sqrt{\psi_{max}^{''\sharp}} \sqrt{n - h_{max}} + (\psi_{max}^{''\sharp} \log n + \psi_{max}^{'\sharp}) h_{max})}{n \psi_{min}^{''\sharp}} {\color{black} \gamma_{1p}}) := \delta^{\ast}.
\end{eqnarray}
It is clear that the bound $\delta^{\ast}$ is tighter than $\delta_n$.
Especially, when $\psi_{max}^{''\sharp}, \psi_{min}^{''\sharp}$ and $\psi_{max}^{'\sharp}$ are bounded and $\gamma_{1p}$ is $O(1)$, then $\delta^{\ast}$ can be simplified as $\frac{\sqrt{p}(\sqrt{n - h_{max}} + (\log n) h_{max})}{n}$.

\newpage

\subsection{Difference of $\Lambda(\mathbf I, \hat {\mathbf b}) - \Lambda(\Pi, \hat{\mathbf b})$}

By straightforward calculation, we get
\begin{eqnarray}
& & \Lambda(\mathbf I, \hat {\mathbf b}) - \Lambda(\Pi, \hat{\mathbf b}) \nonumber \\
&=& (\Lambda(\mathbf I, \hat {\mathbf b}) - \Lambda(\mathbf I, \mathbf b^{\sharp})) - (\Lambda(\Pi, \hat{\mathbf b}) - \Lambda( \Pi, \mathbf b^{\sharp})) + (\Lambda(\mathbf I, \mathbf b^{\sharp}) - \Lambda(\Pi, \mathbf b^{\sharp})) \nonumber \\
&\geq&  - 4 h_{max} \lambda_{max}^{\sharp} x_{max} \delta^{\ast} + \Lambda(\mathbf I, \mathbf b^{\sharp}) - \Lambda(\Pi, \mathbf b^{\sharp}), \label{ineq:continuity}
\end{eqnarray}
where the last inequality depends on the following fact
\begin{eqnarray}
& & (\Lambda(\mathbf I, \hat {\mathbf b}) - \Lambda(\mathbf I, \mathbf b^{\sharp})) - (\Lambda(\Pi, \hat{\mathbf b}) - \Lambda( \Pi, \mathbf b^{\sharp})) \nonumber \\
&=& (\Lambda(\mathbf I, \hat {\mathbf b}) - \Lambda(\Pi, \hat{\mathbf b})) - (\Lambda(\mathbf I, \mathbf b^{\sharp}) - \Lambda( \Pi, \mathbf b^{\sharp})) \nonumber \\
&=& \sum_{i \leq d(\mathbf I, \Pi)} \bigg \{ - \psi(\xx_i^T \hat{\mathbf b}) + \psi'(\xx_i^T {\mathbf b^{\sharp}}) \xx_i^T \hat{\mathbf b} + \psi(\xx_i^T {\mathbf b^{\sharp}}) - \psi'(\xx_i^T {\mathbf b^{\sharp}}) \xx_i^T {\mathbf b^{\sharp}} \nonumber \\
& & + \psi(\xx_{\Pi(i)}^T \hat{\mathbf b}) - \psi'(\xx_i^T {\mathbf b^{\sharp}}) \xx_{\Pi(i)}^T \hat{\mathbf b} - \psi(\xx_{\Pi(i)}^T {\mathbf b^{\sharp}}) + \psi'(\xx_i^T {\mathbf b^{\sharp}}) \xx_{\Pi(i)}^T {\mathbf b^{\sharp}} \bigg \} \nonumber \\
&\leq& \sum_{i \leq d(\mathbf I, \Pi)} 4 \psi_{max}^{'\sharp} \cdot x_{max} \cdot \|\hat {\mathbf b} - \mathbf b^{\sharp}\| \leq 4  d(\mathbf I, \Pi) \psi_{max}^{'\sharp} x_{max} \delta^{\ast},
\end{eqnarray}
where $\psi_{max}^{'\sharp} = \max_{i} \psi'(\xx_i^T  \mathbf b^{\sharp})$ and $x_{max} = \max_i \|\xx_i\|$.

By \eqref{ineq:continuity} and summing over $l \in [m]$, we have that
\begin{eqnarray}
& & \Lambda(\mathbf I, \hat {B}) - \Lambda(\Pi, \hat{B}) \nonumber \\
&\geq&  - 4 m d(\mathbf I, \Pi) \psi_{max}^{'\sharp} x_{max} \delta^{\ast} +
\Lambda(\mathbf I, B^{\sharp}) - \Lambda(\Pi, B^{\sharp}), \label{ineq:continuity:m} \nonumber \\
&\gtrsim& x v_{\Pi,partial},
\end{eqnarray}
when
\begin{eqnarray}
\Lambda(\mathbf I, B^{\sharp}) - \Lambda(\Pi, B^{\sharp}) \gtrsim x v_{\Pi,partial}
\end{eqnarray}
and
\begin{eqnarray}
\Lambda(\mathbf I, B^{\sharp}) - \Lambda(\Pi, B^{\sharp}) \gtrsim m d(\mathbf I, \Pi) \psi_{max}^{'\sharp} x_{max} \delta^{\ast}.
\end{eqnarray}
This completes the proof of Theorem~\ref{thm:known}.

\subsection{On $\Lambda(\mathbf I, B^{\sharp}) - \Lambda(\Pi, B^{\sharp})$}
We investigate the lower bound of $\Lambda(\mathbf I, B^{\sharp}) - \Lambda(\Pi, B^{\sharp})$.
\begin{eqnarray}
& & \Lambda(\mathbf I, B^{\sharp}) - \Lambda(\Pi, B^{\sharp}) \nonumber \\
&=& \sum_{l \in [m]} \sum_{i \leq d(\mathbf I, \Pi)}  \bigg\{ - \psi(\xx_i^T  {\mathbf b_l^{\sharp}}) + \psi'(\xx_i^T  {\mathbf b_l^{\sharp}})\xx_i^T {\mathbf b_l^{\sharp}} \nonumber \\
& &- ( \psi(\xx_{\Pi(i)}^T {\mathbf b_l^{\sharp}}) + \psi'(\xx_i^T  {\mathbf b_l^{\sharp}}) \xx_{\Pi(i)}^T {\mathbf b_l^{\sharp}} ) \bigg\} \nonumber \\
&\geq& \sum_{l \in [m]} \sum_{i \leq d(\mathbf I, \Pi)} \frac{1}{2} \psi_{min}^{''\sharp} \big((\xx_i^T  - \xx_{\Pi(i)}^T) \mathbf b_l^{\sharp} \big)^2. \label{eq:lambda:true}
\end{eqnarray}
Under sub-Gaussian design, $\sum_{l \in [m]} \big((\xx_i^T  - \xx_{j}^T) \mathbf b_l^{\sharp} \big)^2$ is $\Omega(m)$ for any pair of $i,j \in [m]$ when $m \gtrsim \log n$ and $p \gtrsim \log n$.
Thus, $\Lambda(\mathbf I, B^{\sharp}) - \Lambda(\Pi, B^{\sharp})$ is $\Omega(m d(\mathbf I, \Pi))$.

\newpage

\section{Proof of Results without any knowledge of $B^{\sharp}$ and $\Pi^{\sharp}$: Theorem~\ref{thm:unknown:main}}\label{app:unknown}

In this section, we provide the proof when we do not have any knowledge of $B^{\sharp}$ and $\Pi^{\sharp}$.
For each fixed permutation $\Pi$, we define
\begin{eqnarray}
v_{\Pi,B}&=& \sum_{i=1}^n \sum_{l=1}^m \psi''(\xx_{\Pi^{\sharp}(i)}^T \bb_l^{\sharp}) (\xx_{\Pi(i)}^T \bb_l)^2. 
\end{eqnarray}
It can be easily checked that $v_{\Pi,B}$ is the variance of $L(\Pi, B)$.

We compute the deviation  $ |\langle L(\Pi, B) - \Lambda(\Pi, B) \rangle|$.
The moment generating function of $\langle L(\Pi, B) - \Lambda(\Pi, B) \rangle$ is
\begin{eqnarray}
& & \mathbb E \exp\{t \langle L(\Pi, B) - \Lambda(\Pi, B) \rangle \} \nonumber \\
&=& \mathbb E \exp\{t \sum_{i} \langle (Y[i,:] - \psi'(\llam_{i})) \circ \llam_{\pi_i} \rangle \} \nonumber \\
&=& \prod_{i=1}^n \prod_{l=1}^m \exp\{\psi(\llam_i^{\sharp}[l] + t \llam_i[l])
- \psi(\llam_i^{\sharp}[l]) - t \psi'(\llam_i^{\sharp}[l])\llam_{\pi_i}[l]\}  \\
&\leq& \prod_{i=1}^n \prod_{l=1}^m \exp\{\psi''(\llam_{i}^{\sharp}[l]) (\llam_{\pi_i}[l])^2 t^2\} \\
& = & \exp\{ t^2 (\sum_{i,l} \psi''(\llam_{i}^{\sharp}[l]) (\llam_{\pi_i}[l])^2)\} = \exp\{v_{\Pi,B} t^2\}
\end{eqnarray}
for any small $t$ satisfying
$\frac{1}{2}\psi''(\llam_i^{\sharp}[l]) > |t \psi'''(\llam_i^{\sharp}[l]) \llam_{\pi_i}[l]|$.
Thus, we have
\begin{eqnarray}
& & P(|\langle L(\Pi, B) - \Lambda(\Pi, B) \rangle| \geq v_{\Pi,B} x) \nonumber \\
&\leq& \inf_t \exp\{ - t v_{\Pi,B} x \} \exp\{v_{\Pi,B} t^2\} \\
&\leq& \exp\{ - v_{\Pi,B} x^2 / 4\}
\end{eqnarray}
for any $\psi''(\llam_i^{\sharp}[l]) > |x \psi'''(\llam_i^{\sharp}[l]) \llam_{\pi_i}[l]|$.
We also have
\begin{eqnarray}
& & P(|\langle L(\Pi, B) - \Lambda(\Pi, B) \rangle| \geq v_{\Pi,B} x) \nonumber \\
&\leq& \inf_t \exp\{ - t v_{\Pi,B} x \} \exp\{v_{\Pi,B} t^2\} \\
&\leq& \exp\{ v_{\Pi,B}/(x_{max})^2\} \exp\{- v_{\Pi,B} x / x_{max}\}
\end{eqnarray}
for any $x$. Here we only need to consider the situations that $ \llam_i[l]$'s are $O(x_{max})$ according to \eqref{bound:xb}.

\vspace{0.1in}
\noindent \textbf{Two situations}~
In order to prove the results, we consider the following two situations, 1. $d(\Pi, \Pi^{\sharp}) \leq h_c$
2. $d(\Pi, \Pi^{\sharp}) \geq h_c$ where $h_c = c_0 \frac{n}{p \log n}$.

\subsection{Situation 1}
We first show the difference between $B(\Pi)$ and $B^{\sharp}$.
By the definition of $B(\Pi)$, we know
\[B(\Pi) = \arg\max_B \Lambda(\Pi,B).\]
Note that $\Lambda(\Pi,B)$ is separable for each column of $B$, i.e.,
\[\bb_j(\Pi) = \arg\max_{\mathbf b} \sum_{i=1}^n \{\psi'(\xx_{\Pi^{\sharp}(i)}^T \bb_j^{\sharp}) (\xx_{\Pi(i)}^T \mathbf b) - \exp\{\xx_{\Pi(i)}^T \mathbf b\} \}. \]
(Here, we assume $\Pi^{\sharp} = \mathbf I$ without loss of generality.)
Therefore, we only need to bound the difference between $\|\bb_j(\Pi) - \bb_j^{\sharp}\|$.

By the optimality of $\bb_j(\Pi)$, we then have that
\begin{eqnarray}
\sum_{i=1}^n \psi'(\xx_i^T \bb_j^{\sharp}) (\xx_{\Pi(i)}^T \bb_j(\Pi)) - \psi(x_{\Pi(i)}^T \bb_j(\Pi)) \geq \sum_{i=1}^n \psi'(\xx_i^T \bb_j^{\sharp}) (\xx_{\Pi(i)}^T \bb_j^{\sharp}) - \psi(\xx_{\Pi(i)}^T \bb_j^{\sharp}) \nonumber
\end{eqnarray}
which can be written as
\begin{eqnarray}\label{eq:diff:B}
& & \sum_{i: \Pi(i) \neq i} \bigg\{ \psi'(\xx_i^T \bb_j^{\sharp}) (\xx_{\Pi(i)}^T \bb_j(\Pi)) - \psi(x_{\Pi(i)}^T \bb_j(\Pi))
- (\psi'(\xx_i^T \bb_j^{\sharp}) (\xx_{\Pi(i)}^T \bb_j^{\sharp}) - \psi(\xx_{\Pi(i)}^T \bb_j^{\sharp}))
\bigg\} \nonumber \\
&\geq& \sum_{i: \Pi(i) = i} \bigg\{ \psi'(\xx_i^T \bb_j^{\sharp}) (\xx_{\Pi(i)}^T \bb_j^{\sharp}) - \psi(\xx_{\Pi(i)}^T \bb_j^{\sharp})
- (\psi'(\xx_i^T \bb_j^{\sharp}) (\xx_{\Pi(i)}^T \bb_j(\Pi)) - \psi(x_{\Pi(i)}^T \bb_j(\Pi)))
\bigg\}.  \nonumber \\
\end{eqnarray}
The right hand side of \eqref{eq:diff:B} is bounded below by
\begin{eqnarray}\label{eq:diff:B:rhs}
RHS & \geq & \frac{1}{2} \sum_{i: \Pi(i) = i} \psi''(\xx_i^T \tilde \bb) (\xx_i^T(\bb_j^{\sharp} - \bb_j(\Pi)))^2 \nonumber \\
&\geq& \frac{1}{4} c n/ {\color{black} \gamma_{1p}} \psi_{min}^{''\sharp}  \|\bb^{\sharp}_j - \bb_j(\Pi)\|^2,
\end{eqnarray}
where the last inequality depends on the curvature property which will be described later.
The left hand side of \eqref{eq:diff:B} is bounded above by
\begin{eqnarray}\label{eq:diff:B:lhs}
LHS &\leq& \sum_{i: \Pi(i) \neq i} \bigg\{ \psi'(\xx_i^T \bb^{\sharp}_j) ( \xx_i^T \bb^{\sharp}_j) - \psi(\xx_i^T \bb^{\sharp}_j)
- (\psi'(\xx_i^T \bb^{\sharp}_j) (\xx_{\Pi(i)}^T \bb^{\sharp}_j) - \psi(\xx_{\Pi(i)}^T \bb^{\sharp}_j))
\bigg\} \nonumber \\
&\leq& \sum_{i: \Pi(i) \neq i} (\psi'(\xx_i^T \bb^{\sharp}_j) |\xx_i^T \bb^{\sharp}_j - \xx_{\Pi(i)}^T \bb^{\sharp}_j| + \max\{\psi(\xx_i^T \bb^{\sharp}_j),\psi(\xx_{\Pi(i)}^T \bb^{\sharp}_jj)\}).
\end{eqnarray}
Combining \eqref{eq:diff:B:rhs} and \eqref{eq:diff:B:lhs}, we have
\begin{eqnarray}\label{eq:diff:B:together}
\|\bb_j^{\sharp} - \bb_j(\Pi)\|^2 \leq C \frac{p d(\mathbf I, \Pi) (\psi_{max}^{'\sharp} + \psi_{max}^{\sharp}) {\color{black} \gamma_{1p}}}{n \psi_{min}^{''\sharp}},
\end{eqnarray}
by adjusting the constant $C$.

Given a fixed $\Pi$, we next calculate the bound of $\|\hat B(\Pi) - B(\Pi)\|$.
By the definition of $\hat B(\Pi)$ and convexity of negative log-likelihood function, we have that
\begin{eqnarray}
L(\Pi, B^{\sharp}) &\leq& L(\Pi, \hat B(\Pi)) =
L(\Pi, B^{\sharp}) +  \langle \nabla L(\Pi, B^{\sharp}), \hat B(\Pi) - B^{\sharp} \rangle \nonumber \\
& & + \frac{1}{2}  (\hat B(\Pi) - B^{\sharp})^T \nabla^2 L(\Pi, \tilde B) (\hat B(\Pi) - B^{\sharp}),
\end{eqnarray}
which gives us that
\begin{eqnarray}
\frac{1}{2}  (\hat B(\Pi) - B^{\sharp})^T \nabla^2 L_{neg}(\Pi, \tilde B) (\hat B(\Pi) - B^{\sharp})
&\leq& |\langle \nabla L(\Pi, B^{\sharp}), \hat B(\Pi) - B^{\sharp} \rangle| \nonumber \\
\frac{1}{2}  (n \psi_{min}^{''\sharp} / {\color{black} \gamma_{1p}}) \|\hat B(\Pi) - B^{\sharp}\|^2
&\leq& \| \hat B(\Pi) - B^{\sharp}\| \|\nabla L(\Pi, B^{\sharp})\| \label{eq:curvature} \\
(n \psi_{min}^{''\sharp}/ {\color{black} \gamma_{1p}}) \|\hat B(\Pi) - B^{\sharp}\|^2 &\leq& C (\psi_{max}^{\sharp} + \psi_{max}^{'\sharp}) \| \hat B(\Pi) - B^{\sharp}\| {\color{black} \sqrt{p}} (C h \log n   + \sqrt{n \log p} )  \nonumber  \\
\label{uniform:bound}\\
\|\hat B(\Pi) - B^{\sharp}\| &\leq&  C (\psi_{max}^{\sharp} + \psi_{max}^{'\sharp}) \frac{{\color{black} \gamma_{1p}} \sqrt{p} ( C h \log n   + \sqrt{n \log p} )}{n \psi_{min}^{''\sharp}} \label{bound:important1},
\end{eqnarray}
where we define $L_{neg}(\Pi, B) = - L(\Pi,B)$.
This tells us that
\[\|\hat B(\Pi) - B(\Pi)\| \leq \|B(\Pi) - B^{\sharp}\| + \|\hat B(\Pi) - B^{\sharp}\| = o(1)\]
as long as $p h \ll n/ (\gamma_{1p} \log n)$ and $p = n^{a}$ ($0 < a < \frac{1}{2}$).

\vspace{0.1in}
\noindent \textbf{Curvature:} Here \eqref{eq:curvature} depends on the following observations on the curvature of log-likelihood function.
Since
\begin{eqnarray}
\nabla^2 L_{neg}(\Pi, B) = (\Pi X)^T \textrm{diag}(\psi''(\Pi X B)) \Pi X,
\end{eqnarray}
then
\begin{eqnarray}
& & (\hat B(\Pi) - B^{\sharp})^T\nabla^2 L_{neg}(\Pi, \tilde B)(\hat B(\Pi) - B^{\sharp}) \nonumber \\
&=& (\hat B(\Pi) - B^{\sharp})^T(\Pi X)^T \textrm{diag}(\psi''(\Pi X \tilde B)) \Pi X (\hat B(\Pi) - B^{\sharp})^T \label{eq:expand:curvature} .
\end{eqnarray}
Let $r = \|\hat B(\Pi) - B^{\sharp}\|$ and $\mathbf b = \hat B(\Pi) - B^{\sharp}$.
For any monotonically increasing $\psi''$, by assumption \textit{A2}, we have that the cardinality of set
$\mathcal I = \{i | \xx_i^T \bb \geq c_1 r\}$ is greater than $n / {\color{black} \gamma_{1p}}$.
For index $i$ in $\mathcal I$, we can find that
\begin{eqnarray}
\psi''(\xx_i^T \tilde B) &\geq& \min\{\psi''(\xx_i^T \hat B(\Pi)), \psi''(\xx_i^T B^{\sharp})\} \label{eq:curve}\\
&=&  \min\{\psi''(\xx_i^T (\bb + B^{\sharp})), \psi''(\xx_i^T B^{\sharp})\} \nonumber \\
&\geq& \psi_{min}^{''\sharp}, \nonumber
\end{eqnarray}
since $\tilde B(\Pi)$ takes form of $t \hat B(\Pi) + (1-t) B^{\sharp}$.
(Similarly, \eqref{eq:curve} also holds for monotonically decreasing or bounded $\psi''$'s.)

Thus, the right hand side of \eqref{eq:expand:curvature} can be lower bounded by
\begin{eqnarray}
& & (\hat B(\Pi) - B^{\sharp})^T(\Pi X)^T \textrm{diag}(\psi(\Pi X \tilde B)) \Pi X (\hat B(\Pi) - B^{\sharp}) \nonumber \\
&\geq& c n \psi_{min}^{''\sharp} /{\color{black} \gamma_{1p}} r^2 \nonumber \\
& = & c n \psi_{min}^{''\sharp} / {\color{black} \gamma_{1p}} \|\hat B(\Pi) - B^{\sharp}\|^2. \label{eq:curvature:1}
\end{eqnarray}
Similarly, we have
\begin{eqnarray}
B^T \nabla^2 L_{neg}( \Pi, \bar B) B &=& B^T(\Pi X)^T \textrm{diag}(\psi''(\Pi X \bar B)) \Pi X B \nonumber \\
&\geq& c n / \gamma_{1p} \|B\|^2 \label{eq:curvature:general}
\end{eqnarray}
for any $B$ and $\bar B = t \mathbf 0 + (1 - t) B$, $(0 \leq t \leq 1)$.
We call \eqref{eq:curvature:1} and \eqref{eq:curvature:general} as \textit{curvature inequalities} for log-likelihood at $B = B^{\sharp}$ and $B = \mathbf 0$ correspondingly.

Inequality \eqref{uniform:bound} comes from the following fact.
For each $l \in [p]$, we consider to compute the following bound, i.e.,
\begin{eqnarray}
\sup_{\Pi: d(\Pi, \mathbf I) \leq h} |\nabla L(\Pi, B^{\sharp})[l]|
&\leq&
\sup_{\Pi: d(\Pi, \mathbf I) \leq h} \{ |\nabla L(\Pi, B^{\sharp})[l] - \nabla L(\mathbf I, B^{\sharp})[l]| \}
+ |\nabla L(\mathbf I, B^{\sharp})[l]| \nonumber \\
&\leq& C (\psi_{max}^{\sharp} + \psi_{max}^{'\sharp}) ( C h \log n   + \sqrt{n \log p} ),
\end{eqnarray}
by noticing that the entry of design matrix is bounded.

In situation 1, we are going to show that
\[L(\mathbf I, \hat B) > L(\Pi, \hat B(\Pi))  \]
with high probability. It suffices to show
\[L(\mathbf I, \hat B(\Pi)) > L(\Pi, \hat B(\Pi)) \]
for any $\Pi$ with $d(\mathbf I, \Pi) \leq h_c$.

\vspace{0.1in}
\noindent \textbf{Uniform Bound of $\|\hat B(\Pi) - \hat B\|$} ~

\vspace{0.1in}
\noindent By \eqref{bound:important1}, we then have
\[\|\hat B(\Pi) - \hat B\| \leq 2 C (\psi_{max}^{\sharp} + \psi_{max}^{'\sharp}) \frac{\sqrt{p} {\color{black} \gamma_{1p}} (h_{max} \log n + \sqrt{n \log p})}{n \psi_{min}^{''\sharp}}\]
held for any $\Pi$ with $d(\Pi, \Pi^{\sharp}) \leq h_{max}$ by adjusting the constant $C$.

\vspace{0.1in}
\noindent \textbf{Difference of $|(L(\mathbf I, B) - L(\Pi, B)) - (\Lambda(\mathbf I, B) - \Lambda(\Pi, B))|$}

\vspace{0.1in}

\noindent By similar calculation as we have done in \eqref{concen:1}, we can obtain that
\begin{eqnarray}
& &P( |(L(\mathbf I, B) - L(\Pi, B)) - (\Lambda(\mathbf I, B) - \Lambda(\Pi, B))| \geq v_{\Pi, partial} x) \nonumber \\
&\leq& \exp\{-\frac{1}{4} v_{\Pi, partial} x^2 \}
\end{eqnarray}
for any fixed $B \in \mathcal B(B^{\sharp}, \delta)$ with $\delta = o(1/\sqrt{p})$.
By using this, we can further obtain the uniform inequality, i.e.,
\begin{eqnarray}
& &P( \sup_{\Pi: d(\mathbf I, \Pi) \leq h_c} \frac{1}{v_{\Pi, partial} }|(L(\mathbf I, \hat B(\Pi)) - L(\Pi, \hat B(\Pi))) - (\Lambda(\mathbf I, \hat B(\Pi)) - \Lambda(\Pi, \hat B(\Pi)))| \geq x) \nonumber \nonumber \\
&\leq& \sum_{\Pi: d(\mathbf I, \Pi) \leq h_c} \exp\{-\frac{1}{4} v_{\Pi, partial} x^2 \} \nonumber \\
&\leq& \sum_{h=2}^{h_c} \sum_{\Pi: d(\mathbf I, \Pi) = h} \exp\{-\frac{1}{4} v_{\Pi, partial} x^2 \} \nonumber \\
&\leq& \sum_{h=2}^{h_c} n^h \exp\{-\frac{1}{4} h v_{min} x^2 \} \nonumber \\
&\leq& \frac{-2(v_{min} x^2 - \log n)}{1 - \exp\{-(v_{min} x^2 - \log n)\}}. \label{uniform:concentration}
\end{eqnarray}

On the other hand, we could compute the difference
\begin{eqnarray}
& & |(\Lambda(\mathbf I, \hat B(\Pi)) - \Lambda(\Pi, \hat B(\Pi))) - (\Lambda(\mathbf I, B(\Pi)) - \Lambda(\Pi, B(\Pi)))| .\nonumber \\
&\leq& \bigg |\sum_{i: \Pi(i) \neq i} \sum_{j=1}^m (\psi'(\xx_i^T \bb^{\sharp}_j) (\xx_i^T \hat \bb_j(\Pi)) - \psi(\xx_i^T \hat \bb_j(\Pi))) - (\psi'(\xx_{\Pi(i)}^T \bb^{\sharp}_j) (\xx_i^T \hat \bb_j(\Pi)) - \psi(\xx_{\Pi(i)}^T \hat \bb_j(\Pi)))  \nonumber \\
& & -  \bigg( (\psi'(\xx_i^T \bb^{\sharp}_j) (\xx_i^T \bb_j(\Pi)) - \psi(\xx_i^T \bb_j(\Pi))) - (\psi'(\xx_{\Pi(i)}^T \bb_j^{\sharp}) (\xx_i^T \bb_j(\Pi)) - \psi(\xx_{\Pi(i)}^T  \bb_j(\Pi))) \bigg) \bigg| \nonumber \\
&\leq& 2 \sum_{i: \Pi(i) \neq i} \sum_{j=1}^m (\psi_{max}^{\sharp} + \psi_{max}^{'\sharp}) x_{max} \|\hat \bb_j(\Pi) - \bb_j(\Pi)\|. \label{diff:compensator1}
\end{eqnarray}

Lastly, we can compute the lower bound of
\begin{eqnarray}
& & \Lambda(\mathbf I, B(\Pi)) - \Lambda(\Pi, B(\Pi)) \nonumber \\
&=& \Lambda(\mathbf I, B^{\sharp}) - \Lambda(\Pi, B^\sharp) - \bigg( \Lambda(\mathbf I, B^{\sharp}) - \Lambda(\Pi, B^\sharp) - (\Lambda(\mathbf I, B(\Pi)) - \Lambda(\Pi, B(\Pi))) \bigg)\nonumber \\
&\geq&  \Lambda(\mathbf I, B^{\sharp}) - \Lambda(\Pi, B^\sharp) -
2 \sum_{i: \Pi(i) \neq i} \sum_{j=1}^m (\psi_{max}^{\sharp} + \psi_{max}^{'\sharp}) x_{max} \| \bb^{\sharp}_j - \bb_j(\Pi)\| \nonumber \\
&\geq& \frac{1}{2} (\Lambda(\mathbf I, B^{\sharp}) - \Lambda(\Pi, B^\sharp)) \label{gap:compensator}
\end{eqnarray}
by using assumption that
\[\Lambda(\Pi^{\sharp}, B^{\sharp}) - \Lambda(\Pi, B^{\sharp}) \gtrsim m d(\Pi, \Pi^{\sharp}) (\psi_{max}^{\sharp} + \psi_{max}^{'\sharp}) x_{max} \delta^{\ast}.\]
Combining \eqref{uniform:concentration}, \eqref{diff:compensator1} and \eqref{gap:compensator}, we have
\[L(\mathbf I, \hat B(\Pi)) - L(\Pi, \hat B(\Pi)) \geq \frac{1}{2}(\Lambda(\mathbf I, B^{\sharp}) - \Lambda(\Pi, B^\sharp)) > 0\]
with probability going to 1.

\subsection{Situation 2}
In situation 2, for any fixed $\Pi$ with $d(\Pi, \Pi^{\sharp}) \geq h_c$, we are going to bound the difference between $L(\Pi,B)$ and $\Lambda(\Pi,B)$ uniformly over all permutation matrices and the restricted parameter space.

\subsubsection{On restricted space $\mathcal B_0$}
In this section, we will first determine the restricted parameter space $\mathcal B_0$.
First, we know that $\hat B(\Pi)$ is the maximizer of $L(\Pi,B)$.
We have that
\begin{eqnarray}
\langle Y \circ (\Pi X \hat B(\Pi)) - \psi(\Pi X \hat B(\Pi)) \rangle \geq
\langle Y \circ (\Pi X \mathbf 0) - \psi(\Pi X \mathbf 0) \rangle. \label{bound:0}
\end{eqnarray}
{\color{black} By curvature property, we have}
\begin{eqnarray}
& & \langle Y \circ (\Pi X \hat B(\Pi)) - Y \circ (\Pi X \mathbf 0) -
\psi'(\Pi X \mathbf 0) \circ (\Pi X \mathbf 0) \rangle \nonumber \\
&\geq& \frac{1}{2} \psi_{min}^{''0} n \gamma_{1p} \|\hat B(\Pi)\|^2.
\end{eqnarray}
This implies that with high probability that
\begin{eqnarray}\label{bound:xb}
\|\hat B(\Pi)\| \leq C \frac{\sigma(X)}{\psi_{min}^{''0}} =: \delta_{b2}, 
\end{eqnarray}
since $\|Y\|$ is $O_p(\sqrt{n})$.

Then the restricted parameter space $\mathcal B_0$ can be taken as
\[\mathcal B_0 := \{ B ~ |~ \|B\| \leq \delta_{b2} \}.\]

\subsubsection{Upper bound of $v_{\Pi,B}$}
For any $B \in \mathcal B_0$ and $i \in [n]$, we consider to compute the upper bound of $|\xx_i^T B|$. In fact, by Cauchy-Schwartz inequality, we have
\begin{eqnarray}
|\xx_i^T B| \leq \|\xx_i\| \|B\| \leq x_{max} \delta_{b2}.
\end{eqnarray}

By the formula of $v_{\Pi,B}$, we have that
\begin{eqnarray}\label{upperbound:variance}
v_{\Pi,B} = \sum_i \psi''(\lambda_i^{\sharp}) (\xx_{\pi_i}^T B)^2 \leq n \psi_{max}^{''\sharp} x_{max}^2 \delta_{b2}^2 = O(V_2).
\end{eqnarray}

\subsubsection{Lower bound of $v_{\Pi,B}$}
We consider to obtain the lower bound of $v_{\Pi,B}$ over the restricted parameter space $\mathcal B_0 \cap B(\mathbf 0, \delta_{b1})^c$, where $\delta_{b1}$ is determined in \eqref{def:deltab1}.
By the formula of $v_{\Pi,B}$, we know that
\begin{eqnarray}
v_{\Pi,B} = \sum_i \psi''(\lambda_i^{\sharp}) (x_{\pi_i}^T B)^2.
\end{eqnarray}
According to assumption \textit{A2}, we can see that there exist a constants $c_a$ and $c_b$ such that
\[\sharp \{ i | |x_{\pi_i}^T B| \geq  c \|B\| \} \geq n / \gamma_{1p}.\]
Thus, we can have that
\begin{eqnarray}
v_{\Pi,B} \geq c^2 n/\gamma_{1p} \|B\|^2 \psi_{min}^{''\sharp} \geq
c^2 n/\gamma_{1p} \delta_{b1}^2 \psi_{min}^{''\sharp}.
\end{eqnarray}

\subsubsection{Bound of $L(\Pi, \hat B(\Pi)) - \Lambda(\Pi, \hat B(\Pi))$ }

\vspace{0.1in}

\noindent
When $\|\hat B(\Pi)\| < \delta_{b1}$ (see \eqref{def:deltab1}), we then know that
\begin{eqnarray}\label{eq:small:gap}
& & |L(\Pi,\hat B(\Pi)) - \Lambda(\Pi,\hat B(\Pi))| \nonumber \\
& = & |\sum_{i,l} (Y[i,l] - \psi'(\llam_i^{\sharp}[l])) (\xx_{\Pi(i)}^T \hat \bb_l)| \nonumber \\
&\lesssim& \log(mn) mn (\psi_{max}^{'\sharp} + \psi_{max}^{''\sharp}) \sqrt{p} \delta_{b1}
\end{eqnarray}
for any $\Pi$.
We then know that
\[|L(\Pi,B) - \Lambda(\Pi,B)| \leq K (n \log n + mp) \]
holds for some large constant $K$,
when
\begin{eqnarray}\label{def:deltab1}
\delta_{b1} := (n \log n + mp) / (\log(mn) mn (\psi_{max}^{'\sharp} + \psi_{max}^{''\sharp}) \sqrt{p}).
\end{eqnarray}

\vspace{0.1in}
\noindent
When $\|\hat B(\Pi)\| \geq \delta_{b1}$,
for any $B, B^{'} \in \mathcal B_0 \cap B(\mathbf 0, \delta_{b1})^c$ with $\|B - B^{'}\| \leq \delta$, we have that
\begin{eqnarray}
& &|L(\Pi,B') - \Lambda(\Pi,B') - (L(\Pi,B) - \Lambda(\Pi,B))| \nonumber \\
&=& |\sum_{i,l} (Y[i,l] - \psi'(\llam_{i}^{\sharp}[l])) \big(\llam_{\pi_i l}(B') - \llam_{\pi_i l}(B))\big)| \nonumber \\
&\leq& C m n (\log n) \sqrt{p} \psi_{cb}^{\sharp} \delta \nonumber \\
&\leq& \frac{1}{2} v_{\Pi,B} x,
\end{eqnarray}
where $x$ in the last inequality is sufficiently large that the lower bound of $v_{\Pi,B} x$ dominates the term $m n (\log n) \sqrt{p} \psi_{cb}^{\sharp} \delta$ whenever
\begin{eqnarray}\label{eq:require:x1}
\delta \lesssim \delta_0 := x \delta_{b1}^2 \psi_{min}^{''\sharp} /(\psi_{cb}^{\sharp} (\log n) \sqrt{p}).
\end{eqnarray}

We then have the following concentration inequality,
\begin{eqnarray}\label{eq:uniform:large}
& & P( \sup_{B \in \mathcal B_0 \cap B(\mathbf 0, \delta_{b1})^c} \frac{1}{v_{\Pi,B}} |\langle L(\Pi, B) - \Lambda(\Pi, B) \rangle| \geq  x) \nonumber \\
&=& P( \sup_{B \in \mathcal B_g} \frac{1}{v_{\Pi,B}} |\langle L(\Pi, B) - \Lambda(\Pi, B) \rangle| \geq  x / 2) \nonumber \\
& \leq & |\mathcal B_g| \max_{B \in \mathcal B_g} P(\frac{1}{v_{\Pi,B}} |\langle L(\Pi, B) - \Lambda(\Pi, B) \rangle| \geq  x / 2) \nonumber \\
& \leq & |\mathcal B_g| \max_{B \in \mathcal B_g} \exp\{- v_{\Pi,B} x^2 / 16\},
\end{eqnarray}
where $\mathcal B_g$ is the $\delta$-covering net of $\mathcal B_0 \cap B(\mathbf 0, \delta_{b1})^c$ with $\delta = \delta_0$. Here we consider infinity norm on parameter space for constructing $\delta$-covering net.
By straightforward calculation, the cardinality of $\mathcal B_g$ is bounded by
$(C \frac{p}{\delta})^{mp}$. From \eqref{eq:uniform:large}, we can obtain the uniform concentration inequality,
\begin{eqnarray}\label{eq:uniform:large2}
& & P( \max_{\Pi} \sup_{B \in \mathcal B_0} \frac{1}{v_{\Pi,B}} |\langle L(\Pi, B) - \Lambda(\Pi, B) \rangle| \geq  x) \nonumber \\
& \leq & n! (C \frac{p}{\delta})^{m p} \exp\{- v_{\Pi,B} x^2 / 16\}
\end{eqnarray}
for $v_{\Pi,B} \geq 2  (n \log n + mp \log(p/\delta)) x_{max}^2$.

Similarly, we also have the second type concentration inequality
\begin{eqnarray}\label{eq:uniform:large2:type2}
& & P( \max_{\Pi} \sup_{B \in \mathcal B_0} \frac{1}{v_{\Pi,B}} |\langle L(\Pi, B) - \Lambda(\Pi, B) \rangle| \geq  x) \nonumber \\
& \leq & n! (C \frac{p}{\delta})^{m p} \exp\{ v_{\Pi,B}/(x_{max})^2 \}
\exp\{ - v_{\Pi,B} x / x_{max}\}
\end{eqnarray}
for $v_{\Pi,B} < 2  (n \log n + mp \log(p/\delta)) x_{max}^2$.

Then by \eqref{eq:uniform:large2} and \eqref{eq:uniform:large2:type2} with choice of
$\delta = \frac{1}{n^2}$, we know that
\[L(\Pi, \hat{B}(\Pi)) \leq
\Lambda(\Pi, \hat{B}(\Pi)) + O_p(\max\{x_1,x_2\}) \leq \Lambda(\Pi) + O_p(\max\{x_1,x_2\}).\]
We also know that
\[L(I, \hat{B}) \geq L(I, B^{\sharp}) \geq
\Lambda(\Pi, B^{\sharp}) - O_p(\max\{x_1,x_2\}) \geq \Lambda(\mathbf I) - O_p(\max\{x_1,x_2\}).\]
By above facts, we have that
\begin{eqnarray}
& & L(I, \hat{B}) - L(\Pi, \hat{B}(\Pi)) \nonumber \\
&\geq& \Lambda(\mathbf I) - \Lambda(\Pi) - O_p(\max\{x_1,x_2\}), \label{eq:final}
\end{eqnarray}
where $x_1 = \sqrt{(n \log n +  m p \log n) v_{\Pi,B}}$
and $x_2 = \max\{n \log n x_{max}, m p x_{max}\}$.

By condition that
$\Delta(X,B^{\sharp},\Pi,\Pi^{\sharp})\gtrsim \max\{x_1,x_2\}$, we get \eqref{eq:final} is greater than 0 for all $\Pi \neq I$ satisfying $d(\mathbf I, \Pi) > h_c$ with high probability.
We then have $\hat \Pi \neq \Pi$ for any $\Pi$ with $d(\mathbf I, \Pi) > h_c$.
This concludes the proof of Theorem~\ref{thm:unknown:main}.

\subsubsection{On assumption \textit{A2}}
At the end of appendix, we show that assumption \textit{A2} is automatically satisfied for sub-Gaussian design setting.
For simplicity, we take the Gaussian design for example, i.e., each entry of $X$ is sampled from standard normal distribution independently.
Fix $p_0 > 1/2$ and take any $b$ with $\|b\| = 1$, find $c_0$ such that $\Phi(c_0) = p_0$.
Therefore, \[ |P(|\sharp\{x_i^T b> c_0\} - p_0 n| \geq n t) \leq
2 \exp\{ - \frac{2 n t^2}{p_0(1 - p_0)}\}.\]	
Find $\epsilon$-cover of sphere, we then have that
\[ \sharp\{ |\xx_i \delta b| \leq C_1 \epsilon \} \geq n - \frac{n}{C_1}, \]
for any $\|\delta b\| \leq \epsilon$.
The size of $\epsilon$-cover is bounded by $(2/\epsilon + 1)^p$.
We then have that
\begin{eqnarray}
P(\sharp\{\xx_i^T \boldsymbol \beta > c_0 - C_1 \epsilon\} \geq p_0 n - n t - n/C_1; ~~ \forall \boldsymbol \beta) \leq (2/\epsilon + 1)^p 2 \exp\{- \frac{2 n t^2}{p_0(1 - p_0)}\}.
\end{eqnarray}
We than can choose $t$, $p_0$, $C_1$ and $\epsilon$ such that
\begin{eqnarray}
p \log(2/\epsilon + 1) &<& 2 t^2 / (p_0( 1 - p_0)) n \\
p_0 n - nt - n/C_1 &\geq& p \\
C_1 \epsilon &=& o(1).
\end{eqnarray}
Then we have that
\[\sharp \{ i | |x_{\pi_i}^T b| \geq  c_0/2 \} \geq (p_0 - t - 1/C_1) n.\]
Thus, we can see that assumption \textit{A2} is satisfied by letting
$c_1 = c_0/2$ and $\gamma_{1p} = p_0 - t - 1/C_1 = \Theta(1)$.

\clearpage
\section{Proof of Results in the Missing Case}\label{app:missing}

For the purpose of completeness, we provide proof under missing observation cases.
The proof strategy is similar to that of the previous setting.

\subsection{Size of $\mathcal S_l$}

By the definition, we know that $\mathcal S_l = \{i: E[i,l] = 1\}$. Next, we give the upper and lower bound of $\mathcal S_l$.
By Bernstein inequality, we have that
\begin{eqnarray}\label{E:bernstein}
P( |\sum E[i,l] - qn| \geq n x ) \leq \exp\{- \frac{n^2 x^2}{2(n q(1-q)  + nx/3)}\}.
\end{eqnarray}
Thus
\begin{eqnarray}\label{E:bernstein:2}
P( |\sum E[i,l] - qn| \geq n q / 2 ) \leq \exp\{- \frac{3 nq}{7}\}.
\end{eqnarray}
In other words,
\begin{eqnarray}\label{bound:S}
P( qn / 2 \leq \min_l |\mathcal S_l| \leq \max_l |\mathcal S_l| \leq 3qn/2) \leq m \exp\{- \frac{3 nq}{7}\},
\end{eqnarray}
which means the sizes of $S_l$'s are around $qn$ with high probability.

\subsection{When $B^{\sharp}$ is known}

We next calculate the variance of $\langle E[i,:] \circ (\yy_i \circ \llam_i - \psi(\llam_i)) \rangle - \langle E[i,:] \circ (\yy_i \circ  \llam_j - \psi(\llam_j)) \rangle$.
\begin{eqnarray}
& & \mathrm{var}(\langle E[i,:] \circ (\yy_i \circ \llam_i - \psi(\llam_i)) \rangle - \langle E[i,:] \circ (\yy_i \circ \llam_j - \psi(\llam_j)) \rangle) \nonumber \\
& = & q \sum_{l=1}^m \mathrm{var} (Y[i,l] (\llam_i[l] - \llam_j[l]) - (\psi(\llam_i[l]) - \psi(\llam_j[l]))) \nonumber \nonumber \\
& & + q(1-q) (\sum_{l=1}^m \mathbb E [Y[i,l]] (\llam_i[l] - \llam_j[l]) - (\psi(\llam_i[l]) - \psi(\llam_j[l])))^2 \nonumber \\
&=& q \sum_{l=1}^m \psi''(\llam_i[l]) (\llam_i[l] - \llam_j[l])^2 + q(1-q)\sum_{l=1}^m( \psi'(\llam_{i}[l])(\llam_i[l] - \llam_j[l]) - (\psi(\llam_i[l]) - \psi(\llam_j[l])))^2 \nonumber \\
&=& \sum_l \{q x_{ij,2}[l] + q(1-q) (x_{ij,1}[l])^2\}\\
&:=& v_{ij}(q),
\end{eqnarray}
where, for simplicity, we let $x_{ij,1}[l] = \psi'(\llam_{i}[l])(\llam_i[l] - \llam_j[l]) - (\psi(\llam_i[l]) - \psi(\llam_j[l]))$ and $x_{ij,2}[l] = \psi''(\llam_i[l]) (\llam_i[l] - \llam_j[l])^2$.
We may also suppress subscript $i,j$ in $x_{ij,1}$ or $x_{ij,2}$ in the following calculations.

The moment generating function of $\langle E[i,:] \circ (\yy_i \circ \llam_i - \psi(\llam_i)) \rangle - \langle E[i,:] \circ (\yy_i \circ \llam_j - \psi(\llam_j)) \rangle - \Delta_{ij}(q)$ is
\begin{eqnarray}\label{eq:q:mgf}
& & \mathbb E \exp\{t (\langle E[i,:] \circ (\yy_i \circ \llam_i - \psi(\llam_i)) \rangle - \langle E[i,:] \circ (\yy_i \circ \llam_j - \psi(\llam_j)) \rangle - \Delta_{ij}(q))\} \nonumber \\
&=&  \prod_{l=1}^m \mathbb E \exp \bigg \{t \bigg (E[i,l] (Y[i,l] ( \llam_j[l]  - \llam_j[l]) - (\psi(\llam_i[l])  - \psi(\llam_j[l]))) - \Delta_{ij}(q)[l] \bigg) \bigg\} \nonumber \\
&=& \prod_{l=1}^m \bigg \{ \big((1-q) + q \mathbb E \exp\{t (Y[i,l] (\llam_i[l] - \llam_j[l])- (\psi(\llam_i[l]) - \psi(\llam_j[l])))\} \big) \exp\{-t\Delta_{ij}(q)[l]  \} \bigg \} \nonumber \\
&=& \prod_{l=1}^m \bigg \{ \big((1-q) + q \exp\{\psi(\llam_i[l] + t(\llam_i[l] - \llam_j[l])) - \psi(\llam_i[l]) - t (\psi(\llam_{i}[l]) - \psi(\llam_j[l]))\} \big) \exp\{-t \Delta_{ij}(q)[l] \} \bigg \}  \nonumber \\
&\leq& \prod_{l=1}^m (1 + q x_1[l] t + a q x_2[l] t^2 + b q (x_1[l] t + ax_2[l] t^2)^2)(1 - q x_1[l] t + c (q x_1[l] t)^2) \nonumber \\
&\leq& \prod_{l=1}^m (1 + 2 (a q x_2[l] t^2 + b q x_1^2[l] t^2 +c q^2 x_1^2[l] t^2 - q^2 x_1^2[l] t^2)) \nonumber \\
&\leq& \prod_{l=1}^m (1 + 2 (a q x_2[l] t^2 + b q(1-q) x_1^2[l] t^2)) \nonumber \\
&\leq& \prod_{l=1}^m \exp\{ 2 (q x_2[l] + q(1-q)x_1^2[l])t^2 \} \nonumber \\
&=&\exp\{2 v_{ij}(q) t^2\},
\end{eqnarray}
where we suppress symbols $x_{ij,1}, x_{ij,2}$ to $x_1$ and $x_2$ respectively. We also use the Taylor expansion for multiple times in the above inequalities which depend on the following fact,
\[\exp\{x\} \leq 1 + x + (\frac{1}{2} + \frac{1}{5}x)x^2\]
for any $|x| < 0.5$.
In \eqref{eq:q:mgf}, we specifically take $a = 1$, $c = 1 - b$, $b = \frac{1}{2} + \frac{1}{10}(q x_1t + q ax_2t^2) < 1$.
This choice is possible and \eqref{eq:q:mgf} holds for any $t \lesssim \min \{1/\llam_j[l], 1/(q x_1[l]), 1/\sqrt{q x_2[l]}\}$ for any $l$.

Thus, we have
\begin{eqnarray}
& & P(|\langle E[i,:] \circ (\yy_i \circ \llam_i - \psi(\llam_i)) \rangle - \langle E[i,:] \circ (\yy_i \circ \llam_j - \psi(\llam_j)) \rangle - \Delta_{ij}(q)| \geq v_{ij}(q) x) \nonumber \\
&\leq& \inf_t \exp\{2 v_{ij}(q) t^2 \} \exp\{-v_{ij}(q) x t\} \nonumber  \\
&\leq& \exp\{- \frac{1}{8} v_{ij}(q) x^2\} \nonumber
\end{eqnarray}
as long as $x \lesssim 4 \min_l \{1/\llam_j[l], 1/(qx_1[l]), 1/\sqrt{qx_2[l]}\}$.

Similar to no missing observation case, we can obtain that the requirement for permutation recovery is
\begin{eqnarray}
\Delta_{ij}(q) \gtrsim \sqrt{(\log n) v_{ij}(q)}. \nonumber
\end{eqnarray}
That is,
\[\Delta_{ij}(q)^2 \gtrsim  q \sum_{l=1}^m \psi''(\llam_i[l]) (\llam_i[l] - \llam_j[l])^2 + q(1-q)\sum_{l=1}^m( \psi'(\llam_{i}[l])(\llam_i[l] - \llam_j[l]) - (\psi(\llam_i[l]) - \psi(\llam_j[l])))^2.\]

Especially, when $\lambda_{il}^{\sharp}$'s are bounded and $\min_{i,j} \sum_{l \in [m]} (\lambda_{il}^{\sharp} - \lambda_{jl}^{\sharp})^2 = \Omega(m)$,
the above inequality becomes
\[q^2 m^2 \gtrsim \log n (q m + q(1 - q) m).\]
Thus $q \geq \frac{\log n}{m}$ is required for the perfect permutation recovery.

\subsection{With knowledge that $d(\mathbf I, \Pi^{\sharp})$ is small}

Again, in order to prove the recovery consistency, we need to control the following quantities,
$\|B - B^{\sharp}\|$
and
$(L(\mathbf I, B, E) - L(\Pi, B, E)) - (\Lambda(\mathbf I, B, q) - \Lambda(\Pi, B, q))$.

Suppose we have already known that the estimator $\hat {B}$ which is close to the truth $B^{\sharp}$, i.e., in the $\delta$-neighborhood of $B^{\sharp}$.
Then we let $B_{\delta}(B^{\sharp}) := \{B : \|B - B^{\sharp}\|_{2} \leq \delta\}$ and $\delta$ is a sufficiently small constant which will be determined later.
For any fixed $\Pi$, we have
\begin{eqnarray}\label{eq:goal:concentration:single2:qq}
& & P( \sup_{\mathbf B \in B_{\delta}(B^{\sharp})}  |(L(\mathbf I, B, E) - L(\Pi, B, E)) - (\Lambda(\mathbf I, B, q) - \Lambda(\Pi, B, q))| \geq 2 v_{\Pi,partial,q} x ) \nonumber \\
&\leq& P( |(L(\mathbf I, B^{\sharp},E) - L(\Pi, B^{\sharp},E)) - (\Lambda(\mathbf I, B^{\sharp},q) - \Lambda(\Pi, B^{\sharp},q))| \geq v_{\Pi,partial,q}x ) \nonumber \\
&\leq& \exp\{ - \frac{1}{4} v_{\Pi,partial,q} x^2 \}.
\end{eqnarray}
Hence, we get
\begin{eqnarray}\label{eq:goal:concentration:combine2:qq}
& & P( \max_{\Pi} \sup_{B \in B_{\delta}(B^{\sharp})}  |(L(\mathbf I, B, E) - L(\Pi, B, E)) - (\Lambda(\mathbf I, B, q) - \Lambda(\Pi, B, q))| \geq 2 v_{\Pi,partial,q} x ) \nonumber \\
&\leq& \sum_{\Pi} P(  |(L(\mathbf I, B^{\sharp}) - L(\Pi, B^{\sharp})) - (\Lambda(\mathbf I, B^{\sharp}) - \Lambda(\Pi, B^{\sharp}))| \geq v_{\Pi,partial,q} x ) \nonumber  \\
& = & \sum_h \sum_{\Pi: d(\Pi,\mathbf I) = h} \cdot \exp\{ - v_{\Pi,partial,q} x^2 / 4 \} \nonumber \\
&\leq& \sum_h n!/(n-h)! \cdot \exp\{ - v_{\Pi,partial,q} x^2 / 4 \} \nonumber \\
&\leq& \sum_h n^h \cdot \exp\{ - h v_{min,q} x^2 / 4 \} \nonumber \\
&=& \sum_h \exp\{ - h (v_{min,q} x^2 - \log n)\}  \nonumber \\
&\leq& \frac{\exp\{ - 2 (v_{min,q} x^2 - \log n)\}}{1 - \exp\{ - (v_{min,q} x^2 - \log n)\}}, \label{eq:close:rate:combine:qq}
\end{eqnarray}
where we recall that $v_{min,q} = \min_{i,j} \sum_l \bigg \{q  \psi''(\llam_{i}^{\sharp}[l]) (\llam_{j}^{\sharp}[l] - \llam_{i}^{\sharp}[l])^2 + q(1-q) (\psi'(\llam_{i}^{\sharp}[l]) (\llam_{i}^{\sharp}[l] - \llam_{j}^{\sharp}[l]) - (\psi(\llam_{i}^{\sharp}[l]) - \psi(\llam_{k}^{\sharp}[l])))^2 \bigg\}$.

By \eqref{eq:close:rate:combine:qq}, we know that $L(\Pi, \hat{B},E) \leq
L(\mathbf I, \hat{B},E) - \Lambda(\mathbf I, \hat{B},q) + \Lambda(\Pi, \hat{B},q) + O_p(x v_{\Pi,B^{\sharp},q})$ with $x = \sqrt{\log n / v_{min}}$.
This tells us that if we can show that
\begin{eqnarray}\label{eq:goal:onestep:qq}
\Lambda(\mathbf I, \hat B,q) - \Lambda(\Pi, \hat{B},q) - O_p(x v_{\Pi,B^{\sharp},q}) > 0.
\end{eqnarray}
Then we can conclude that $\hat \Pi = \mathbf I$ which leads to the desired result.

\subsubsection{First bound of $\|B - B^{\sharp}\|_F$}
Again, without loss of generality, we first assume $m=1$. Then the estimator $\hat {\mathbf b}$ is
\begin{eqnarray}
\hat {\mathbf b} = \arg \max_{\mathbf b} \{ \langle E \circ (- \psi(X \mathbf b) + Y \circ X \mathbf b) \rangle \}.
\end{eqnarray}
When $d(\mathbf I, \Pi^{\sharp}) \leq h_{max}$, we aim to show $\hat{\mathbf b}$ is a consistent estimator of $\mathbf b^{\sharp}$.
For simplicity, we assume $\Pi^{\sharp}(i) = i$ for $i > h_{max}$ and let $L(\mathbf b) = \langle E \circ (- \psi(X \mathbf b) + Y \circ X \mathbf b) \rangle $.
In the following, we construct a $\delta_n$ such that
for any $\mathbf b$ with $\|\mathbf b - \mathbf b^{\sharp}\| \geq \delta_n$, it holds
$L(\mathbf b) < L(\mathbf b^{\sharp})$.
By the definition that
$L(\hat {\mathbf b}) \geq L(\mathbf b^{\sharp})$, we will arrive at $\|\hat {\mathbf b} - \mathbf b^{\sharp}\| \leq \delta_n$.

We compute that
\begin{eqnarray}
& & L(\mathbf b^{\sharp}) - L(\mathbf b) \nonumber \\
&=& (L(\mathbf b^{\sharp}) - \Lambda(\mathbf b^{\sharp},q)) - (L(\mathbf b) - \Lambda(\mathbf b, q)) +  \Lambda(\mathbf b^{\sharp},q) - \Lambda(\mathbf b,q) \nonumber \\
&=& (L(\mathbf b^{\sharp}) - \Lambda(\mathbf b^{\sharp},q)) - (L(\mathbf b) - \Lambda(\mathbf b,q)) +  \Lambda_1(\mathbf b^{\sharp},q) - \Lambda_1(\mathbf b,q)
+ \Lambda_2(\mathbf b^{\sharp},q) - \Lambda_2(\mathbf b,q) \nonumber \\
&\geq& - O_p(\sqrt{v_{b^{\sharp}, q}}) - O_p(\sqrt{v_{b,q}}) - q h_{max} C_{max}  + \sigma_{min}(H_q) \delta_n^2. \label{eq:small1:qq}
\end{eqnarray}
For the last inequality, we use the following facts. Note that $L(\mathbf b) - \Lambda(\mathbf b,q) = O_p( \sqrt{v_{b,q}})$ with
\[v_{b,q} = \sum_{i=1}^n \{ q \psi''(\llam_{i}^{\sharp}) (\llam_i(\mathbf b))^2 + q(1-q) \big(\psi'(\llam_{i}^{\sharp}) \llam_i(\mathbf b) - \psi(\llam_i(\mathbf b))\big) \}\] for any $\mathbf b$. Additionally note that
\begin{eqnarray}
& & |\Lambda_1(\mathbf b^{\sharp},q) - \Lambda_1(\mathbf b,q)| \nonumber \\
&=& q |\sum_{i \leq h_{max}} \{- \psi(\xx_i^T \mathbf b^{\sharp}) + \psi'(\xx_i^T \mathbf b^{\sharp}) \xx_i^T \mathbf b^{\sharp} - (- \psi(\xx_i^T \mathbf b) + \psi'(\xx_i^T \mathbf b^{\sharp}) \xx_i^T \mathbf b) \} | \nonumber \\
&\leq& q \sum_{i \leq h_{max}} |\psi_{max}(\delta_n)| + q \psi'(\xx_i^T \mathbf b^{\sharp}) |\xx_i^T(\mathbf b^{\sharp} - \mathbf b)| \nonumber \\
&\leq& q h_{max} \psi_{max}(\delta_n) +  q h_{max} \psi_{max}^{'\sharp} \max_{i,\mathbf b} |\xx_i^T(\mathbf b^{\sharp} - \mathbf b)| \nonumber \\
&=& q h_{max} C_{max},
\end{eqnarray}
where $C_{max} = \psi_{max}(\delta_n) + \psi_{max}^{'\sharp} \max_{i, \mathbf b} |\xx_i^T(\mathbf b^{\sharp} - \mathbf b)|$.
Moreover, for any $\mathbf b$ and $\mathbf b^{'}$, we compute
\begin{eqnarray*}
& & \Lambda_2(\mathbf b,q) - \Lambda_2(\mathbf b^{'},q) \nonumber \\
&=& q \sum_{i > h_{max}} \{- \psi(\xx_i^T\mathbf b) + \psi'(\xx_i^T\mathbf b) \xx_i^T \mathbf b - (- \psi(\xx_i^T \mathbf b^{'}) + \psi'(\xx_i^T \mathbf b) \xx_i^T \mathbf b^{'}) \} \\
& = & q \sum_{i > h_{max}} \psi''(\xx_i^T \tilde {\mathbf b})(\xx_i^T \mathbf b - \xx_i^T \mathbf b^{'})^2 / 2 \\
&=&  q (\mathbf b - \mathbf b^{'})^T \sum_{i \geq h_{max}} \psi''(\xx_i^T \tilde {\mathbf b}) \xx_i \xx_i^T (\mathbf b - \mathbf b^{'}) / 2 \\
&=& (\mathbf b - \mathbf b^{'})^T H_q (\mathbf b - \mathbf b^{'}) / 2,
\end{eqnarray*}
where $\tilde {\mathbf b}$ is between $\mathbf b$ and $\mathbf b^{'}$, and
\begin{eqnarray*}
H_q = q \sum_{i > h_{max}} \psi''(\xx_i^T\tilde{\mathbf b}) \xx_i \xx_i^T.
\end{eqnarray*}
Again by curvature technique, we know that
\[(\mathbf b - \mathbf b^{\sharp})^T H_q (\mathbf b - \mathbf b^{\sharp}) \geq c_1 q (n - h_{max}) \psi_{min}^{\sharp} / {\color{black} \gamma_{2p}}\|\mathbf b - \mathbf b^{\sharp}\|^2 \]
for some constant $c_1$.
Specifically, we can take
\[\delta_n^2 = C \frac{\sqrt{v_{b^{\sharp},q}} + q \lambda_{max}^{\sharp} h_{max}}{q (n - h_{max}) \lambda_{min}^{\sharp} / {\color{black} \gamma_{2p}}}\]
with some large constant $C$. With this choice of $\delta_n$, from \eqref{eq:small1}, we can check that
\[L(\mathbf b^{\sharp}) - L(\mathbf b) > 0\]
holds for any $\mathbf b$ with $\|\mathbf b - \mathbf b^{\sharp}\| = \delta_n$ when $ p^2/q < n$ and $ph_{max} < n / \log n$. By the concavity of likelihood function, we then know that $\|\hat{\mathbf b} - \mathbf b^{\sharp}\| \leq \delta_n$.

\subsubsection{Second bound of $\|B - B^{\sharp}\|_F$}
We do the Taylor expansion of $L(\mathbf b)$ at  $\mathbf b = \mathbf b^{\sharp}$. Then, it can be computed that
\begin{eqnarray}
\mathbf 0 = \nabla L(\hat {\mathbf b}) = \nabla L(\mathbf b^{\sharp}) + \nabla^2 L(\bar{\mathbf b}) (\hat{\mathbf b} - \mathbf b^{\sharp}) \nonumber
\end{eqnarray}
where $\bar{\mathbf b}$ is some point between $\hat{\mathbf b}$ and $\mathbf b^{\sharp}$.
Again, by curvature technique under assumption \textit{E2}, we have
\begin{eqnarray}
& & \|\nabla^2 L(\bar{\mathbf b}) (\hat{\mathbf b} - \mathbf b^{\sharp})\| \nonumber \\
&\geq& c q n \psi_{min}^{''\sharp}/{\color{black} \gamma_{2p}} \|\hat{\mathbf b} - \mathbf b^{\sharp}\|.
\end{eqnarray}
We thus have
\begin{eqnarray}
\|\hat{\mathbf b} - \mathbf b^{\sharp}\| = \| (\nabla^2 L(\bar{\mathbf b}))^{-1} \nabla L(\mathbf b^{\sharp})\| \leq \frac{1}{c qn \psi_{min}^{''\sharp}/{\color{black} \gamma_{2p}}} \|\nabla L(\mathbf b^{\sharp})\|.
\end{eqnarray}
For $l$-th element of $\nabla L(\mathbf b^{\sharp})$, we can find that \[\nabla L(\mathbf b^{\sharp})[l] = \sum_{i \leq h_{max}: E[i] = 1}
\nabla L_i(\mathbf b^{\sharp})[l] + \sum_{i > h_{max}: E[i] = 1} \nabla L_i(\mathbf b^{\sharp})[l].\]
The first term is bounded by
\[b_s := |\sum_{i \leq h_{max}} (Y[i] -\psi'(\xx_i^T \mathbf b^{\sharp})) X[i,l] \}|,\]
which is order of $ (\log n) \sum_{i \leq h_{max}} \psi_{max}^{''\sharp} |X[i,l]|$. The second term is bounded by $ C \sqrt{\textrm{var}(\sum_{i > h_{max}} \nabla L_i(\mathbf b^{\sharp})[l])}$ and the upper bound of $\textrm{var}(\sum_{i > h_{max}: E[i]=1} \nabla L_i(\mathbf b^{\sharp})[l])$ can be computed explicitly, i.e.,
\[v_{s,q} := \max_{l \in [p]} \sum_{i > h_{max}} q \psi''(\xx_i^T \mathbf b^{\sharp}) (X[i,l])^2 + q(1-q)( \psi'(\xx_i^T \mathbf b^{\sharp}) X[i,l] - \psi(\xx_i^T \mathbf b^{\sharp}))^2.\]
Thus
\begin{eqnarray}
\|\hat{\mathbf b} - \mathbf b^{\sharp}\| &\leq& \sqrt{p} (\sqrt{v_{s,q}}+b_s) / (c qn \psi_{min}^{''\sharp}/{\color{black} \gamma_{2p}}) = O_p(\frac{\sqrt{p}(\sqrt{v_{s,q}} + \psi_{max}^{''\sharp}h_{max}\log n)}{qn})\nonumber \\
&=& O_p(\frac{\sqrt{p}(\sqrt{q \psi^{''\sharp}_{max} + q(1-q) \psi^{\sharp 2}_{cb}} \sqrt{n - h_{max}} + \psi_{max}^{''\sharp}h_{max}\log n)}{qn \psi_{min}^{''\sharp}/{\color{black} \gamma_{2p}}})  := \delta_q^{\ast}.
\end{eqnarray}
Thus the bound $\delta^{\ast}$ is tighter than $\delta_n$.
Especially, when $\xx_i^T \mathbf b^{\sharp}$ is bounded for all $i$ and $\gamma_{2p} = O(1)$, then $\delta^{\ast}$ can be simplified as $\frac{\sqrt{p}(\sqrt{q (n - h_{max})} + h_{max}\log n)}{q n}$.

\subsubsection{Difference of $\Lambda(\mathbf I, \hat {\mathbf b},q) - \Lambda(\Pi, \hat{\mathbf b},q)$}

By straightforward calculation, we get
\begin{eqnarray}
& & \Lambda(\mathbf I, \hat {\mathbf b},q) - \Lambda(\Pi, \hat{\mathbf b},q) \nonumber \\
&=& q \bigg\{(\Lambda(\mathbf I, \hat {\mathbf b}) - \Lambda(\mathbf I, \mathbf b^{\sharp})) - (\Lambda(\Pi, \hat{\mathbf b}) - \Lambda( \Pi, \mathbf b^{\sharp})) + (\Lambda(\mathbf I, \mathbf b^{\sharp}) - \Lambda(\Pi, \mathbf b^{\sharp})) \bigg\}\nonumber \\
&\geq&  - 4 q h_{max} \psi_{cb}^{\sharp} x_{max} \delta^{\ast} + q (\Lambda(\mathbf I, \mathbf b^{\sharp}) - \Lambda(\Pi, \mathbf b^{\sharp})), \label{ineq:continuity:qq}
\end{eqnarray}
where the last inequality holds due to the same reason as explained in no-missing observation case.
By \eqref{ineq:continuity:qq} and summing over $l \in [m]$, we have that
\begin{eqnarray}
& & \Lambda(\mathbf I, \hat {B},q) - \Lambda(\Pi, \hat{B},q) \nonumber \\
&\geq&  - 4 q m d(\mathbf I, \Pi) \psi_{cb}^{\sharp} x_{max} \delta^{\ast} + q (\Lambda(\mathbf I, \mathbf b^{\sharp}) - \Lambda(\Pi, \mathbf b^{\sharp})), \label{ineq:continuity:m:qq} \nonumber \\
&\gtrsim& x v_{\Pi,partial,q},
\end{eqnarray}
when
\begin{eqnarray}
q (\Lambda(\mathbf I, \mathbf b^{\sharp}) - \Lambda(\Pi, \mathbf b^{\sharp})) \gtrsim x v_{\Pi,partial,q} \nonumber
\end{eqnarray}
and
\begin{eqnarray}
\Lambda(\mathbf I, \mathbf b^{\sharp}) - \Lambda(\Pi, \mathbf b^{\sharp}) \gtrsim m d(\mathbf I, \Pi) \psi_{cb}^{\sharp} x_{max} \delta^{\ast}. \nonumber
\end{eqnarray}
This completes the proof if we take $x = \sqrt{\log n / v_{min,q}}$.

\bigskip

\subsection{With no knowledge of $B^{\sharp}$ and $\Pi^{\sharp}$}
We consider to compute the moment generating function of $\langle L(\Pi,B,E) - \mathbb E \Lambda(\Pi,B,E) \rangle$.
By similar procedure, we can obtain
\begin{eqnarray}
& &\mathbb E \exp\{t\langle L(\Pi,B,E) - \mathbb E \Lambda(\Pi,B,E) \rangle\} \nonumber \\
&\leq& \prod_{i=1}^n \prod_{l=1}^m \mathbb E \exp \bigg \{t \bigg (E[i,l] (Y[i,l] \llam_{\Pi(i)}[l] - \psi(\llam_{\Pi(i)}[l])) - q (\psi'(\llam_i^{\sharp}[l]) \llam_{\Pi(i)}[l] - \psi(\llam_{\Pi(i)}[l])) \bigg) \bigg\} \nonumber \\
&\leq& \exp\{2 v_{\Pi,B,q} t^2\} \nonumber
\end{eqnarray}
for $t \leq \min\{\frac{c}{\psi'(x_{max}\sqrt{p/q}) + \psi''(x_{max}\sqrt{p/q})}\} := 1/g(n,p)$ ($c$ is some small constant, $g(n,p)$ is around of order $\psi_{cb}^{\sharp}$),
where
\begin{eqnarray*}
v_{\Pi,B,q} = q \sum_{i=1}^n \sum_{l=1}^m \psi''(\llam_i^{\sharp}[l]) (\llam_{\Pi(i)}[l])^2 + q(1-q) \sum_{i=1}^n \sum_{l=1}^m( \psi'(\llam_{i}^{\sharp}[l]) \llam_{\Pi}[l] - \psi(\llam_{\Pi(i)}[l]))^2.
\end{eqnarray*}
Thus, we have
\begin{eqnarray*}
& & P(|\langle L(\Pi, B) - \Lambda(\Pi, B) \rangle| \geq v_{\Pi,B,q} x) \nonumber \\
&\leq& \inf_t \exp\{ - t v_{\Pi,B,q} x \} \exp\{v_{\Pi,B,q} t^2\} \\
&\leq& \exp\{ - v_{\Pi,B,q} x^2 / 4\}
\end{eqnarray*}
for any $x \leq \frac{1}{2 g(n,p)}$.
We also have
\begin{eqnarray*}
& & P(|\langle L(\Pi, B) - \Lambda(\Pi, B) \rangle| \geq v_{\Pi,B,q} x) \nonumber \\
&\leq& \inf_t \exp\{ - t v_{\Pi,B,q} x \} \exp\{v_{\Pi,B,q} t^2\} \\
&\leq& \exp\{ v_{\Pi,B,q}/(g(n,p))^2\} \exp\{- v_{\Pi,B,q} x / g(n,p)\}
\end{eqnarray*}
for any $x$.

As before, we still consider to situations, 1. $d(\Pi, \Pi^{\sharp}) \leq h_c$
2. $d(\Pi, \Pi^{\sharp}) \geq h_c$ where $h_c = c_0 qn / p$.
We first show the difference between $B(\Pi)$ and $B^{\sharp}$.
By the definition of $B(\Pi)$, we know
\[B(\Pi) = \arg\max_B \Lambda(\Pi,B,q).\]
Note that $\Lambda(\Pi,B,q)$ is also separable for each column of $B$, i.e.,
\[\bb_j(\Pi) = \arg\max_{\mathbf b} q \bigg\{\sum_{i=1}^n \psi'(\xx_i^T \bb_j^{\sharp}) (\xx_{\Pi(i)}^T \mathbf b) - \psi(\xx_{\Pi(i)}^T \mathbf b) \bigg\}. \]
(Here, we still assume $\Pi^{\sharp} = \mathbf I$ without loss of generality.)
Notice that that the maximizer of $\Lambda(\Pi, B, q)$ remains the same as that of $\Lambda(\Pi,B)$. Thus, the difference $\|\bb_j(\Pi) - \bb_j^{\sharp}\|$ has already been obtained as before.

\subsubsection{Situation 1}
For situation 1, we are going to show
\[L(\mathbf I, \hat B, E) > L(\Pi, \hat B(\Pi),E)  \]
with high probability. It suffices to show
\[L(\mathbf I, \hat B(\Pi), E) > L(\Pi, \hat B(\Pi), E) \]
for any $\Pi$ with $d(\mathbf I, \Pi) \leq h_c$.

Given $\Pi$, we aim to calculate the bound of $\|\hat B(\Pi) - B(\Pi)\|$.
By the definition of $\hat B(\Pi)$ and convexity of negative log-likelihood function, we have that
\begin{eqnarray}
L(\Pi, B^{\sharp},E) \geq L(\Pi, \hat B(\Pi),E) &=&
L(\Pi, B^{\sharp},E) +  \langle \partial L(\Pi, B^{\sharp},E), \hat B(\Pi) - B^{\sharp} \rangle \nonumber \\
& & + \frac{1}{2}  (\hat B(\Pi) - B^{\sharp})^T \nabla^2 L(\Pi, \tilde B, E) (\hat B(\Pi) - B^{\sharp}), \nonumber
\end{eqnarray}
which gives us that
\begin{eqnarray}
\frac{1}{2}  (\hat B(\Pi) - B^{\sharp})^T \nabla^2 L(\Pi, \tilde B, E) (\hat B(\Pi) - B^{\sharp})
&\leq& |\langle \nabla L(\Pi, B^{\sharp}, E), \hat B(\Pi) - B^{\sharp} \rangle| \nonumber \\
\frac{1}{2}  (n q \psi_{min}^{''\sharp}/ {\color{black} \gamma_{3p}}) \|\hat B(\Pi) - B^{\sharp}\|^2
&\leq& \| \hat B(\Pi) - B^{\sharp}\| \|\langle \nabla L(\Pi, B^{\sharp}, E)\| \nonumber \\
(n q \psi_{min}^{''\sharp}/ {\color{black} \gamma_{3p}}) \|\hat B(\Pi) - B^{\sharp}\|^2 &\leq& C \psi_{cb}^{\sharp} \| \hat B(\Pi) - B^{\sharp}\| \sqrt{p}(C h \log n  + \sqrt{n \log p} )  \nonumber \\
\label{uniform:bound:q} \\
\|\hat B(\Pi) - B^{\sharp}\| &\leq&  C \psi_{cb}^{\sharp} \frac{\sqrt{p} ( C h \log n  + \sqrt{nq \log p} )}{\psi_{min}^{''\sharp} nq/{\color{black} \gamma_{3p}}}. \label{bound:important1:q}
\end{eqnarray}
This tells us that
\[\|\hat B(\Pi) - B(\Pi)\| \leq \|B(\Pi) - B^{\sharp}\| + \|\hat B(\Pi) - B^{\sharp}\| = o(1)\]
as long as $p h \ll q n / (\gamma_{3p} \log n)$ and $p = (qn)^{1/2 - o(1)}$.

Here, \eqref{uniform:bound:q} comes from the following fact.
For each $l \in [p]$, we consider to compute the following bound, i.e.,
\begin{eqnarray}
\sup_{\Pi: d(\Pi, \mathbf I) \leq h} |\nabla L(\Pi, B^{\sharp},E)[l]|
&\leq&
\sup_{\Pi: d(\Pi, \mathbf I) \leq h} \{ |\nabla L(\Pi, B^{\sharp},E)[l] - \nabla L(\mathbf I, B^{\sharp},E)[l]| \}
+ |\nabla L(\mathbf I, B^{\sharp},E)[l]| \nonumber \\
&\leq& C \psi_{cb}^{\sharp} ( C h \log n  + \sqrt{n q \log p} ),
\end{eqnarray}
by noticing that the entry of design matrix is bounded.

\vspace{0.1in}
\noindent \textbf{Uniform Bound of $\|\hat B(\Pi) - \hat B\|$}~
By \eqref{bound:important1:q}, we have
\[\|\hat B(\Pi) - \hat B\| \leq 2 C \psi_{cb}^{\sharp} \frac{\sqrt{p}(h \log n + \sqrt{n q \log p})}{\psi_{min}^{''\sharp} nq / {\color{black} \gamma_{3p}}}\]
by adjusting the constant.

\vspace{0.1in}
\noindent \textbf{Uniform Concentration Inequality}
By similar calculation as we have done in \eqref{concen:1}, we can obtain that
\begin{eqnarray}
& &P( |(L(\mathbf I, B, E) - L(\Pi, B, E)) - (\Lambda(\mathbf I, B, q) - \Lambda(\Pi, B, q))| \geq v_{\Pi, partial, q} x) \nonumber \\
&\leq& \exp\{-\frac{1}{4} v_{\Pi, partial, q} x^2 \}
\end{eqnarray}
for any fixed $B \in \mathcal B(B^{\sharp}, \delta)$ with $\delta = o(1/\sqrt{p})$.
By using this, we can further obtain the uniform inequality:
\begin{eqnarray}
& &P( \sup_{\Pi: d(\mathbf I, \Pi) \leq h_c} \frac{1}{v_{\Pi, partial,q} }|(L(\mathbf I, \hat B(\Pi), E) - L(\Pi, \hat B(\Pi), E) - (\Lambda(\mathbf I, \hat B(\Pi), q) - \Lambda(\Pi, \hat B(\Pi),q))| \geq x) \nonumber \nonumber \\
&\leq& \sum_{\Pi: d(\mathbf I, \Pi) \leq h_c} \exp\{-\frac{1}{4} v_{\Pi, partial,q} x^2 \} \nonumber \\
&\leq& \sum_{h=2}^{h_c} \sum_{\Pi: d(\mathbf I, \Pi) = h} \exp\{-\frac{1}{4} v_{\Pi, partial,q} x^2 \} \nonumber \\
&\leq& \sum_{h=2}^{h_c} n^h \exp\{-\frac{1}{4} h v_{min,q} x^2 \} \nonumber \\
&\leq& \frac{-2(v_{min,q} x^2 - \log n)}{1 - \exp\{-(v_{min,q} x^2 - \log n)\}}. \label{uniform:concentration:q}
\end{eqnarray}

On the other hand, we calculate the difference
\begin{eqnarray}
& & |(\Lambda(\mathbf I, \hat B(\Pi),q) - \Lambda(\Pi, \hat B(\Pi),q) - (\Lambda(\mathbf I, B(\Pi),q) - \Lambda(\Pi, B(\Pi),q))| \nonumber \\
&\leq& q \bigg |\sum_{i: \Pi(i) \neq i} \sum_{j=1}^m (\psi'(\xx_i^T \bb_j^{\sharp}) (\xx_i^T \hat \bb_j(\Pi)) - \psi(\xx_i^T \hat \bb_j(\Pi))) - (\psi'(\xx_{\Pi(i)}^T \bb_j^{\sharp}) (\xx_i^T \hat \bb_j(\Pi)) - \psi(\xx_{\Pi(i)}^T \hat \bb_j(\Pi)))  \nonumber \\
& & -  \bigg( (\psi'(\xx_i^T \bb_j^{\sharp}) (\xx_i^T \bb_j(\Pi)) - \psi(\xx_i^T \bb_j(\Pi))) - (\psi'(\xx_{\Pi(i)}^T \bb_j^{\sharp}) (\xx_i^T \bb_j(\Pi)) - \psi(\xx_{\Pi(i)}^T  \bb_j(\Pi))) \bigg) \bigg| \nonumber \\
&\leq& 2 q \sum_{i: \Pi(i) \neq i} \sum_{j=1}^m \psi_{cb}^{\sharp} x_{max} \|\hat \bb_j(\Pi) - \bb_j(\Pi)\|. \label{diff:compensator1:q}
\end{eqnarray}

Moreover, we can compute the
\begin{eqnarray}
& & \Lambda(\mathbf I, B(\Pi),q) - \Lambda(\Pi, B(\Pi),q) \nonumber \\
&=& \Lambda(\mathbf I, B^{\sharp},q) - \Lambda(\Pi, B^\sharp,q) - \bigg( \Lambda(\mathbf I, B^{\sharp},q) - \Lambda(\Pi, B^\sharp,q) - (\Lambda(\mathbf I, B(\Pi), q) - \Lambda(\Pi, B(\Pi)), q) \bigg)\nonumber \\
&\geq&  \Lambda(\mathbf I, B^{\sharp}, q) - \Lambda(\Pi, B^\sharp, q) -
2 \sum_{i: \Pi(i) \neq i} \sum_{j=1}^m \psi_{cb}^{\sharp} x_{max} \| \bb^{\sharp}_j - \bb_j(\Pi)\| \nonumber \\
&\geq& c (\Lambda(\mathbf I, B^{\sharp},q) - \Lambda(\Pi, B^\sharp,q), \label{gap:compensator:q}.
\end{eqnarray}
Combining \eqref{uniform:concentration:q}, \eqref{diff:compensator1:q} and \eqref{gap:compensator:q}, we have
\[L(\mathbf I, \hat B(\Pi), E) - L(\Pi, \hat B(\Pi), E) \geq c(\Lambda(\mathbf I, B^{\sharp}, q) - \Lambda(\Pi, B^\sharp, q)) > 0\]
with probability going to 1.

\subsubsection{Situation 2}
In situation 2, for any fixed $\Pi$ with $d(\Pi, \Pi^{\sharp}) \geq h_c$, we are going to show the following concentration inequality,
\begin{eqnarray}\label{eq:uniform:large:qq}
& & P( \sup_{B \in \mathcal B_0} \frac{1}{v_{\Pi,B,q}} |\langle L(\Pi, B, E) - \Lambda(\Pi, B,q) \rangle| \geq  x) \nonumber \\
&=& P( \sup_{B \in \mathcal B_g} \frac{1}{v_{\Pi,B,q}} |\langle L(\Pi, B, E) - \Lambda(\Pi, B, q) \rangle| \geq  x / 2) \nonumber \\
& \leq & |\mathcal B_g| \max_{B \in \mathcal B_g} P(\frac{1}{v_{\Pi,B,q}} |\langle L(\Pi, B, E) - \Lambda(\Pi, B, q) \rangle| \geq  x / 2) \nonumber \\
& \leq & |\mathcal B_g| \max_{B \in \mathcal B_g} \exp\{- v_{\Pi,B,q} x^2 / 16\},
\end{eqnarray}
where $\mathcal B_0$ is some restricted parameter space which will be specified later and $\mathcal B_g$ is the $\delta$-covering set of $\mathcal B_0$.

\vspace{0.1in}
\noindent \textbf{On $\mathcal B_0$}~
In this part, we first determined the restricted parameter space $\mathcal B_0$.
First, we know that $\hat B(\Pi)$ is the maximizer of $L(\Pi,B, E)$.
We let $\nabla L (\Pi, B, E) = (\Pi X)^T (E \circ Y) - (\Pi X)^T (E \circ \psi'(\Pi X B))$ and then know that
\begin{eqnarray}
\nabla L (\Pi, \hat B(\Pi), E) = \mathbf 0
\end{eqnarray}
and
\begin{eqnarray}
\nabla L (\Pi, \mathbf 0, E) = (\Pi X)^T (E \circ Y) - (\Pi X)^T (E \circ \psi'(\mathbf 1)).
\end{eqnarray}
By Talyor expansion of $\nabla L (\Pi, B, E)$, we have that
\begin{eqnarray}
\nabla L (\Pi, \mathbf 0, E) = \nabla L (\Pi, \hat B(\Pi), E)  + \nabla^2 L (\Pi, \tilde B(\Pi), E) \hat B(\Pi).
\end{eqnarray}

By the formula that $\nabla^2 L (\Pi, B, E) = \Pi^T X^T \textrm{diag}(E \circ \psi''(\Pi X B)) \Pi X$,
we can easily obtain that
\[\|\hat B(\Pi)\| \leq C \sqrt{p} (\log n) {\color{black} \gamma_{3p}} \psi_{cb}^{\sharp} / (x_{max} \psi_{min}^{''0}) := \delta_{b2}^{'}.\]
Then the restricted parameter space $\mathcal B_0$ is taken as
\[\mathcal B_0 := \{ B ~ | \|B\| \leq \delta_{b2}^{'}\}.\]

\vspace{0.1in}
\noindent \textbf{Upper bound of $v_{\Pi,B,q}$}
We need to estimate the upper bound of $v_{\Pi,B}$ over the restricted parameter space. By the formula of $v_{\Pi,B,q}$, we have that
\begin{eqnarray}\label{upperbound:variance:q}
v_{\Pi,B,q} &=& q \sum_i \psi''(\lambda_i^{\sharp}) (\xx_{\Pi(i)}^T B)^2 +  q(1 - q) \sum_i
(\psi'(\lambda_i^{\sharp}) (\xx_{\Pi(i)}^T B) - \psi(\xx_{\Pi(i)}^T B) )^2 \nonumber \\
&\leq& q n \psi_{max}^{''\sharp} (x_{max} \delta_{b2}')^2 \nonumber \\
& & + 2 q(1-q) (\psi_{max}^{'\sharp} x_{max} \delta_{b2}')^2
+ 2 q(1-q) (\psi(x_{max} \delta_{b2}'))^2 := V_2(q) \nonumber \\
\end{eqnarray}

\noindent \textbf{Lower bound of $v_{\Pi,B,q}$}
We consider to obtain the lower bound of $v_{\Pi,B,q}$ over the restricted parameter space $\mathcal B_0$.
By the formula of $v_{\Pi,B,q}$, we know that
\begin{eqnarray}
v_{\Pi,B,q} =  q \sum_i \psi''(\lambda_i^{\sharp}) (\xx_{\Pi(i)}^T B)^2 + q(1 - q) \sum_i (\psi'(\lambda_i^{\sharp}) (\xx_{\Pi(i)}^T B) - \psi(\xx_{\Pi(i)}^T B))^2 .
\end{eqnarray}
Let $\delta_b := \|B\|$ and we have
Then we will have
\[v_{\Pi,B,q} \geq q c / \gamma_{3p} n \psi_{min}^{''\sharp} \delta_{b}^2.\]
For any $B$ with $\delta_b \leq d_0/ (x_{max} \max\{1,\psi_{max}^{'\sharp}\})$ with $d_0$ satisfying $|\psi(x)| > d_0/2$ for any $|x| < d_0$, it then holds
\[\sum_i (\psi'(\lambda_i^{\sharp}) (\xx_{\Pi(i)}^T B) - \psi(\xx_{\Pi(i)}^T B))^2
\geq n d_0^2 / 4. \]
Therefore, $v_{\Pi,B,q} \geq \min\{c qn \psi_{min}^{''\sharp} \delta_{b1}^2 / \gamma_{3p}, q(1-q) n / 4\} := v_{lb,q}$,
where $\delta_{b1} := d_0 / (x_{max}\max\{1, \psi_{max}^{'\sharp}\})$.

\vspace{0.1in}
\noindent
\textbf{Bound of $|\langle L(\Pi, B, E) - \Lambda(\Pi, B, q) \rangle|$}~

For any $B, B^{'}$ with $\|B - B^{'}\| \leq \delta$, we have that
\begin{eqnarray}
& &|\langle L(\Pi, B', E) - \Lambda(\Pi, B', q) - (\langle L(\Pi, B, E) - \Lambda(\Pi, B, q))| \nonumber \\
&=& \bigg|\sum_{i,l} \bigg\{
E[i,l] (Y[i,l] \llam_{\pi_i l}(B') - \psi(\llam_{\pi_i l}(B')))
-q (\psi'(\llam_{il}^{\sharp}) \llam_{\pi_i l}(B') - \psi(\llam_{\pi_i l}(B'))) \nonumber \\
& & - \big ( E[i,l] (Y[i,l] \llam_{\pi_i l}(B) - \psi(\llam_{\pi_i l}(B)))
-q (\psi'(\llam_{il}^{\sharp}) \llam_{\pi_i l}(B) - \psi(\llam_{\pi_i l}(B))) \big)
\bigg\}
\bigg| \nonumber \\
&\leq& C q m n^2 (\log n) x_{max} \delta \nonumber \\
&\leq& 1/2 v_{\Pi,B,q} x,
\end{eqnarray}
where the last inequality uses the fact that the lower bound of $v_{\Pi,B,q} x$ dominates the term $q m n^2 \log n x_{max} \delta$, when we take $\delta \lesssim \delta_0 := \frac{v_{lb,q} x}{q m n^2 (\log n) x_{max} \delta}$.

Furthermore, the cardinality of $\mathcal B_g$ is bounded by
$(C \frac{p}{\delta})^{m p}$. Thus we can obtain the uniform concentration inequality,
\begin{eqnarray}\label{eq:uniform:large2:qq}
& & P( \max_{\Pi} \sup_{B \in \mathcal B_0} \frac{1}{v_{\Pi,B,q}} |\langle L(\Pi, B, E) - \Lambda(\Pi, B, q) \rangle| \geq  x) \nonumber \\
& \leq & n! (C \frac{p}{\delta})^{m p} \exp\{- v_{\Pi,B,q} x^2 / 16\},
\end{eqnarray}
for $v_{\Pi,B,q} \geq (n \log n + mp \log (p/\delta)) (g(n,p))^2$ by using \eqref{eq:uniform:large:qq}.
Similarly, we also have the second type of concentration inequality,
\begin{eqnarray}
& & P( \max_{\Pi} \sup_{B \in \mathcal B_0} \frac{1}{v_{\Pi,B,q}} |\langle L(\Pi, B, E) - \Lambda(\Pi, B, q) \rangle| \geq  x) \nonumber \\
&\leq& n! (C \frac{p}{\delta})^{m p} \exp\{ v_{\Pi,B,q}/(g(n,p))^2\} \exp\{- v_{\Pi,B,q} x /g(n,p)\}  \nonumber
\end{eqnarray}
for $v_{\Pi,B,q} \leq (n \log n + mp \log (p/\delta)) (g(n,p))^2$.

Then by \eqref{eq:uniform:large2:qq}, we know that
\[L(\Pi, \hat{B}(\Pi), E) \leq
\Lambda(\Pi, \hat{B}(\Pi),q) + O_p(x_1) \leq \Lambda(\Pi,q) + O_p(x_0).\]
We also know that
\[L(I, \hat{B},E) \geq L(I, B^{\sharp},E) \geq
\Lambda(\Pi, B^{\sharp},q) - O_p(x_1) \geq \Lambda(\mathbf I,q) - O_p(x_0).\]
By above facts, we have that
\begin{eqnarray}
& & L(I, \hat{B}, E) - L(\Pi, \hat{B}(\Pi), E) \nonumber \\
&\geq& \Lambda(\mathbf I, q) - \Lambda(\Pi,q) - O_p(x_0), \nonumber
\end{eqnarray}
where $x_0 = \max\{ \sqrt{(n \log n + m p \log (n)) v_{\Pi,B,q}}, (n \log n + mp) (\log n) g(n,p) \}$.
Finally, noting that $g(n,p) = O(\psi_{cb}^{\sharp})$ and $V_{\Pi,B,q} \leq V_2(q)$, then it holds that
$\Delta_2(X, B^{\sharp}, \Pi, \Pi^{\sharp}) \gtrsim x_0$ according to Assumption \eqref{cond:1:q}.
It then implies $\hat \Pi \neq \Pi$ for any $\Pi$ with $d(\mathbf I, \Pi) > h_c$ with high probability.
This completes the proof.

\section{ADMM Computational Approach}\label{sec:comp_mtd}

For completeness,
in this section, we discuss the computational aspects of the problem in linear models.
We relax the ML estimation problem to a bi-convex problem and solve it via an ADMM algorithm
proposed in~\cite{zhang2019permutation}.
A detailed description is given in the sequel.
\par
\paragraph{ADMM formulation}
First, we are trying to solve
\begin{equation}
\label{eq:num_simul_p2_ml}
\min_{\Pi, B}~\|Y - \Pi X  B\|_F^2 =
\|P_{\Pi X}^{\perp} Y\|_F^2
\end{equation}
where projection matrix $P_{\Pi X}^{\perp}$ is defined as
$I - \Pi X \left(X^{T}X\right)^{-1}X^{T}\Pi^{T}$.
Note that we can decompose $Y$ as
$P_{\Pi X}^{\perp} Y + P_{\Pi X} Y$. Since $\|Y\|_F^2 = \|P_{\Pi X}^{\perp} Y\|^2 +
\|P_{\Pi X} Y\|^2$ can be treated as a constant,
minimizing $\|P_{\Pi X}^{\perp} Y\|^2$ is equivalent
to maximizing $\|P_{\Pi X} Y\|^2$.

By introducing two redundant variables $\Pi_1$ and $\Pi_2$, we formulate \eqref{eq:num_simul_p2_ml} as
\begin{equation}\label{eq:num_simul_biconvex}
\min_{\Pi_1,~\Pi_2} -
\text{trace}\left( \Pi_1 P_{X}
\Pi_2^{T} Y Y^{T}\right),~
s.t. \Pi_1 = \Pi_2,
\end{equation}
where $P_{X} := X(X^{T}X)^{-1}X^{T}$.
We propose to solve \eqref{eq:num_simul_biconvex} with
the ADMM Algorithm~\citep{boyd2011distributed} and present the
details of the algorithm in Algorithm~\ref{alg:compt_mtd_admm}.

\begin{algorithm}[h!]
	\caption{ADMM algorithm for the recovery of $\Pi$.}
	\label{alg:compt_mtd_admm}
	\begin{algorithmic}[1]
		\STATE
		{\bfseries Input:}
	 Initial estimate for the permutation matrix
		$\Pi^{(0)}$ and create an $n \times n$ matrix $\mu^{(0)} = \mathbf 0$.
		
		\STATE
		\textbf{For time $t+1$:}
		Update $\Pi_1^{(t+1)}, \Pi_2^{(t+1)}$ as
		\begin{align}
		\label{eq:num_simul_admm}
		\Pi_1^{(t+1)} =&\argmin_{\Pi_1}
		\langle \Pi_1,- Y Y^{T}\Pi^{(t)}_2 P_{X}^{T} +
		\mu^{(t)} - \rho \Pi^{(t)}_2\rangle \notag \\
		\Pi_2^{(t+1)} =&
		\argmin_{\Pi_2}\langle \Pi_2,~Y Y^{T} \Pi^{(t+1)}_1 P_{X}
		- \mu^{(t)} - \rho\Pi^{(t+1)}_1\rangle  \notag \\
		\mu^{(t+1)} =&~ \mu^{(t)} +
		\rho\left(\Pi^{(t+1)}_1 - \Pi_2^{(t+1)}\right).\notag
		\end{align}
		\STATE
		\textbf{Termination}: Stop the ADMM algorithm once $\Pi_1^{(t+1)}$
		is identical to $\Pi_2^{(t+1)}$.
		
	\end{algorithmic}
\end{algorithm}

\newpage

Since ADMM may exhibit slow convergence~\citep{boyd2011distributed}, we adopt a warm start strategy to accelerate the algorithm, which consists of two steps:
\begin{itemize}
    \item Compute the average values $\bar X = \frac{1}{p} \sum_{i=1}^p X[:,i]$.
    \item Obtain a rough estimate $\Pi^{(0)}$ by using Algorithm~\ref{alg:compt_mtd_aver_sort} or~\ref{alg:compt_mtd_eig_sort} with $ X = \bar X$.
\end{itemize}

\begin{algorithm}[h!]
	\caption{Averaging estimator.}
	\begin{algorithmic}[1]
		\STATE
		Compute the average $\frac{1}{m}\sum_{i=1}^m Y[:, i]$.
		\STATE
		Compute $\hat \Pi$ by maximizing
		$\left(\langle m^{-1}\sum_{i=1}^m Y[:, i], \Pi X \rangle \right)^2$.
	\end{algorithmic}
	\label{alg:compt_mtd_aver_sort}
\end{algorithm}

\begin{algorithm}[h!]
	\caption{Eigenvalue estimator.}
	\begin{algorithmic}[1]
		\STATE
		Compute the principal eigenvector $\mathbf u$ of
		$m^{-1}\left(\sum_{i=1}^m Y[:, i] Y[:, i]^{T}\right)$.
		\STATE
		Recover $\hat \Pi$ by maximizing
		$\left(\langle \mathbf u, ~\Pi X \rangle \right)^2$.
	\end{algorithmic}\label{alg:compt_mtd_eig_sort}
\end{algorithm}

\end{document}